\documentclass[prd,aps,
nofootinbib,
floatfix,
superscriptaddress]{revtex4}
\usepackage{graphicx}
\usepackage{epsfig}
\usepackage{rotating}
\usepackage{amssymb}
\usepackage{dsfont}
\usepackage{psfrag}
\usepackage{amsmath,euscript,array,mathrsfs}
\usepackage{epstopdf}
\usepackage[latin1]{inputenc}
\newcommand{\EQ}[1]{\begin{equation} #1 \end{equation}}

\newcommand{\SP}[1]{\begin{equation}\begin{split} #1
\end{split}\end{equation}}

\newcommand{\beq}{\begin{equation}}
\newcommand{\eeq}{\end{equation}}
\newcommand{\beqs}{\begin{eqnarray}}
\newcommand{\eeqs}{\end{eqnarray}}

\newcommand{\gsim}{\mathrel{\raisebox{-
.6ex}{$\stackrel{\textstyle>}{\sim}$}}}

\def\hbar{\hspace{0pt}\raisebox{1pt}{$-$} \hspace{-7pt} h}

\def\di{\mbox{d}}
\def\r{\rho}

\newcommand{\be}{\begin{equation}}
\newcommand{\ee}{\end{equation}}
\newcommand{\bea}{\begin{eqnarray}}
\newcommand{\eea}{\end{eqnarray}}

\def\lbldef#1#2{\expandafter\gdef\csname #1\endcsname {#2}}

\def\href#1#2{#2}

\newcommand{\ber}{\begin{eqnarray}}
\newcommand{\eer}{\end{eqnarray}}

\newcommand{\beqar}{\begin{eqnarray}}

\newcommand{\eeqar}{\end{eqnarray}}

\newcommand{\dsl}
  {\kern.06em\hbox{\raise.15ex\hbox{$/$}\kern-.56em\hbox{$\partial$}}}

\newcommand{\eeqarr}{\end{eqnarray}}
\newcommand{\ZZ}{{\rm \kern 0.275em Z \kern -0.92em Z}\;}
\def\CC{{\mathchoice
{\rm C\mkern-8mu\vrule height1.45ex depth-.05ex
width.05em\mkern9mu\kern-.05em}
{\rm C\mkern-8mu\vrule height1.45ex depth-.05ex
width.05em\mkern9mu\kern-.05em}
{\rm C\mkern-8mu\vrule height1ex depth-.07ex
width.035em\mkern9mu\kern-.035em}
{\rm C\mkern-8mu\vrule height.65ex depth-.1ex
width.025em\mkern8mu\kern-.025em}}}

\def\RR{{\rm I\kern-1.6pt {\rm R}}}

\def\ZZ{{\rm Z}\kern-3.8pt {\rm Z} \kern2pt}
\def\IB{\relax{\rm I\kern-.18em B}}
\def\ID{\relax{\rm I\kern-.18em D}}
\def\II{\relax{\rm I\kern-.18em I}}
\def\IP{\relax{\rm I\kern-.18em P}}

\newcommand{\bear}{\begin{eqnarray}}
\newcommand{\eear}{\end{eqnarray}}

\def\arctanh{~{\rm arctanh}~}

\def\r{\rho}                                     %     \varrho
\def\6{\partial}

\def\bea{\begin{eqnarray}}
\def\eea{\end{eqnarray}}

\def\beqx{\begin{displaymath}}
\def\eeqx{\end{displaymath}}

\newcommand{\bmat}{\left(\begin{array}}
\newcommand{\emat}{\end{array}\right)}

\def\r{\rho}

\def\bo{{\raise-.3ex\hbox{\large$\Box$}}}               % D'Alembertian
\def\face{{\raise.2ex\hbox{$\displaystyle \bigodot$}\mskip-2.2mu \llap {$\ddot
        \smile$}}}                                   % happy face
\def\>{\rangle}                                      %right angle
\def\<{\langle}                                      %left angle

\def\leftrightarrowfill{$\mathsurround=0pt \mathord\leftarrow \mkern-6mu
        \cleaders\hbox{$\mkern-2mu \mathord- \mkern-2mu$}\hfill
        \mkern-6mu \mathord\rightarrow$}        % <--> double differential
\def\dvec#1{\vbox{\ialign{##\crcr
        \leftrightarrowfill\crcr\noalign{\kern-1pt\nointerlineskip}
        $\hfil\displaystyle{#1}\hfil$\crcr}}}           % <--> accent
\def\-{\hphantom{-}}

\begin{document}
\title{Light scalars from a compact fifth dimension}

\author{Daniel Elander}
\author{Maurizio Piai}
\affiliation{Swansea University, School of Physical Sciences, \\ Singleton Park, Swansea, Wales, UK}

\date{\today}

%\vspace{6mm}

\begin{abstract}
We consider a general five-dimensional sigma-model coupled to gravity, with any number of scalars and general sigma-model metric and potential. We discuss in detail the problem of the boundary conditions for the scalar fluctuations, in the case where the fifth dimension is compact, and provide a simple (and very general) algorithmic procedure for computing the spectrum of physical scalar fluctuations of the fully back-reacted system. Focusing in particular on the conditions under which the spectrum of scalar excitations (glueballs) contains parametrically  light states, we apply the formalism to some especially simple toy models, which can be  thought of as the gauge/gravity duals of strongly-coupled, non-conformal four-dimensional gauge theories. Our examples are chosen both within the context of phenomenological effective field theory constructions (bottom-up approach), and within the context of  consistent truncations of ten-dimensional string theories in the supergravity limit (top-down approach). 
In one of the examples, a light dilaton is present in the spectrum in spite of the presence of a bad naked singularity
in the deep IR, near which the RG flow of the dual theory is certainly very far away from any fixed point.
If this feature were to persist in a complete model in which the singularity is resolved, this would prove
that a light dilaton is to be expected in at least  certain walking technicolor theories.
We provide here all the technical details for testing this statement, once such a complete model is identified.

\end{abstract}

%\pacs{11.25.Tq, 12.60.Nz.}

\maketitle

\tableofcontents

%%%%%%%%%%%%%%%%%%%%%%%%%%%%%%%%
%%%%%%%%%%%%%%%%%%%%%%%%%%%%%%%%
%\newpage
\section{Introduction}

Many physical systems are  described by  strongly-coupled  field theories,
the dynamics of which is encoded in the fixed points of their renormalization group flow.
Long-distance properties of such systems can be classified in terms of universal coefficients,
which depend only on the properties of the system very close to the fixed points,
but do not depend on the model-dependent features of the flows away from such fixed points.
On the other hand, there are many cases in which the knowledge of the physics at the fixed points 
does not provide  enough information as to allow one to compute phenomenologically important physical quantities that are 
experimentally measurable. In these latter cases,
 traditional field theory techniques are not powerful enough to yield robust predictions,
 which can be compared to the experimental data, mainly because of
 the strongly-coupled nature of the underlying dynamics.
 
 One such example emerges in the context of dynamical electro-weak symmetry breaking,
in particular in walking technicolor~\cite{WTC}. In this case, the underlying strong dynamics is quasi-conformal
 (approaching an IR fixed-point) over a range of energies above the electro-weak scale,
 but ultimately yields to confinement and to the formation of symmetry-breaking condensates
at the electro-weak scale.
One might expect that there exists a sense in which the condensates break spontaneously the (approximate) dilatation symmetry
of the system near the (approximate) fixed point, thus leading to the appearance of 
a light dilaton (the pseudo-Goldstone boson of scale-invariance) in the spectrum of
composite states.
From a phenomenological point of view, this example is of most urgent importance, 
because  such a light dilaton 
might mimic the properties
 of the Higgs particle of weakly-coupled 
models such as the minimal version of the Standard Model~\cite{dilatonpheno}.
Unfortunately, because of the strong dynamics, and because the physics of a massive state such as
the dilaton depends not only on the (universal) properties of the fixed-point, but also on the (non-universal)
RG flow that yields confinement and chiral symmetry breaking itself, 
neither a firm confirmation nor a disproof of the existence of a light dilaton in walking technicolor
 has  been provided
by conventional field-theory methods, in spite of many attempts~\cite{dilatonWTC}. For recent work supporting the idea that such a light dilaton exists, see for instance~\cite{dilatonWTCrecent,ENP}.

In recent years, the discovery of gauge/gravity dualities provided a new tool, that
allows to reformulate field-theory problems emerging within four-dimensional strongly-coupled systems
in terms of weakly-coupled extra-dimensional systems~\cite{AdSCFTreviews}.
In its original formulation~{\cite{AdSCFT}, the idea is to relate a particularly simple and symmetric
10-dimensional background ($AdS_5\times S^5$) to a very special conformal four-dimensional theory
(${\cal N} = 4$  supersymmetric $SU(N)$ gauge theory), in the sense that a prescription 
is given for computing the generating functionals of  correlation functions
on the two sides of the correspondence, and the physical results agree.
More recent developments provided large classes of dual models 
that correspond to non-conformal field theories with much less supersymmetry,
such as those in which the 10-dimensional background is constructed starting from
the conifold and its variations~\cite{conifolds}, and those that are related to controllable deformations
of the ${\cal N}=4$ field theory~\cite{deformingN4}.

It turns out that, for several reasons, all the models that are of phenomenological interest
share some very general properties. In particular,  the 10-dimensional metric is always written  in terms of a 
non-compact five-dimensional part (four directions of which are directly related to the dual four-dimensional space,
with the fifth dimension related to the energy scale at which the dual theory is tested),
and a compact (internal) five-dimensional space, the isometries of which are related to the internal global symmetries of
the field theory.
Formally, this means that it is often possible, and very convenient,
to study the 10-dimensional dynamics by first studying its five-dimensional reduction.
One also implements a consistent truncation that reduces the number (and simplifies the action)
of the resulting active fields.

The problem is thus reduced to finding a background solution of the classical equations of the truncated five-dimensional theory, and then use it to construct the
lift to a background solution of the full 10-dimensional system.
In practice, this means that what one has to solve, in order to determine the background,
are the classical equations of a five-dimensional sigma-model of $n$ scalars coupled to gravity.
Since one usually wants to preserve the Lorentz structure of the four-dimensional space-time,
one also assumes that the background depends only on the radial direction.
With all of this, the complexity of the original problem of finding fully back-reacted 10-dimensional backgrounds
is turned into the more  treatable problem of solving a one-dimensional classical system. 

One can even use this formalism dispensing with
the original problem of finding a consistent truncation, by simply writing a five-dimensional 
phenomenological model  that captures  the most important features
of the dynamics, postponing the problem of its completion (all we are going to say applies also in the fake supergravity context~\cite{Freedman:2003ax}). 
In this spirit, a vast literature of phenomenological models exists which aim at 
capturing  the most important aspects of the gauge/gravity dualities without 
dealing with the technical difficulties
of a complete string-theory construction.
Relevant examples include the Randall-Sundrum model~\cite{RS}, in which the sigma-model consists
just of a cosmological constant, the Goldberger-Wise mechanism~\cite{GW} (see also~\cite{GW2}), 
in which the model contains
only one scalar  (see also~\cite{RSreview}), and many applications, 
such as the Higgsless models~\cite{Higgsless}, the AdS/QCD models~\cite{AdSQCD},
some composite-Higgs models~\cite{compositeHiggs}
and the holographic technicolor models~\cite{AdSTC}.
Recently, a similar strategy has been proposed also in order to study lower-dimensional 
 condensed matter systems (see~\cite{super} for an introduction to the subject).

After reinstating (in the five-dimensional action) 
the dependence on the Minkowski directions, 
one can study the spectrum of classical fluctuations, which can be interpreted as
the composite states of the dual theory (the glueballs, for instance).
From this, one can finally access those very non-trivial properties of the 
strong dynamics that we started by describing in the beginning of this introduction, and ask hard questions such as whether a specific model yields a light dilaton in the spectrum.
Yet, there are still two difficulties to overcome, before a final answer to these questions can be provided.

First of all, because the spectrum of massive states is not a universal property,
one has to construct and study a variety of explicit models, possibly such that a lift to a complete 10-dimensional 
theory exists, and such that the dual field theory has all the properties required by phenomenology.
In the specific context of walking technicolor, this program has recently been initiated, with some very
encouraging results~\cite{NPP,NPR,ENP} (see also~\cite{P} for a more extensive discussion of what
one might want to achieve along this line).
Yet, at present we are nowhere close to having constructed the actual 10-dimensional dual of a 
phenomenologically relevant four-dimensional 
model, as a substantial amount of  model-building is required in order to do so.

A second difficulty is of a more technical nature, and is the main subject of the present paper. When studying the spectrum and the properties of the fluctuations, a long preliminary work appears to be necessary (see for example~\cite{GW3}) --- mainly because of the non-trivial mixing between fluctuations of the bulk scalars and the five-dimensional metric. One hence would need a formalism that is general and simple enough to correctly incorporate the relevant dynamics without having to analyze all the details on a model-by-model basis. In the case of a single scalar with trivial sigma model, this program has been addressed time ago by several collaborations  (see for example~\cite{Kofman:2004tk}). Yet, the formalism developed by these authors needs to be extended far beyond the level needed for a single scalar, in order for it to apply to a realistic gravity dual of a strong dynamics. 

In~\cite{BHM}, a step towards a systematic resolution of this technical problem 
was taken. 
By introducing appropriate gauge-invariant combinations of the original 
fluctuations  (see also~\cite{Giovannini:2001fh} and~\cite{Kofman:2004tk}), 
which from the four-dimensional point of view correspond to scalars, vectors and tensors,
it was shown that, given a completely general (two-derivatives) 
sigma-model with $n$ scalars, and with a superpotential $W$ (from which the 5d potential $V$ can be derived), it is possible
to algebraically manipulate the system of linear differential equations so as to rewrite it in a sigma-model covariant form
and reduce it to a set of $n$ second-order equations for the same number of physical fluctuations, from which the spectrum of the spin-0 sector of the theory can be derived.
It was also shown in~\cite{Elander:2009bm} that the formalism can easily be generalized to the case
where the superpotential $W$ is not known (or does not exist), 
in which case one needs to know the sigma-model metric and the potential.

This important formal tool works thanks to the fact that one can use the five-dimensional diffeomorphism 
invariance in order to remove some of the unphysical fluctuations. However, in practical applications one has 
to generalize this instrument further, so that it applies to the case where the radial direction is not
infinite and, hence, boundary actions may need to be added. 
For example, in many cases an IR boundary is present because of an end-of-space in the geometry
(which must be the case when discussing the dual of a confining gauge theory).
Also, in the UV it is often necessary to work with a finite cutoff, for three possible generic reasons. 
 It might be known that the dual field theory requires a UV completion above a given scale,
 so that the UV cutoff is actually physical.
Retaining a UV cutoff is also necessary for technical reasons related to holographic renormalization~\cite{HR}.
And finally, it may be that the strongly-interacting dual theory is (weakly) coupled with 
an external (weakly-coupled) four-dimensional sector, modeled by UV-boundary interactions.
In all these three cases, the boundary terms are going to break the five-dimensional diffeomorphism invariance, and hence
some caution has to be used when applying the gauge-invariant formalism.

In this paper, we provide the general form of the boundary conditions for the scalar fluctuations,  
both in the case in which the five-dimensional dynamics is known in terms of a superpotential,
and when only the potential exists, without any restriction on the number of sigma-model scalars
or on the sigma-model metric.
We discuss the residual freedom in the form of the boundary conditions, in particular in relation to the light scalars in the spectrum, one
 (linear combination) 
of which may be interpreted as a dilaton,
while any others correspond either to ordinary pseudo-Nambu-Goldstone bosons of 
approximate global symmetries of the dual theory, or are the result of accidental cancellations.
We illustrate the formalism hence derived by applying it to a set of simple phenomenological examples,
for which we provide both (approximate) analytical results for the spectrum, and
(exact) numerical studies.

We do not treat here the problem of holographic renormalization, which we postpone to future work.
In particular, we will always consider the models as defined on a compact fifth-dimension, away from any possible
singularities. In this way, all the states are going to be physical. In practical examples, one has also to 
decide how to couple the model to possible external (weakly-coupled) systems, and how to take the
limits in which the IR and UV boundaries are removed. We will briefly comment 
on these issues in due time, but we are not going to provide a systematic prescription for doing so.

\subsection{The algorithm}

The main purpose of the paper is to provide the reader with a simple algorithmic procedure
for computing the spectrum of scalar excitations.
Assuming that a five-dimensional model is of interest (irrespective of the fact that it is
built either as a phenomenological model, or as the consistent truncation of a given 
supergravity, or superstring, or M-theory), one should go through the following steps.

\begin{itemize}

\item One must first ensure that 
the model can be written in the general form we provide, i.~e. as a two-derivative, 5-dimensional action involving 
 $n$ real scalars coupled to gravity. We do not consider the case where higher-derivative terms are
 present, and ignore the possibility that higher-spin fields (such as gauge bosons or fermions, for instance)
 are relevant in determining the background.
 
\item One must assume that two boundaries are present in the fifth dimension, representing the UV and IR cutoffs of the dual theory. 
These cutoffs may have a physical meaning, in which case all the results will explicitly depend on the dual scales.
Or they may be thought of as regulators, in which case the physical results should be  obtained by extrapolating the final results of the calculation
to the actual physical case (typically, one wants to remove the UV cutoff completely, while the IR cutoff will approach the end-of-space).
The presence of the boundaries means that boundary actions must be added. We provide the most general form of such boundary actions, subject to the 
limitation that we do not include terms that depend on the Minkowski four-momentum $q^2$.
As we will see, this form is very constrained, although we restrict ourselves to the quadratic order. 

\item One has to solve the system of background equations, and find a suitable background,
with the ansatz that all the background functions depend only on the radial direction.
 We provide the general form of the bulk equations and their boundary conditions derived from the complete action.
We do so both in the case in which the sigma-model is described in terms of a potential, but also in the case in which
a superpotential description is known. The latter has the advantage that the background is completely determined by a set of first-order 
(coupled and non-linear) differential equations.

\item One has to solve the linearized, second-order equations for the fluctuations around the background.
The spectrum is determined by solutions that satisfy the boundary conditions both in the UV and in the IR. 
We provide the complete set of bulk equations~\cite{BHM}, directly in the physical basis, written in terms of the background solution. And we provide the  boundary conditions,
the form of which depends again on the background functions, but also 
on a set of parameters that incorporate the residual freedom in the choice 
of boundary actions. These parameters should be chosen on the basis of physical
principles, and are hence model-dependent. 
However, in the absence of symmetry reasons, there is only one such choice 
that ensures the absence of any fine-tuning in the physical results,
and in this limit the boundary conditions depend, again, only on the functions determining the background,
evaluated at the boundaries.

 \end{itemize}
 
This procedure is very general, and since we write the fluctuation equations and boundary conditions
already in terms of physical fields, no algebraic manipulations are needed. 
The reader who wants to apply this procedure can directly write the
final system in terms of  $n$ gauge-invariant fluctuations 
and solve it  (in most of the cases, this must be done numerically).
 The formalism of~\cite{BHM} allows to write all the relevant equations in
 an elegant form that is fully covariant with respect not only to the space-time,
 but also the internal sigma-model geometry.
One  should be careful in correctly using all the covariant derivatives,
 which are determined both by the space-time and sigma-model connections: we will 
 present all the relevant (heavy) notation in Section~\ref{Sec:2}.

 Again, we must stress that the generality of the boundary terms  for the fluctuations 
 is limited  by the fact that we do not
 include $q^2$-dependent terms. These are model-dependent, and important in the
 context of holographic renormalization, when trying to take the UV cutoff to infinity,
  a problem that we postpone to future work.

 \subsection{Reader's guide}

The paper is organized as follows.
In Section~\ref{Sec:2} we summarize the basic formalism we use. This is a rather technical section, which may be 
skipped at first reading. However, all the material contained is necessary in order to
correctly interpret and use the basic equations appearing in the paper,
and we find it convenient to group together all the necessary
 definitions in one place.
 Also, we discuss here some subtleties emerging in the introduction of gauge-invariant variables
 in the presence of boundaries, which clarify  and complete the literature on the subject.

In Section~\ref{Sec:3}
we write the action of the sigma-model coupled to gravity, the equations determining the background, 
and the final differential equations and boundary conditions satisfied by the physical fluctuations.
The derivation of these results is summarized in the appendices.
All the relevant equations are
written in terms of the sigma-model metric, the background fields, 
and the potential (and, when available, superpotential).

In Section~\ref{Sec:4} we present three particularly interesting  examples, and derive some analytical results.
In Section~\ref{Sec:5} we apply the mid-point determinant method~\cite{BHM} to study these examples numerically.
The main purpose of the examples is to illustrate the procedure, and hence we choose them to be particularly simple.
However, it turns out that their physical interpretations are quite interesting,
and that this set of exercises also provides some important insight 
into how the regulation procedure may work in non-trivial physical cases.

The first example is based on the same action used in the GW mechanism, and allows us to compare 
our results to the literature, but also to generalize the results and discuss many interesting subtleties that
have been ignored in the past.
In particular, we explicitly show that some freedom in the definition of the boundary terms 
results in the possible appearance of additional light scalars besides the dilaton, and that hence one has to exercise some
caution in interpreting the results.
We also show that a light dilaton is present (in great generality)
not only when the scaling dimension $\Delta$ of the field-theory deformation encoded in the background
is small ($\Delta \ll 1$), but also for any $\Delta\gsim 2$.

The second example is taken from a peculiarly simple 
five-dimensional model constructed by consistent truncation of type-IIB supergravity.
The dual gauge theory has many properties that resemble those of a QCD-like theory, 
in the sense that the formation of a condensate in the IR
takes the theory away from its fixed-point, presumably leading to confinement.
Unfortunately, the model (studied here at zero temperature) 
suffers from the appearance of a naked singularity in the background,
which limits its physical meaning.
Yet, it is  interesting to study what happens to the  spectrum in the limit where
 the IR cutoff approaches the singularity. As we will see, the procedure adopted here
 yields a spectrum that, while distorted by the presence of the singularity,
 does not show any signs of pathologies, suggesting that the procedure that we follow removes some of the unpleasant features of the background at the singularity.

The third example is a phenomenological model yielding a background that can be 
interpreted in terms of the RG flow between a UV fixed-point and an IR fixed point.
We study in some detail what happens to the spectrum by comparing several backgrounds
that differ only by the value of the scale at which the transition from the proximity to
one fixed point to the other takes place.

In Section~\ref{Sec:6} we present a set of field-theory arguments aimed at explaining 
the results of the previous two sections. We elaborate on possible interpretations
of the five-dimensional models in terms of dual, strongly-coupled theories,
and derive some lessons about the physics of the three models
we considered. These lessons  extend to any model the examples somehow approximate,
and are hence of general interest.

In Section~\ref{Sec:7} we conclude, by summarizing the most important equations needed
in the proposed algorithmic procedure, by commenting on the limitations and subtleties
involved in using the algorithm itself, by summarizing briefly the
physics lessons we learned, and finally by outlining some possible future applications.

%%%%%%%%%%%%%%%%%%%%%%%%%%%%%%%%
%%%%%%%%%%%%%%%%%%%%%%%%%%%%%%%%
%\newpage
\section{Formalism}
\label{Sec:2}

We introduce here the main definitions and conventions we use in the paper.
We do not make explicit use of supergravity transformation and other supersymmetric properties. We start from the definition of the geometric properties of a sigma-model of $n$ scalars
coupled to gravity in five dimensions.
Most of the notation and conventions we use are taken from~\cite{BHM}.
The notation is somewhat heavy, and hence we devote some time to explain it,
explicitly showing all the definitions used in the whole paper.
We then explicitly discuss the effect of gauge transformations and introduce the
gauge invariant variables that will be used throughout the paper.

\subsection{Geometry}

All the equations we write make use of the geometric properties
of the sigma-model, and are hence completely covariant.
The space-time, sigma-model and background covariant derivatives are 
written in terms of the space-time and sigma-model metric 
and of the background fields. 

We use the following conventions. 
Capital roman indices $M=0,1,2,3,5$ are five-dimensional space-time indexes,
while greek indexes $\mu=0,1,2,3$ are restricted to the 4-dimensional Minkowski slices of the space.
In this way, we label the space-time coordinates as $x^M=(x^{\mu},r)$, with $r$ the radial (fifth) direction.
Lower-case roman indexes  $a=1,\ldots,n$ refer to the sigma-model (internal) space.

We write the five-dimensional metric $g_{MN}$ with signature $-++++$. We define the five-dimensional connection as
\beqs
\Gamma^{P}_{\,\,MN}&\equiv&\frac{1}{2}g^{PQ}\left(\frac{}{}\partial_M g_{QN}+\partial_N g_{QM}-\partial_Q g_{MN}\right)\,,
\eeqs
and hence the covariant derivatives are of the form
\beqs
\nabla_M T^{P}_{\,\,N}&\equiv&\partial_M T^{P}_{\,\,N}+\Gamma^{P}_{\,\,MQ}T^{Q}_{\,\,N}-\Gamma^{Q}_{\,\,MN}T^{P}_{\,\,Q}\,,
\eeqs
for a $(1,1)$-tensor, and analogous for other tensors, in such a way as to ensure compatibility with $\nabla_P g_{MN}=0$.
The Riemann tensor, Ricci tensor and Ricci scalars are defined, respectively, as
\beqs
R_{MRN}^{\,\,\,\,\,\,\,\,\,\,\,\,\,\,\,\,P}&\equiv&\partial_R\Gamma^{P}_{\,\,MN}-\partial_M\Gamma^{P}_{\,\,RN}
+\Gamma^{Q}_{\,\,MN}\Gamma^{P}_{\,\,RQ}-\Gamma^{Q}_{\,\,MR}\Gamma^{P}_{\,\,NQ}\,,\\
R_{MN}&=&R_{MRN}^{\,\,\,\,\,\,\,\,\,\,\,\,\,\,\,\,R}\,,\\
R&\equiv&g^{MN}R_{MN}\,.
\eeqs

One important fact that we will use in this paper is that we will assume the space-time
to have one compact dimension. It is convenient to choose the coordinates in such a way that
the radial direction $r_1<r<r_2$ is compact, with the slices of space-time with constant $r$ 
supporting a Minkowski four-dimensional metric. 
The presence of four-dimensional boundaries means that 
we will need to use the induced four-dimensional analogs of all of the above geometric objects,
which we will label  as $\tilde{g}_{\mu\nu}$, $\tilde{\nabla}_{\mu}$ and so on.
The boundary terms are built starting from the ortho-normalized vector $N^{M}$, defined so that
\beqs
g^{MN}N_NN_M&=&1\,,\\
\tilde{g}_{MN}N^{N}&=&0\,,
\eeqs
which implies that 
$
\tilde{g}_{MN}=g_{MN}-N_MN_N$. The extrinsic curvature is defined from 
\beqs
K_{MN}&\equiv&\nabla_MN_N\,,
\eeqs
as the contraction with the bulk metric
\beqs
K&=&g^{MN}K_{MN}\,.
\eeqs
The boundary actions will contain terms proportional to $K$.

The field content of the 5-dimension action comprises
a set of real scalar fields that we label as $\Phi^a$ with $a=1,\ldots,n$.
In a way that is analogous to the space-time metric,
we indicate with $G_{ab}$  the sigma-model metric, which
ultimately encodes the geometric properties of the internal space spanned
by the scalars. In the case where the five-dimensional system is obtained by consistently truncating 
some higher-dimensional theory, the structure of the sigma-model is determined 
unambiguously by the fact that the scalars parameterize some coset space, which is in general non-compact,
and which emerges from the fact that the compactification of the internal space results in the breaking of some global symmetry
of the underlying theory. We will keep the sigma-model structure as general as possible, hence
not committing ourselves to any specific realization of this structure.

Sigma-model indexes are lowered and raised by the sigma-model metric $G_{ab}$ and its inverse $G^{ab}$
defined by $G^{ab}G_{bc}=\delta^a_{\,\,\,c}$.
When unambiguous, we use lower indices to denote field derivatives with respect to the $\Phi^a$, so that for example 
given a scalar function $V$ we define
\beqs
V_a &\equiv& \partial_a V \,\equiv\, \partial V /\partial \Phi^a\,.
\eeqs
The sigma model connection is given by
\beqs
{\cal G}^d_{\,\,ab}&\equiv&\frac{1}{2}G^{dc}\left(\frac{}{}\partial_aG_{cb}+\partial_bG_{ca}-\partial_cG_{ab}\frac{}{} \right)\,,
\eeqs
and the Riemann tensor with respect to the non-linear sigma-model metric is given by
\EQ{
	\mathcal R^a_{\ bcd} \equiv \partial_c \mathcal G^a_{\ bd} - \partial_d \mathcal G^a_{\ bc} + \mathcal G^a_{\ ce} \mathcal G^e_{\ bd} - \mathcal G^a_{\ de} \mathcal G^e_{\ bc}.
}

Using the sigma-model connection, we define the sigma-model covariant derivative.
It is convenient to introduce also another convention. 
When a sigma-model index is placed after a ``$|$'', it means that the
sigma-model covariant derivative with respect to $G_{ab}$ should be taken, 
which is defined as acting on a $(1,1)$ sigma-model tensor $X^{d}_{\,\,\,a}$ by
\EQ{
		X^d_{\,\,\,a|b} \equiv D_b X^d_{\,\,\,a} \equiv \partial_b X^d_{\,\,\,a} + \mathcal{G}^d_{\ cb} X^c_{\,\,\,a} - \mathcal{G}^c_{\ ab} X^d_{\,\,\,c}\,,
}
and analogous expressions for other tensors.

Finally, one needs to rewrite in a covariant form those objects that have indexes both on the space-time and on the sigma model,
making them into generalized tensors.
However, we do not need to write the general form of the covariant derivative for this case, because our theory is written only in terms of space-time scalars (that carry sigma-model indexes) and 
the metric (which is a sigma-model scalar), so that the only object we actually need is
the background-covariant derivative ${\cal D}_M$ defined for a five-dimensional scalar, 
sigma-model $(1,0)$-tensor $ \mathfrak a^a$, via
\EQ{
		{\cal D}_M \mathfrak a^a \equiv \partial_M \mathfrak a^a + \mathcal{G}^a_{\ bc} \partial_M\bar{\Phi}^{\,b} \mathfrak a^c,
}
where $\bar{\Phi}^{\,b}$ means that $\Phi^{\,b}$ is evaluated on the classical background
(as is ${\cal G}={\cal G}(\bar{\Phi}^a)$). Notice that in the following we will assume the background 
functions to depend only on the radial direction $r$, and hence only the fifth component of the
background-covariant derivative has a non-trivial connection contribution, while the other components reduce to
ordinary derivatives. 

\subsection{ADM formalism}
We derive all the relevant equations using the ADM formalism, the basic idea being that we will rewrite the metric $g_{MN}$ and the scalars $\Phi^a$
as a background function plus general fluctuations, and then decompose the metric in terms of 
four-dimensional tensors by slicing the space-time along the radial direction. 
We start by writing the metric in the form
\SP{
	g_{MN} =
	\left(
	\begin{array}{ll}
	 	\tilde g_{\mu\nu} & \nu_\nu \\
	 	\nu_\mu & \nu_\mu \nu^\mu + (1+\nu)^2
	\end{array}
\right).
}
Because we singled out the radial direction as orthogonal to the boundaries,
the normal vector is defined by $N_M = (0,(1+\nu))$ and $N^M = (1+\nu)^{-1} (-\nu_\mu,1)$,
so that $\tilde{g}^{MN}={\rm diag}\{\tilde{g}^{\mu\nu},0\}$ (notice that this tensor is not the inverse of $\tilde{g}_{MN}$).

We assume that the background metric satisfies the ansatz
\beqs
\label{Eq:bgmetric}
\di s^2&=&e^{2A}\eta_{\mu\nu}\di x^{\mu}\di x^{\nu}+\di r^2\,,
\eeqs
with $A=A(r)$. Similarly for the scalar, we assume that $\langle \Phi^a \rangle=\bar{\Phi}^a(r)$, so that the background
depends on the radial direction $r$, but not on $x^{\mu}$.
We fluctuate the whole system by expanding the scalars as (using the exponential map)
\SP{\label{eq:exponentialmap}
	\Phi^a = \rm{exp}_{\bar{\Phi}} (\varphi)^a \equiv \bar{\Phi}^a + \varphi^a - \frac{1}{2} \mathcal G^a_{\ bc} \varphi^b \varphi^c + \ldots,
}
and the metric (to first order in the fluctuations) as
\SP{
	\tilde{g}_{\mu\nu} &= e^{2A} (\eta_{\mu\nu} + h_{\mu\nu}), }
with
\SP{
	h^\mu_{\,\,\,\nu} = {h^{TT}}^\mu_{\,\,\,\nu} + \partial^\mu \epsilon_\nu + \partial_\nu \epsilon^\mu + \frac{\partial^\mu \partial_\nu}{\Box} H + \frac{1}{3} \delta^\mu_{\,\,\,\nu} h,
}
where ${h^{TT}}^\mu_{\,\,\,\nu}$ is traceless and transverse, and $\epsilon^\mu$ is transverse. 
Altogether, we have the fluctuation variables $\{ \varphi^a, \nu, \nu^\mu, {h^{TT}}^\mu_{\,\,\,\nu}, h, H, \epsilon^\mu \}$.\footnote{Notice that, 
with some abuse of notation, we identify the fluctuations $\nu$ and $\nu^{\mu}$ with the components of the metric in the ADM formalism.}

\subsection{Diffeomorphism invariance}

Here we write explicitly the gauge transformations,
and discuss the fact that the boundary actions restrict their general form compared to what is allowed in the bulk.
The starting point is the five-dimensional diffeomorphisms
\beqs
\delta x^{M}&=&-\xi^{M}\,,
\eeqs
which imply 
\beqs
\delta \Phi^a&=&\xi^{M}\partial_{M}\Phi^a\,,\\
\delta g_{MN}&=&\partial_M\xi^{R}g_{RN}+\partial_N\xi^{R}g_{MR}+\xi^{R}\partial_{R}g_{MN}\,.
\eeqs
To first-order in the fluctuations, this yields the gauge transformations for all the fluctuations
\SP{
	\delta \varphi^a &= \bar{\Phi}^{\prime\,a} \xi^r, \
	\delta \nu = \partial_r \xi^r, \
	\delta \nu^\mu = \partial^\mu \xi^r\, +\, e^{2A}\partial_r\xi^{\mu}, \
	\delta {h^{TT}}^\mu_\nu = 0, \\
	\delta \epsilon^\mu &= \Pi^\mu_{\,\,\,\nu} \xi^\nu, \
	\delta H = 2 \partial_\mu \xi^\mu, \
	\delta h = 6A' \xi^r,
}
where we defined the projector $\Pi^\mu_{\,\,\,\nu}\equiv\delta^\mu_{\,\,\,\nu}- \frac{\partial^\mu \partial_\nu}{\Box} $.
In all of this, all the functions depend on the five coordinates $x^M$.
The five-dimensional part of the action is going to be invariant under all of these transformations.

However, we do have four-dimensional boundary actions, the very existence of which is not
compatible with all of the above.
To be more specific, the action is still symmetric under these transformations for generic $\xi^{\mu}(x^{\mu},r)$,
but we have to specify how to treat the diffeomorphisms in the fifth direction.
Because we will explicitly write the boundaries to support localized actions at the points $r=r_i$, with $i=1,2$,
one must require\footnote{This expression could be made covariant, if one wanted to manifestly show that one does not need to commit to a specific choice of coordinates.}
\beqs
\label{Eq:rest}
\xi^{r}(x^{\mu},r_i)&=&0\,.
\eeqs
This observation plays an important role
in the subsequent discussion about gauge-invariance and gauge-fixing.

\subsection{Gauge-invariant formalism}
\label{Sec:2d}

Generalizing the notation of~\cite{BHM}, we define the following  variables 
\SP{
\label{eq:gaugeinvariantvariables}
	\mathfrak a^a &= \varphi^a - \frac{\bar{\Phi}^{\prime\,a}}{6 A'} h, \\
	\mathfrak b &= \nu - \frac{\partial_r (h/A')}{6}, \\
	\mathfrak c &= e^{-2A} \partial_\mu \nu^\mu - \frac{e^{-2A} \Box h}{6 A'} - \frac{1}{2} \partial_r H, \\
	\mathfrak d^{\mu} &= e^{-2A} \Pi^\mu_{\,\,\,\nu} \nu^\nu - \partial_r \epsilon^\mu, \\
	\mathfrak e^{\mu}_{\,\,\,\nu} &= {h^{TT}}^\mu_\nu.
}
These are a generalization of the Mukanov-Sasaki variable~\cite{MS}.
By inspection, one can verify that these new variables are 5d gauge invariant.

Before proceeding, let us go through a counting exercise.
Besides the $n$ scalars in the sigma-model, the off-shell degrees of freedom derived from 
the dimensional reduction of the five-dimensional fluctuations of $g_{MN}$ comprise another 15 components for a total of $15+n$ components. Counting in the basis of original fluctuations $\{ \varphi^a, \nu, \nu^\mu, {h^{TT}}^\mu_{\,\,\,\nu}, h, H, \epsilon^\mu \}$ yields the same number
$n+1+4+5+1+1+3=15+n$, as it should (we counted $3$ for transverse vectors and 4 for generic vectors). The counting of the gauge-invariant variables $\{ \mathfrak a^a, \mathfrak b, \mathfrak c, \mathfrak d^\mu, \mathfrak e^\mu_{\ \nu} \}$, however, yields $n+1+1+3+5=n+10$ off-shell components. The five extra components of the original fluctuations are pure gauge, corresponding to the diffeomorphisms $\xi^M$.

One needs to show explicitly that these gauge-invariant variables are physically equivalent to the original set of fluctuations, so that
the full set of equations can be rewritten directly in this form, hence removing all the possibly spurious gauge artifacts, while retaining all the physical information.
In the absence of the boundary action, this is straightforward, because one can make a choice of $\xi^M$ such as to 
set $h=0=H=\epsilon^{\mu}$, and hence by simple counting one can see that 
the whole system can be rewritten in equivalent form
in terms of the gauge invariant variables. 
This can be verified explicitly to hold for the bulk equations (which hence are the ones 
derived in~\cite{BHM}).

In our specific case, though, the boundaries restrict the gauge transformations allowed.
One has to show that the thus constrained system can still be completely rewritten
in terms of the variables in Eq.~(\ref{eq:gaugeinvariantvariables}).
In order to do so, we use a different strategy. 
First, we observe that because of the restriction on $\xi^r$ in Eq.~(\ref{Eq:rest}), 
at the boundaries one can define two independent 4d gauge invariant variables, which replace $\mathfrak{c}$:
\beqs
\mathfrak{c}_1&\equiv& - \frac{e^{-2A} \Box h}{6 A'}\,, \\
\mathfrak{c}_2&\equiv& e^{-2A} \partial_\mu \nu^\mu - \frac{1}{2} \partial_r H\,,
\eeqs
so that $\mathfrak{c}=\mathfrak{c}_1+\mathfrak{c}_2$. The original system of fluctuations is certainly equivalent to the set $\{\mathfrak{a},\mathfrak{b},\mathfrak{c}_1,\mathfrak{c}_2,\mathfrak{d},\mathfrak{e}\}$.
Then, we have to show that the boundary conditions actually remove the extra degrees of freedom, hence 
proving that the presence of the boundary actions consistently restricts both the fluctuations and the gauge transformations,
so that the whole system (bulk and boundaries) can be fully expressed in terms of the variables in Eq.~(\ref{eq:gaugeinvariantvariables}).

More precisely, one can show that one of the boundary conditions can be written in a 4d gauge invariant form as
\beqs
\label{eq:c2equals0}
\left.\frac{}{}\mathfrak{c}_2\right|_{r_i}&=&0\,.
\eeqs
This, together with the fact that in the bulk the gauge freedom allows to always set $\mathfrak{c}=\mathfrak{c}_1$,
means that $\mathfrak{c}_2$ is actually not a physical degree of freedom, and it can be set to zero everywhere,
hence allowing for the whole set of fluctuation equations and boundary conditions to be written purely in terms
of the variables in Eq.~(\ref{eq:gaugeinvariantvariables}). In the appendices, the boundary condition \eqref{eq:c2equals0} is derived by first choosing a gauge transformation $\xi^{\mu}$
such that $\nu^{\mu}(x,r) = 0$ everywhere, then showing that at the boundaries $\partial_r H |_{r_i} = 0$. Finally, we can use the residual gauge $\xi^r$ to set $\partial_r H=0$ also in the bulk 
(together with $\mathfrak{c}_2$).

Working in the gauge $\nu^\mu(x,r) = 0$, and making use of the boundary condition \eqref{eq:c2equals0}, there is a straightforward and natural one-to-one map between the fluctuations $\{ \varphi^a, \nu, h, \epsilon^\mu, {h^{TT}}^\mu_{\ \nu} \}$ and the gauge invariant variables $\{ \mathfrak a^a, \mathfrak b, \mathfrak c, \mathfrak d^\mu, \mathfrak e^\mu_{\ \nu} \}$. The main advantage of the gauge-invariant formalism is that it allows to decouple
the equations in a very simple way~\cite{BHM}, hence rendering 
the calculation of the spectrum much easier.
In practice, this means that the equations for $\mathfrak{b}$ and $\mathfrak{c}$ 
are algebraic equations relating them to the dynamical variables $\mathfrak{a}^a$,
while the equations for $\mathfrak{d}^{\mu}$ and $\mathfrak{e}^{\mu}_{\,\,\,\nu}$ decouple.
Hence, the spectrum of scalar fluctuations can be identified by solving a set of $n$ second-order differential equations involving only the
variables $\mathfrak{a}^a$, subject to boundary conditions that, again, involve only the variables $\mathfrak{a}^a$.

%%%%%%%%%%%%%%%%%%%%%%%%%%%%%%%%
%%%%%%%%%%%%%%%%%%%%%%%%%%%%%%%%
%\newpage
\section{Dynamics}
\label{Sec:3}

In this section we summarize all the important equations that one has to 
solve in order to study the spectrum of a given model.

%%%%%%%%%%%%%%%%%
\subsection{The complete action}

We are now ready to write explicitly the complete action.
Our  starting point is the general definition of the five-dimensional sigma-model coupled to gravity.
We write the action as
\beqs
{\cal S}&\equiv&\int\di^4x\di  r \left\{
\sqrt{-g}
\Theta
\left[\frac{1}{4}R+{\cal L}_5(\Phi^a,\partial_M\Phi^a,{g}) \right] \right.\nonumber\\
&&\left.
+\sqrt{-\tilde{g}}\delta( r- r_1) \left[c_KK+{\cal L}_1(\Phi^a,\partial_{\mu}\Phi^a,\tilde{g}) \right] \right.\nonumber\\
&&\left.
- \sqrt{-\tilde{g}} \delta( r- r_2)\left[c_KK+{\cal L}_2(\Phi^a,\partial_{\mu}\Phi^a,\tilde{g}) \right] \right\}\,,
\label{Eq:action}
\eeqs
where  $R$ is the Ricci scalar, where
$K$ is the extrinsic curvature, where the coupling $c_K=-1/2$ is fixed by consistency and where
${\cal L}_i$ are the sigma-model actions.
The step function is defined by $\Theta\equiv\Theta( r- r_1)-\Theta( r- r_2)$.

We define the action of the matter fields in terms of the real scalar fields $\Phi^a=\Phi^a(x^{\mu}, r)$ as
\beqs
\label{Eq:L5}
{\cal L}_5&\equiv&-\frac{1}{2}G_{ab}g^{MN}\partial_M\Phi^a\partial_N\Phi^b-V(\Phi^a)\,,\\
\label{Eq:L1}
{\cal L}_1&\equiv&-\lambda_{(1)}(\Phi^a)\,,\\
\label{Eq:L2}
{\cal L}_2&\equiv&-\lambda_{(2)}(\Phi^a)\,.
\eeqs
Hence, we assume that no kinetic term is present at the boundaries, but only localized potential terms.
We will provide the explicit forms of the $\lambda_{(i)}$ terms later. Notice
the different sign with which they enter the complete action.

%%%%%%%%%%%%%%
\subsection{Background equations}
The background is determined by solving the classical equations, assuming that all the functions defining the background depend only
on the radial direction $ r$, and not on the $x^{\mu}$.
We take the variation of the complete action in order to determine the equations of motion for $A(r)$ and $\bar \Phi(r)$.\footnote{See Appendix~\ref{App:A} for details. An alternative way is to take the variation of the action by linearizing in $\{h, \nu, \varphi^a\}$. The terms proportional to $\Theta$ then give us the equations of motion satified by the background in the bulk, whereas the terms proportional to $\delta = \partial_r \Theta$ give us the boundary conditions.} The equation of motion for the scalars is
\SP{
\label{eq:scalarsEOM}
	\bar \Phi''^a + 4 A' \bar \Phi'^a + \mathcal{G}^a_{\ bc} \bar \Phi'^b \bar \Phi'^c - V^a = 0,
}
while Einstein's equations read
\SP{
\label{eq:einsteinEOM}
	6 A'^2 + 3 A'' &= - G_{ab} \bar \Phi'^a \bar \Phi'^b - 2V, \\
	6 A'^2 &= G_{ab} \bar \Phi'^a \bar \Phi'^b -2V.
}
The boundary conditions satisfied by the background, at the UV ($r=r_2$) and IR ($r=r_1$) are
\SP{\label{Eq:BOUN1}
	\bar \Phi'^a \Big|_{r_i} &= \lambda_{(i)}^{\,\,\,\,a} \Big|_{r_i}\,\equiv\,G^{ab}\partial_b \lambda_{(i)} \Big|_{r_i}\,,}
\SP{\label{Eq:BOUN2}
	A' \Big|_{r_i} &= - \frac{2}{3} \lambda_{(i)} \Big|_{r_i}.
}
Thus we see that to quadratic order the localized potentials $\lambda_{(i)}(\Phi^a)$ are constrained to have the following form\footnote{Note that the covariant derivatives of $\lambda_{(i)}$ arise due to expanding $\Phi^a$ according to the exponential map Eq.~\eqref{eq:exponentialmap}.}
\SP{
	\lambda_{(1)} &= \left.-\frac{3}{2} A'\right|_{r_1} + \left.\frac{}{} G_{ab} \bar \Phi'^a\right|_{r_1} (\Phi^b -  \Phi^b_1) + \frac{1}{2} D_a D_b \lambda_{(1)} (\Phi^a -  \Phi^a_1) (\Phi^b -  \Phi^b_1), \\
	\lambda_{(2)} &=  \left.-\frac{3}{2} A'\right|_{r_2} +  \left.\frac{}{}G_{ab} \bar \Phi'^a\right|_{r_2} (\Phi^b -  \Phi^b_2) + \frac{1}{2} D_a D_b \lambda_{(2)} (\Phi^a -  \Phi^a_2) (\Phi^b -  \Phi^b_2)\,,
}
were $\Phi^a_{1,2}\equiv\bar{\Phi}^a|_{r_{1,2}}$ are the values assumed by the scalars at the boundaries. In effect, this is equivalent to imposing the condition that the scalars have fixed boundary values $\Phi^a_i$.

%%%%%%%%%%%%%%%%%%%%%%%%%%%%%%%
%%%%%%%%%%%%%%%%%%%%%%%%%%%%%%%
\subsubsection{First-order formalism}

Assuming that there exists a superpotential $W$ such that the potential $V$
 can be rewritten as
\beqs
V&=&\frac{1}{2}G^{ab}W_aW_b\,-\,\frac{4}{3}W^2\,,
\eeqs
with $W_a=\partial_a W=\partial W/\partial\Phi^a$, 
then the system can be reduced to a
set of $n+1$  first-order equations
\beqs
\label{Eq:BPS1}
A^{\prime} &=& -\frac{2}{3} W\,,\\
\label{Eq:BPS2}
\bar{\Phi}^{\prime\,a}&=&G^{ab}W_b\,=\,W^a\,,
\eeqs
in the sense that all solutions to Eqs.~(\ref{Eq:BPS1})-(\ref{Eq:BPS2}) are also solutions to the second-order equations Eqs.~\eqref{eq:scalarsEOM}-\eqref{eq:einsteinEOM}.
However, for these to provide solutions to the original system, also the boundary conditions must be satisfied,
implying a set of constraints on the form of the localized potentials $\lambda_{(i)}$.
The general form of the localized potentials must be
\beqs
\lambda_{(1)}&=&W(\Phi_1)\,+\,W_c(\Phi_1)({\Phi}^c-\Phi_1^c)\,+\,\frac{1}{2} D_d D_c\lambda_{(1)}({\Phi}^c-\Phi_1^c)({\Phi}^d-\Phi_1^d)\,,\\
\lambda_{(2)}&=&W(\Phi_2)\,+\,W_c(\Phi_2)({\Phi}^c-\Phi_2^c)\,+\,\frac{1}{2} D_d D_c\lambda_{(2)}({\Phi}^c-\Phi_2^c)({\Phi}^d-\Phi_2^d)\,.
\eeqs

Before we move onto studying the spectrum of fluctuations, a brief comment is needed.
One might, legitimately, wonder why we allow ourselves the freedom to add localized potential terms for the scalars,
but not localized kinetic terms, and/or more general functions not only of the fields but also of their (four-dimensional) derivatives.
The fact that we truncate $\lambda_{(i)}$ at the quadratic order is just due to the fact that we are interested 
here only in the equations for the background and in the linearized equations for the fluctuations, so that higher-order terms 
have no effect.
On the other hand, while terms that depend explicitly on the four-momentum $q^2$ do not enter the background equations, they do enter into the boundary conditions for the fluctuations. However, their presence and structure in entangled with
the problem of holographic renormalization, in the sense that in order to systematically identify  what such terms are needed,
one has to study the (model-dependent) structure of divergences in the two-point functions. This suggests using some caution when discussing the spectrum.

%%%%%%%%%%%%%%%%%%%%%%%%%%%%%%%
%%%%%%%%%%%%%%%%%%%%%%%%%%%%%%%
\subsection{Boundary conditions for the fluctuations}

As discussed in Section~\ref{Sec:2d}, the equations for the fluctuations of the scalars and the metric can be rewritten in terms of gauge-invariant 
fields, up to some subtleties that we will discuss later. 
Most important, we anticipate here that there are two possible ambiguities in this procedure, which is not
well-defined for the (interrelated) cases when $\bar{\Phi}'^a=0$ and/or when the four-dimensional momentum 
$\Box=-K^2=m^2\equiv q^2$ vanishes. On the one hand, the procedure is rigorous even for the case in which these are infinitesimal,
and on the other hand all the examples one might think of that have any phenomenological relevance will
not contain exactly massless scalars, none of the global symmetries being exact.
We will come back to these questions later in the next section.

With all of these caveats, the final result is that the spectrum can be obtained
by solving the following second-order differential equation for
a set of $n$  gauge-invariant scalar fluctuations denoted by $\mathfrak{a}^a$:
\SP{
\label{eq:diffeq}
	\Big[ {\cal D}_r^2 + 4 A' {\cal D}_r + e^{-2A} \Box \Big] \mathfrak{a}^a - \Big[ V^a_{\ |c} - \mathcal{R}^a_{\ bcd} \bar \Phi'^b \bar\Phi'^d + \frac{4 (\bar \Phi'^a V_c + V^a \bar \Phi'_c )}{3 A'} + \frac{16 V \bar \Phi'^a \bar \Phi'_c}{9 A'^2} \Big] \mathfrak{a}^c = 0,
}
with boundary conditions (suppressing the index $i$ of $\lambda_{(i)}$)
\SP{
\label{eq:BCa}
	&\left[ \delta^a_{\ b} + e^{2A} \Box^{-1} \left( V^a - 4 A' \Phi'^a - \lambda^a_{\ |c} \bar \Phi'^c \right) \frac{2 \bar \Phi'_b}{3 A'} \right] {\cal D}_r \mathfrak a^b \Big|_{r_i} = \\&  \left[ \lambda^a_{\ |b} + \frac{2 \bar \Phi'^a \bar  \Phi'_b}{3 A'} + e^{2A} \Box^{-1} \frac{2}{3 A'} \left( V^a - 4 A' \bar \Phi'^a - \lambda^a_{\ |c} \bar\Phi'^c \right) \left( \frac{4 V \bar \Phi'_b}{3 A'} + V_b \right) \right] \mathfrak a^b \Big|_{r_i}.
}

\subsubsection{Superpotential formalism}

In the special case where there is a superpotential $W$, we have that Eqs.~\eqref{eq:diffeq} and \eqref{eq:BCa} become
\SP{\label{Eq:diffeqW}
	\Bigg[ \left( \delta^a_b \mathcal D_r + W^a_{|b} - \frac{W^a W_b}{W} - \frac{8}{3} W \delta^a_b \right) \left( \delta^b_c \mathcal D_r - W^b_{|c} + \frac{W^b W_c}{W} \right) + \delta^a_c e^{-2A} \Box \Bigg] \mathfrak{a}^c = 0,
}
and
\SP{\label{Eq:BCW}
	&\left[ \delta^a_{\ b} + e^{2A} \Box^{-1} \left( \lambda^a_{\ |c} - W^a_{\ |c} \right) \frac{W^c W_b}{W} \right] \mathcal D_r \mathfrak a^b \Big|_{r_i} = \\& \left[ \lambda^a_{\ |b} - \frac{W^a W_b}{W} + e^{2A} \Box^{-1} \left( \lambda^a_{\ |c} - W^a_{\ |c} \right) \frac{W^c W_d}{W} \left( W^d_{\ |b} - \frac{W^d W_b}{W} \right) \right] \mathfrak a^b \Big|_{r_i},
}
respectively.

%%%%%%%%
\subsection{Boundary masses}
When the superpotential is known, it is convenient to write
\beqs
N^d_{\,\,b}&\equiv&W^d_{\ |b} - \frac{W^d W_b}{W} \,,\\
\lambda_{(1)\,|c}^{\,\,\,\,\,\,\,a}&\equiv&\left.W^{a}_{\,\,\,|c}\right|_{r_1}\,+\,\left(m^{2}_{1}\right)^{a}_{\,\,c}\,,\\
\lambda_{(2)\,|c}^{\,\,\,\,\,\,\,a}&\equiv&\left.W^{a}_{\,\,\,|c}\right|_{r_2}\,-\,\left(m^{2}_{2}\right)^{a}_{\,\,c}\,.
\eeqs
Notice the different sign in the definition.
It is  convenient to rewrite the bulk equation for the fluctuations in the compact form
\SP{\label{Eq:diffeqWN}
	\Bigg[ e^{-4A} \left( \delta^a_b \mathcal D_r + N^a_{\,\,\,\,b} \right) e^{4A}\left( \delta^b_c \mathcal D_r - N^b_{\,\,\,\,c} \right) + \delta^a_c e^{-2A} \Box \Bigg] \mathfrak{a}^c = 0\,,
}
and the boundary conditions as
\SP{\label{Eq:BCWinftyIR}
	&\left[ \delta^a_{\ b} + e^{2A} \Box^{-1} \left(m_1^2\right)^a_{\,\,c} \frac{W^c W_b}{W} \right] \mathcal D_r \mathfrak a^b \Big|_{r_1} = \\& \left[ \left(m_1^2\right)^a_{\,\,b} +N^a_{\,\,b} + e^{2A} \Box^{-1} \left(m_1^2\right)^a_{\,\,c}  \frac{W^c W_d}{W} N^d_{\,\,b} \right] \mathfrak a^b \Big|_{r_1}\,,\\
}
and
\SP{\label{Eq:BCWinftyUV}
	&\left[ \delta^a_{\ b} - e^{2A} \Box^{-1}\left(m_2^2\right)^a_{\,\,c} \frac{W^c W_b}{W} \right] \mathcal D_r \mathfrak a^b \Big|_{r_2} = \\& \left[  -\left(m_2^2\right)^a_{\,\,b} +N^a_{\,\,b}- e^{2A} \Box^{-1} \left(m_2^2\right)^a_{\,\,c}  \frac{W^c W_d}{W} N^d_{\,\,b} \right] \mathfrak a^b \Big|_{r_2}\,,\\
}
respectively. 

The matrices $m_i^2$ encode the degree of arbitrariness in the definition of the boundary terms.
These matrices do not necessarily preserve the same amount of  (internal) global symmetries as the bulk sigma-model action.
In particular, they provide a source of explicit breaking for possible residual internal symmetries of the 
sigma-model action. Taking their entries to be large ensures that possible pseudo-Goldstone bosons 
(arising by the spontaneous breaking of such internal symmetries) are effectively removed from the spectrum.
Hence, it is convenient to take the limit in which $m_i^2$ are diagonal, and their eigenvalues are all positive and divergent,
so that the boundary terms reduce to
\SP{\label{Eq:BCWinfty}
	&\left[ e^{2A} \Box^{-1}\frac{W^c W_d}{W}\right] \left(\delta^d_{\,\,\,\,b} \mathcal D_r - N^d_{\,\,\,\,b} \right)\mathfrak a^b \Big|_{r_i} 
	=   \delta^c_{\,\,\,\,b} \mathfrak a^b \Big|_{r_i}\,.
}

One can think of using this procedure also in the case when the physical system does not have explicit physical cutoffs.
In this case, this procedure can be thought of as a regulator: one computes the spectrum at finite cutoffs, and then studies
how the physical spectrum changes when extrapolating the values of $r_i$ to their natural boundaries (either towards the UV, or towards a singularity/end-of-space boundary in the IR).

%\newpage
%%%%%%%%%%%%%%%%%%%%%%%%%%%%%%%
%%%%%%%%%%%%%%%%%%%%%%%%%%%%%%%
\section{Applications}
\label{Sec:4}

In this section, we apply the formalism developed in the previous sections to a few examples, in order to demonstrate how the algorithm that we propose works, as well as to verify that our results agree with the literature when applicable. Although the examples we study here are all simple, in the sense that they consist of only one scalar with a canonical sigma-model metric, the formalism is also applicable to more general cases with several scalars whose sigma-model metric is non-trivial. Indeed, most applications, such as various consistent truncations from supergravity, will fall into this more general category, and it was with this in mind that the formalism of the previous sections was developed.

%%%%%%%%%%%%%%%%%%%%%%%%%%%%%%%
%%%%%%%%%%%%%%%%%%%%%%%%%%%%%%%
\subsection{Example A: quadratic superpotential and Goldberger-Wise mechanism}
\label{Sec:4a}

The first example we consider is related to 
the Goldberger-Wise stabilization mechanism~\cite{GW},
written in the form of~\cite{GW2}.
This example has two advantages. It is peculiarly simple, in that there is only one scalar $\Phi$
with canonical kinetic term, and has a very simple superpotential.
And it has been extensively studied in the literature, thus providing us with a way to check that the formalism 
used here yields results that are consistent with other approaches. Also, there is a sense in which, in some limit, any physical system near a fixed point resembles it.

The superpotential is
\beqs
\label{Eq:GWsuperpotential}
W&=&-\frac{3}{2}-\frac{\Delta}{2}\Phi^2\,,
\eeqs
so that the potential is
\beqs
\label{Eq:GWpotential}
V&=&-3+\frac{1}{2}(\Delta^2-4\Delta)\Phi^2-\frac{1}{3}\Delta^2\Phi^4\,.
\eeqs
Notice the form of the quadratic term in the potential as a function of the parameter $\Delta$,
which yields the well-known result that the five-dimensional mass is
related to the scaling dimension of the dual operators by $M_5^2L^2=\Delta (\Delta-4)$,
and hence provides a natural interpretation for the parameter $\Delta$.
The boundary potentials are
\beqs
\lambda_{(1)}&=&-\frac{3}{2}-\frac{\Delta}{2}\Phi_1^2-\Delta\Phi_1(\Phi-\Phi_1)-\frac{1}{2}\left(\Delta-m_1^2\right)(\Phi-\Phi_1)^2\,,\\
\lambda_{(2)}&=&-\frac{3}{2}-\frac{\Delta}{2}\Phi_2^2-\Delta\Phi_2(\Phi-\Phi_2)-\frac{1}{2}\left(\Delta+m_2^2\right)(\Phi-\Phi_2)^2\,.
\eeqs
The differential equations and boundary terms for the background are
\beqs
\bar{\Phi}^{\prime}&=&-\Delta\bar{\Phi}\,,\\
A^{\prime}&=&1+\frac{\Delta}{3}\bar{\Phi}^2\,,\\
\left.A^{\prime}-1-\frac{\Delta}{3}\Phi_1^2 \right|_{ r_1}&=&0\,,\\
\left.A^{\prime}-1-\frac{\Delta}{3}\Phi_2^2 \right|_{ r_2}&=&0\,,\\
\left.\bar{\Phi}^{\prime}+\Delta\Phi_1 \right|_{ r_1}&=&0\,,\\
\left.\bar{\Phi}^{\prime}+\Delta\Phi_2 \right|_{ r_2}&=&0\,.
\eeqs

The solution is hence 
\beqs
\bar{\Phi}( r)&=&\Phi_1e^{-\Delta( r- r_1)}\,,\\
 r_2- r_1&=&-\frac{1}{\Delta}\ln\frac{\Phi_2}{\Phi_1}\,,\\
A&=&a_0+ r-\frac{1}{6}\Phi_1^2e^{-2\Delta( r- r_1)}\, ,
\eeqs
where $a_0$, and $\Phi_1$ are integration constants.
As a result, a big hierarchy between the (physical) UV and IR scales can originate from a  small value of $\Delta$ (which is protected),
and with natural choices of the unprotected $\Phi_2/\Phi_1 \sim {\cal O}(1)$, hence removing possible sources
of fine-tuning.
The constant $a_0$ can always be reabsorbed into a redefinition of the four-dimensional units, and hence
 can be chosen  to be $a_0=0$, so that when $\Phi_{1,2} \rightarrow 0$ 
one recovers exactly the standard form of the AdS case with unit curvature.
It is often convenient to change variable according to $ r=-\ln z$, so that 
with $L_0<z<L_1$ one finds that
\beqs
\frac{L_0}{L_1}&=&e^{- r_2+ r_1}\,=\,\left(\frac{\Phi_2}{\Phi_1} \right)^{\frac{1}{\Delta}}\,.
\eeqs
The value of $L_0$ and $L_1$ are the UV and IR (length) scales, and in this form it is manifest that
an exponential hierarchy is naturally generated.

%%%%%%%%%%%%%%%%%%%%%%%%%%%%%%
%%%%%%%%%%%%%%%%%%%%%%%%%%%%%%
\subsubsection{Spectrum}
\label{Sec:4.1.1}

Before discussing the spectrum, let us make two observations. 
In the limit in which $\Delta \rightarrow 0$ the fluctuations $\varphi$ and $h$
decouple from each other. 
Furthermore, in this limit the field $\Phi$ has trivial bulk dynamics,
while the background becomes exactly AdS.
As a consequence of these observations, the vacuum is determined by two arbitrary,
non-dynamical quantities $\bar{\Phi}={\Phi}_1={\Phi}_2$ and $r_2- r_1$. They correspond to two moduli.
In this limit, one expects (at least in this semi-classical analysis) 
 the presence of two massless states in the scalar spectrum,
associated with these moduli parameterizing  the space of (inequivalent) vacua.

The non-vanishing of $\Delta$ yields a background which is not AdS, at least in the IR.
Effectively, this corresponds to an explicit breaking of scale invariance, and is ultimately 
responsible for the dynamical stabilization of the finite hierarchy $ r_2- r_1$.
As a consequence, one expects the second scalar to stay light even for very large choices of $m^2_i$,
and its mass to vanish with $\Delta$ and $\Phi_1$. In the language of the AdS/CFT correspondence,
$4-\Delta$ is the dimension of a dual operator with coupling proportional to $\Phi_1$, the insertion of which
breaks explicitly scale invariance.\footnote{Notice that the unitarity bounds imply that the five-dimensional mass is $M_5^2 L^2 \geq -4$, which is automatically true provided that the superpotential is given by Eq.~\eqref{Eq:GWsuperpotential}. This means that when $\Delta<1$, the only possible interpretation is that the dual operator has dimension $4-\Delta$.}

Because we are interested in the case in which the background is at least approximately AdS,
we choose $\Phi_1$ to be small. 
By inspection of the bulk equation, of  the background warp factor $A$, of the superpotential $W$, 
and of its first and second  field derivatives $W_{\Phi}$ and $W_{\Phi\Phi}$,
it is apparent that one can expand the fluctuations in powers of 
\beqs
x&\equiv&\Delta^2 \Phi_1^2e^{2\Delta r_1}\,,
\eeqs
by writing
\beqs
\mathfrak{a}(r)&=&\mathfrak{a}_0(r)+x\,\mathfrak{a}_1(r)\,+\cdots\,,
\eeqs
and replacing in Eq.~(\ref{Eq:diffeqWN}). At the leading ${\cal O}(x^0)$, the bulk equation becomes
\beqs
0&=&\left[\frac{}{}\left({\partial}_{ r}-\Delta+4\right)
\left({\partial}_{ r}+\Delta\right)
+e^{-2r}q^2\frac{}{} \right]\mathfrak{a}^k_0(r)\,,
\label{Eq:GWeq}
\eeqs 
where the superscript $k$ refers to the heavy KK-modes,
and the boundary conditions
\beqs
0&=&
\left.\frac{}{}-\left(m_1^2-{\partial}_{ r}
-\Delta \right)\mathfrak{a}^k_0 \right.
-\left.\frac{2}{3}m_1^2\frac{e^{2r}W_{\Phi}^{\,2}}{q^2}
\left({\partial}_{ r}+\Delta\right)\mathfrak{a}^k_0
 \right|_{ r_1}
 \label{Eq:GWbc1}
\,,\\
0&=&
\left.\frac{}{}-\left(-m_2^2-{\partial}_{ r}
-\Delta \right)\mathfrak{a}^k_0 \right.
+\left.\frac{2}{3}m_2^2\frac{e^{2r}W_{\Phi}^{\,2}}{q^2}
\left({\partial}_{ r}+\Delta\right)\mathfrak{a}^k_0
 \right|_{ r_2}
 \label{Eq:GWbc2}
\,,
\eeqs
where we kept explicitly terms proportional to $W_{\Phi}=-\Delta\Phi_1e^{-\Delta(r-r_1)}$ for reasons 
that will become clear soon.

The heavy states can be discussed by looking at the solutions to these equations setting $W_{\Phi}|_{r_i}=0$.
The solution to the bulk equations is
\beqs
\mathfrak{a}^k_0&=&e^{-2r}\left[c_1 J_{2-\Delta}\left(e^{-r}q\right)+c_2 Y_{2-\Delta}\left(e^{-r}q\right)\right]\,,
\eeqs
with $c_i$ determined by the boundary conditions
\beqs
0&=&
\left.\frac{}{}-\left(m_1^2-{\partial}_{ r}
-\Delta \right)\mathfrak{a}_0 
 \right|_{ r_1}
\,,\\
0&=&
\left.\frac{}{}-\left(-m_2^2-{\partial}_{ r}
-\Delta \right)\mathfrak{a}_0 
 \right|_{ r_2}
\,.
\eeqs
The details of the spectrum depend on the specific choice of the $m^2_i$ terms.
Yet, in general the mass gap is related to the zeros of the Bessel functions $J_{2-\Delta}$, 
at least for $r_2\gg r_1$. The mass gap is hence $\pi e^{-r_1}=\pi/L_1$, as it is sensible to expect.

Conversely, the lightest states must be treated by keeping the term depending on $1/q^2$ in the boundary conditions,
while dropping the one proportonal to $q^2$  in the bulk equations.
The reason is that in the limit where $W_{\Phi}\rightarrow 0$ we expect a massless state to be present.
Hence, the lightest state will itself have mass ${\cal O}(x)$.
It is then convenient to write $q^2\equiv x\tilde{q}^2$ and $W_{\Phi}^{\,2}=x e^{-2\Delta r}$.
The bulk equation hence becomes, at ${\cal O}(x^0)$: 
\beqs
0&=&\left[\frac{}{}\left({\partial}_{ r}-\Delta+4\right)
\left({\partial}_{ r}+\Delta\right)\frac{}{}\right]\mathfrak{a}^d_0\,,
\eeqs 
and the boundary conditions
\beqs
0&=&
\label{Eq:77}
\left.\frac{}{}-\left(m_1^2-{\partial}_{ r}
-\Delta \right)\mathfrak{a}^d_0 \right.
-\left.\frac{2}{3}m_1^2\frac{e^{2r(1-\Delta)}}{\tilde{q}^2}
\left({\partial}_{ r}+\Delta\right)\mathfrak{a}^d_0
 \right|_{ r_1}
\,,\\
0&=&
\label{Eq:78}
\left.\frac{}{}-\left(-m_2^2-{\partial}_{ r}
-\Delta \right)\mathfrak{a}^d_0 \right.
+\left.\frac{2}{3}m_2^2\frac{e^{2r(1-\Delta)}}{\tilde{q}^2}
\left({\partial}_{ r}+\Delta\right)\mathfrak{a}^d_0
 \right|_{ r_2}
\,,
\eeqs
where the superscript $\mathfrak{a}^d_0$ indicates that we interpret this state as a light (pseudo-)dilaton.
The bulk equation is satisfied by 
\beqs
\mathfrak{a}_0^d&=&c_1 e^{-\Delta r}\,+\,c_2 e^{-(4-\Delta) r}\,.
\eeqs
Notice that, because of the presence of $q^2$ in a denominator in the boundary conditions,
it is sufficient to solve the leading-order bulk equation in order to derive the (subleading) 
${\cal O}(x)$ value for the mass of the dilaton.

One has to solve two algebraic equations to determine the ratio $c_1/c_2$ and $\tilde{q}$.
We do so for the extreme case $m_i^2 \rightarrow \infty$.
The result is
\SP{
	\tilde q^2 = \frac{4 e^{-2 \left(r_1+r_2\right)} \left(e^{2 r_1}-e^{2
   r_2}\right) (\Delta -2)}{3 \left(e^{2 (\Delta -2)
   r_1}-e^{2 (\Delta -2) r_2}\right)},
}
which for $r_1=0$ yields the dilaton mass
\beqs
m_d^2&=&4 \Delta^2\Phi_1^2\frac{2-\Delta}{3}\,\frac{1-e^{-2r_2}}{1-e^{2(\Delta-2)r_2}}\,.
\label{Eq:dilatonmass}
\eeqs
This result, obtained so easily, is in splendid agreement with the literature~\cite{GW3, Kofman:2004tk}.

Notice that in  deriving the mass of the light  (pseudo-)dilaton
we did not make any assumptions about $\Delta$, aside from requiring it to
be positive. Hence all of the above holds for generic $\Delta$, not just for the $\Delta \ll 1$ case.
Also,  the last factor in Eq.~(\ref{Eq:dilatonmass}), 
dependent on $r_2$, assumes unit value for large $r_2\gg 0$,
provided $\Delta<2$. It turns negative when $\Delta>2$, at which point however
the negative sign is compensated by the $\Delta-2$ factor, ensuring that the mass is positive.
In this case the mass vanishes for asymptotically large values of $r_2$.

%%%%%%%%
\subsubsection{Zero modes}

We devote this short subsection to analyzing more in detail the limits in which
$m^2_d\rightarrow 0$.
For $m_i^2=0$, 
the boundary conditions in Eq.~(\ref{Eq:77}) and Eq.~(\ref{Eq:78}) reduce to 
\beqs
\left.\frac{}{}(\partial_r+\Delta) \mathfrak{a}_0^d\right|_{r_i}&=&0\,,
\eeqs
and there is a massless state with bulk profile $\mathfrak{a}_0^d\propto e^{-\Delta r} = z^{\Delta}$.
In the $\Delta\rightarrow 0$ case this profile becomes constant.

The existence of this zero-mode for $m_i^2\rightarrow 0$ is a very general property for any
 system of $n$ scalars. If a superpotential description exists,
and we apply the formalism of Eq.~(\ref{Eq:diffeqWN}), (\ref{Eq:BCWinftyIR}) and (\ref{Eq:BCWinftyUV}),
 notice that
\beq
\left(\frac{}{}\delta^a_{\,\,\,\,c}{\cal D}_r -N^a_{\,\,\,\,c}\right)\frac{W^c}{W} = 0.
\eeq 
This implies that $\tilde{\mathfrak{a}}^a=W^a/W$ always solves Eq.~(\ref{Eq:diffeqWN})  for $q^2=0$.

$\tilde{\mathfrak{a}}^a$ satisfies the boundary conditions obtained by setting $m_i^2=0$, which reduce precisely to
\beqs
\left.\left(\frac{}{}\delta^a_{\,\,\,\,c}{\cal D}_r -N^a_{\,\,\,\,c}\right)\tilde{\mathfrak{a}}^c\right|_{r_i}&=&0\,.
\eeqs
This observation shows explicitly that there is always a massless state when $m_i^2=0$, which is the one discussed at
the beginning of this subsection.
Notice however that  this is in general the result of fine-tuning of  $m_i^2$, 
and caution must be used in interpreting this result.

In the other extreme, more physical  case,  in which $m_i^2\rightarrow +\infty$, the boundary conditions 
in Eq.~(\ref{Eq:77}) and Eq.~(\ref{Eq:78}) become (for $x \ll 1$)
\beqs
\left.\frac{2e^{2r(1-\Delta)}}{3\tilde{q}^2}(\partial_r+\Delta) \mathfrak{a}_0^d\,+\,\mathfrak{a}_0^d\right|_{r_i}&=&0\,.
\eeqs
Setting $r_1=0$ for simplicity, the solution is
\beqs
\label{Eq:zero}
\mathfrak{a}_0^d\propto 
e^{(\Delta-4) r}-\frac{e^{-\Delta r} \left(1-e^{2 (\Delta-1)
   r_2}\right)}{1-e^{2 r_2}}\,,
\eeqs
which reduces to $\mathfrak{a}_0^d\propto e^{-4r} +e^{-2r_2}$ in the $\Delta\rightarrow 0$ case (in which this
is a massless state).

Notice a very interesting fact: taking the limit $r_2\rightarrow +\infty$ in Eq.~(\ref{Eq:zero})
automatically yields a profile that corresponds to keeping only the subleading behavior
in the generic solution. This shows for a concrete example that the procedure we are implementing automatically
reproduces the results obtained with the more widely adopted idea of defining the spectrum 
only in the absence of a UV boundary, by imposing that the solutions to the fluctuation equations vanish at infinity
as fast as possible with $r$.
Finally notice one important fact about Eq.~(\ref{Eq:zero}). The bulk profile of the massless state that is present 
when $\Delta\rightarrow 0$ is correctly identified by first studying the $\Delta \neq 0$ limit, in which case
the corresponding state is light but not massless, and then taking the $\Delta\rightarrow 0$ limit 
at the end of the calculation. 

%%%%%%%%%%%%%%%%%%%%%%%%%%%%%%%%%%%%
%%%%%%%%%%%%%%%%%%%%%%%%%%%%%%%%%%%%
\subsubsection{Discussion}

Let us  discuss now  what happens for generic values of $m^2_i$.
We keep working under the assumption that $x\ll 1$, so that Eqs.~(\ref{Eq:GWeq}),  (\ref{Eq:GWbc1}) and
(\ref{Eq:GWbc2}), which 
describe the generic bulk profiles,  still hold.
Let us consider first the lightest state.
Solving for generic values of $m_i^2$, and for simplicity setting $r_1=0$, yields
\SP{
\label{Eq:GWgeneral}
	m_d^2 = \frac{4 \Delta^2 \Phi_1^2 \left(1-e^{-2 r_2}\right) (\Delta -2)
   m_1^2 m_2^2}{3 \left(e^{2 (\Delta
   -2) r_2} m_1^2 \left(m_2^2+2 \Delta
   -4\right)-\left(m_1^2-2 \Delta +4\right)
   m_2^2\right)}
}
This result is completely general (i.e. valid for any $m_i^2$ and $\Delta$).
Notice that, as we already know, for $m_i^2=0$ one finds a massless state,
and for $m_i\rightarrow +\infty$ the finite result in Eq.~(\ref{Eq:dilatonmass}), suppressed by $x$. The former choice should be avoided, if what one is trying to understand is whether the scenario in question predicts the existence of a light scalar, irrespectively of the details of the UV dynamics and of the coupling to other sectors of the complete theory of interest (such as the SM).

Also,  a pathology appears when taking $|m_i^2|=4-2\Delta$.
The reason for this lies in the way in which we wrote the boundary terms.
Consider for example the UV term: looking at the coefficient of the $(\Phi-\Phi_1)^2$ term,
replacing such pathological choice one has
$m_1^2-\Delta=-4+\Delta$.
In this case, what is happening is that this is the choice that would render massless the excitations
around the other solution to the (second-order) bulk equations for the background, which 
has scaling dimension $4-\Delta$ rather than $\Delta$. Therefore, this choice should also be avoided.

One special comment about unitarity.
The way in which we wrote the system, in terms of a quadratic superpotential,
yields a potential in which the (five-dimensional) mass term $M_5^2=\Delta^2-4\Delta>-4$ is always 
above the unitarity bounds, irrespective of the value of $\Delta$.
The bound is saturated for $M_5^2=-4$, or $\Delta=2$, in which case the theory is close to a special transition 
point that we will discuss at length elsewhere. In proximity of this point, the mass is anomalously suppressed.
Yet, we did add boundary terms
with arbitrary parameters $m_i^2$, 
which distort the spectrum,
and one has to check that in doing so no tachyon state has been added.
One can easily verify that for $r_2\rightarrow +\infty$, and $\Delta<2$ this implies that one must enforce the choice $m_1^2\geq 0$.
For  $r_2\rightarrow +\infty$, and $\Delta>2$ one conversely must impose $m^2_2\geq 0$.
In general, given a choice of $\Delta$ and $m_i^2$ one has to verify that no tachyon is present.

Finally, we turn to the heavy modes, which have bulk profile  $\mathfrak{a}_0^k$ up to ${\cal O}(x)$ corrections.
We already stated (in Section~\ref{Sec:4.1.1}) that the heavy modes form towers with separation $\pi e^{r_1}$.
One has to clarify where the second light state mentioned several times ends up, for general values of $m_i^2$.
The general solution can be obtained by simply solving for the integration constants and for $q^2$ in the equations for $\mathfrak{a}_0^k$.
This is a somewhat intricate exercise, which is not very illuminating. 
Yet, there exists an interesting limiting case: for $m_1^2\rightarrow +\infty$ and $m_2^2\rightarrow 0$ one finds that
one light state has the mass $m_d^2$ discussed earlier on in \eqref{Eq:dilatonmass}, while at the same time an exactly massless state also is present.
Hence, in this case it is clear that two light scalars are present, while the heavy states start appearing with
masses proportional to ${\cal O}(\pi e^{-r_1})$.
Effectively, what is happening is that the whole tower is shifted down in this limit, and the first excited state in the tower  
becomes parametrically light. The numerical study performed in the next section will make these observations more clear.

%%%%%%%%%%%%%%%%%%%%%%%%%%%%%%%
%%%%%%%%%%%%%%%%%%%%%%%%%%%%%%%
\subsection{Example B: a consistent truncation of type IIB supergravity}

Let us now consider a different example. We still consider the case where
only one scalar $\Phi$ is present, and the sigma-model is trivial. 
But now the superpotential is
\beqs
W&=&-\frac{3}{4}\left(\frac{}{}1+\cosh 2\sqrt{\frac{\Delta}{3}} \Phi\frac{}{}\right)\,,
\eeqs
so that the potential is
\beqs
V&=&
-3 \cosh\left[\sqrt{\frac{\Delta}{3}}\Phi\right]^4 + 
 \frac{3}{8} \Delta \sinh\left[2\sqrt{\frac{\Delta}{3}}\Phi\right]^2\\
 &\simeq&
 -3 + \left(-2 \Delta + \frac{\Delta^2}{2}\right) \Phi^2+\cdots\,,
\eeqs
where in the last expression we expanded for small $\Phi$.
Notice how this expansion is in agreement with the potential 
of the previous sections, at leading order. 
The difference is important only away from the AdS fixed point $\Phi=0$.

The solution to the bulk equation is
\beqs
\bar{\Phi}&=&\sqrt{\frac{3}{\Delta}}\arctanh e^{-\Delta r +c_1}\,,
\eeqs
and we can always choose $c_1=0$ for simplicity, setting the radial coordinate in such 
a way that $\bar{\Phi}$ diverges for $r\rightarrow 0$.
The warp factor is
\beqs
A&=&\frac{1}{2\Delta}\ln\left(-1+e^{2\Delta r}\right)\,.
\eeqs
Replacing $r=-\log z$, and expanding for small $z\rightarrow 0$,
\beqs
\bar{\Phi}&=&\sqrt{\frac{3}{\Delta}}\,z^{\Delta}\,,
\eeqs
not surprisingly. Notice however that, as opposed to the what we did in the previous sections,
the coefficient in front of $z^{\Delta}$ is now fixed. There is no free parameter analogous to $\Phi_1$. 
The integration constant that we set to zero 
would  change this coefficient,
but it would also change the position of the singularity in the  IR.
In this sense, if we think of $\bar{\Phi}\neq 0$ in terms of spontaneous symmetry-breaking, 
in this model there is a direct link between the formation of such a symmetry-breaking condensate 
and the end of space in the IR (which one would like to associate with confinement).

For $\Delta=3$ the potential becomes
\beqs
V&=&
\frac{3}{4} \cosh^2 \Phi (-5 + \cosh 2 \Phi)\,,
\eeqs
which is the five-dimensional potential obtained by consistently truncating type IIB supergravity
on a Sasaki-Einstein manifold discussed in~\cite{Gubser} in the context 
of holographic superconductivity, and that has a very long history in the context of truncations of type IIB 
to 5D supergravity intended to yield the duals of controlled deformations of ${\cal N}=4$ SYM, being a special case of the
GPPZ flows~\cite{GPPZ}. 
In this case the full lift to 10-dimensional type-IIB supergravity is known. Specifying $\Delta=3$, the background scalar is
\beqs
\bar{\Phi}&=&\arctanh e^{-3 r}\,,
\label{Eq:cubicbg}
\eeqs
and the five-dimensional warp factor is
\beqs
A&=&{\frac{1}{6}}\log\left(-1+e^{6 r} \right)\,,
\eeqs
where we chose an integration constant in such a way that $A\rightarrow r$ 
far in the UV.
Notice how this background is practically the same as the one discussed in the previous example,
with the choices $\Delta=3$, $r_1=0$ and $\Phi_1=1$, aside from a very narrow region near the IR
boundary, where a singular behavior appears.

For our purposes, it is  interesting to study the spectrum with the background solution given by the first-order equations,
mainly because of its simplicity, which will help us elucidate on the effect of 
performing the calculation with the explicit boundary terms in presence of singular backgrounds. 
Unfortunately, because of the singularity in the IR, there is no small parameter
controlling the VEV to expand in.
As a consequence, one must rely on numerically solving the fluctuation equations.
Yet, we can obtain some semi-quantitative information from our previous results.
We will perform the numerical calculations at finite $r_2\gg r_1>0$, and then extrapolate 
to the cases where the IR and UV cutoffs are removed ($r_1\rightarrow 0$ and $r_2\rightarrow +\infty$).

As long as we choose $r_1 \gg 0$, the background being hardly different from the previous case,
we expect the same results to hold. In particular, because
the non-trivial departure is localized very close to the IR boundary, we do not expect any interesting changes in the spectrum of the heavy modes, for which all the approximations we made  
should still hold, up to a possible overall shift of the spectrum.
However, more interesting is the case of the light state. In this case, the fact that the VEV of $\Phi$ diverges
near the IR boundary means that the approximations we made might not hold. 
It  is hence very interesting to see what happens when $r_1\rightarrow 0$.
We will study this numerically in the next section.

We conclude with a few comments on the (real) supergravity background generated by this action.
It must be noted that this system has been extensively studied in the literature, and yields a background
that is well-known to be a  badly singular limit of the GPPZ system, which fails
to satisfy even the modest demands of~\cite{G}.

One way of seeing explicitly that a problem is present is the following.
The lift to 10 dimensions yields the metric~\cite{Gubser}\footnote{The dilaton is constant hence there is no real 
difference between string frame and Einstein frame.}
\beqs
\di s^2_{10}&=&\cosh \Phi \left(e^{2A}\di x_{1,3}^2+\di r^2\right)+\di \Omega_5\,,
\eeqs
with $\di \Omega_5$ the internal metric, which depends in general on $\Phi$ and $r$, besides the coordinates of 
the five-dimensional internal manifold.

The internal-space structure of the metric is not very important for the present discussion,
what matters is that we know the warp factor  $\cosh\Phi$ needed to  lift the five-dimensional metric
to ten dimensions.
This information allows to use the background and compute the Wilson loop. 
One must solve the classical problem of
determining the configuration of a (probe) string, the end-points of which are bounded to a D3-brane
fixed at some radial position, that we can identify with the UV cutoff $r_2$~\cite{wilson},
by minimizing the Nambu-Goto action.

Following the standard procedure (see also~\cite{NPR}), one first defines the functions $f^2=g_{tt}g_{xx}$ and 
$g^2=g_{tt}g_{rr}$, in terms of the elements of the 10-dimensional metric. The separation between the end-points of the string $L_{QQ}$
can be computed, by using the auxiliary effective potential $V_{eff}^2(r)=f^2(r)(f^2(r)-f^2(r_0))/(g^2(r)f^2(r_0))$,
as a function of the minimum value of the radial direction $r_0$ reached by the string in its fall into the radial direction, via
\beqs
L_{QQ}&=&2 \int_{r_0}^{r_2}\frac{\di \r}{V_{eff}(\r)}\,. 
\eeqs
We are not going to do this exercise here, but we want to observe the fact that
in the case we are discussing one finds that
\beqs
f^2&=&(1-e^{-6r})^{-1}(-1+e^{6r})^{2/3}\,
\eeqs
is not monotonic, but rather has a minimum at $\bar{r}=(1/6)\ln(3/2)$.
(Also, $f^2$ diverges at the singularity, which means that the model fails to satisfy any of the criteria in~\cite{G}, as anticipated).
In turns, this means that the string cannot fall all the way into the bulk towards the singularity,
but can at most reach down to $\bar{r}$.
Ultimately, this means that the singularity 
is bad enough that probing the system with extended objects is going to yield unphysical results,
and hence 
one should not think of this as a complete model, in which the  background
captures all the physics of the dual confining theory.
A resolution of the IR singularity would be needed.

\subsubsection{The $\Delta=1$ case}

The spectrum of the actual GPPZ system for less pathological cases
has been discussed for instance in~\cite{GPPZspectrum}, and 
the spectrum of this model for $\Delta=1$ is discussed for instance in the first reference in~\cite{BHM}.
We briefly digress here and redo this last calculation, which can be performed analytically, and which is of marginal
relevance to the rest of the paper.
We apply our procedure, keeping $r_2\gg r_1$ fixed, and considering values of $r_1 \ll 1$, very close to the singularity.
We limit ourselves to the $m_i\rightarrow +\infty$ case.
By using the fact that $N\equiv W_{\Phi\Phi}-(W_{\Phi})^2/W = -\Delta$, and specifying $\Delta=1$, the bulk equation for the fluctuations is
\beqs
\left(-1+e^{2r}\right)   \mathfrak{a}^{\prime\prime}(r)
+4 e^{2r} \mathfrak{a}^{\prime}(r)
+\left(q^2+1+3e^{2r}\right)
\mathfrak{a}(r) 
&=&0\,,
\eeqs
subject to the boundary conditions
\beqs
\left.-\frac{2}{q^2}\left(\partial_r+1\right)\mathfrak{a}-\mathfrak{a}\right|_{r_i}&=&0\,.
\eeqs
By solving this equation and imposing the boundary conditions, and then taking the $r_2\rightarrow +\infty$ limit, and $r_1\rightarrow 0$,
one finds that the spectrum is given by
\beqs
m^2_n&=&4n(n+1)\,, \,\,\,\,\,\,\,\,{\rm for }\,\,\,n=1,\cdots\,\infty\,,
\eeqs
in agreement with the literature.\footnote{Note, however, that for generic values of $m_i^2$, the spectrum would in general be different.}
Notice in particular that there are no parametrically light states: the lightest state has a mass $m^2=8$,
while extrapolating (outside its regime of validity)  Eq.~(\ref{Eq:dilatonmass}) computed  in Example A, with $\Delta=1$ and $\Phi_1=\sqrt{3}$ would yield
$m^2_d=4$.

Concluding this short exercise, let us make two comments inspired by the fact that these results agree with the literature.
First of all, the boundary conditions we use do not rely on the concepts of {\it normalizability} and/or {\it regularity}.
They are simply defined algebraically, in terms of the background functions, and there is no need to analyze on a model-by-model basis the 
fluctuations near special points, in order to decide what is physically acceptable and what not: the whole procedure is automatically taking care of this,
because the boundary actions implement (both on the background and on the fluctuations) all the physical requirements.
Second, for this particular example, the procedure we follow, in which two cutoffs are explicitly present, ultimately yields the same physical results as other procedures, once the limit of removing the cutoffs is taken (provided we take $m_i^2 \rightarrow \infty$). This suggests a general procedure for how to calculate the spectrum in the case of when a singularity may be present for the background, assuming the IR and UV behaviors of the background are not too pathological, and the limit of removing the cutoffs can be taken without difficulties.

%%%%%%%%%%%%%%%%%%%%%%%%%%%%%%%
%%%%%%%%%%%%%%%%%%%%%%%%%%%%%%%
\subsection{Example C: a phenomenological model with cubic superpotential}

Consider  now the following superpotential, for one scalar field $\Phi$ with trivial sigma-model:
\beqs
W&=&-\frac{3}{2}-\frac{\Delta}{2}\Phi^2+\frac{\Delta}{3\Phi_I}\Phi^3\,.
\eeqs
This amounts to including a cubic correction to the quadratic potential we studied earlier.
This superpotential admits two fixed points for $\Phi=\Phi_U=0$ and $\Phi=\Phi_I$.

The potential,  when expanded near the two fixed points, yields respectively
\beqs
V_U&=&-3+\frac{1}{2}\Delta(\Delta-4)\Phi^2\,+\cdots\,,\\
V_I&=&-3-\frac{2 \Delta \Phi_I^2}{3} - \frac{\Delta^2 \Phi_I^4}{27}+\frac{1}{2}\Delta\left(\Delta+4+\frac{4\Delta\Phi_I^2}{9}\right)\left(\Phi-\Phi_I\right)^2\,+\cdots\,.
\eeqs
In order for the $\Phi_I$ to be an attractive IR fixed point, the cosmological constant must be negative and have larger absolute value than it has
at the UV fixed point. 
Which is  true for $\Delta \Phi_I^2> - 18$, and in particular for any positive value of $\Delta$.

The solution of the background equations can be obtained by simply integrating 
\beqs
\partial_r \Phi&=&\frac{ \di W}{\di \Phi}\,=\,-\frac{\Delta}{\Phi_I} \Phi(\Phi_I- \Phi)\,.
\eeqs
We conventionally choose $\Phi_I>0$. Then the solution for $\Phi>\Phi_I$ of $\Phi<0$ is
\beqs
\Phi_Q&=&\frac{\Phi_I}{e^{\Delta (r-r_0)}-1}\,,
\eeqs
where $r_0$ is an integration constant. For $r>r_0$ the result is a flow away from the UV
fixed point $\Phi=0$, towards a singularity at $r\rightarrow r_0$ at which $\Phi\rightarrow -\infty$.
For $r<r_0$, $\Phi_Q$ describes a flow from asymptotically large values of $\Phi>\Phi_I$ near $r\rightarrow r_0$,
to the IR fixed point $\Phi_I$ when $r\rightarrow -\infty$.

More interesting is the solution obtained when setting the boundary condition so that $0<\Phi<\Phi_I$.
In this case
\beqs
\bar{\Phi}&=&\frac{\Phi_I}{e^{\Delta (r-r_{\ast})}+1}\,,
\eeqs
where again $r_{\ast}$ is an integration constant, which in this case has a very different meaning.
The flow described from $\bar{\Phi}$ connects the two fixed points, running from $\Phi\rightarrow 0$ when $r\gg r_{\ast}$,
to $\Phi\rightarrow \Phi_I$ when $r\ll r_{\ast}$. One can easily see that $\Phi(r_{\ast})=\Phi_I/2$, so that $r_{\ast}$ is
the scale that separates the regimes in which the theory can be described as a deformation of the two fixed points, respectively.

We are interested in studying the spectrum of the theory defined by the background $\bar{\Phi}$.
Strictly speaking, this is continuous. We perform the study by assuming that there exist two hard-wall cutoffs
in the IR and UV, such that $r_I \ll r_{\ast} \ll r_U$.
The scales $r_{I,U}$ are clearly spurious, representing cutoffs put in by hand.
In particular, the IR scale should appear as a consequence of a relevant deformation driving the flow away
from the IR fixed-point. This requires extending the system, with more scalars being included.
Also, the ultimate end-of-space (singularity) will determine the spectrum. In particular, it may be that the physics 
near the singularity, responsible for its resolution, can be described only by embedding the model into a full 10-dimensional supergravity, or even superstring theory. This is well beyond the scopes of this simple phenomenological model, and 
hence one should not be too much concerned about the dependence of the masses (in particular of the lightest modes) on $r_I$.
Yet, it is of interest to perform this calculation in order to understand how the spectrum depends on $r_{\ast}$,
for fixed choices of $r_{I,U}$. 
This will offer some guidance as to what happens in actual string-theory models describing 
the field theory RG flow between UV and IR fixed points.

%%%%%%%%%%%%%%%
%%%%%%%%%%%%%%%
%\newpage
\section{Numerical studies}
\label{Sec:5}

In this section we present a set of numerical studies of the backgrounds introduced in the
previous section. Besides allowing us to check explicitly some of the results, this also 
allows us to understand how good our approximations are.

\subsection{Quadratic superpotential}
\label{Sec:5a}

We start from the system with quadratic superpotential,
considering generic values of $\Delta$.
The first thing we want to do is to understand how precise
our results in Eqs.~(\ref{Eq:dilatonmass}) and (\ref{Eq:GWgeneral}) are.
Figure~\ref{Fig:GWDelta} shows the dependence of 
the mass of the dilaton on the dimension $\Delta$, computed in the limit $m_i^2\rightarrow +\infty$.
For (not necessarily very) small values of $\Phi_1$, the agreement of the numerical results with 
Eq.~(\ref{Eq:dilatonmass}) is very remarkable (see the left panel in Figure~\ref{Fig:GWDelta}).
The agreement deteriorates for larger values of $\Phi_1$, yet the qualitative features are preserved.\footnote{In the right panel of Figure~\ref{Fig:GWDelta}, a couple of extra points can be seen for small $\Delta$. Let us note that these are spurious states, in the sense that if one were to take the limit of $r_2 \rightarrow \infty$, their masses would diverge and they would decouple from the rest of the spectrum.}

A set of remarkable physics lessons that can be read directly off Eq.~(\ref{Eq:dilatonmass}) are confirmed.
\begin{itemize}

\item There are no tachyons. At least at this level, there is no reason to question the stability of 
the backgrounds we are studying.

\item There is unmistakeable evidence that two very different behaviors appear for $\Delta>2$ and $\Delta<2$,
indicating the fact that $\Delta=2$ is a very special point of the parameter space, with peculiar physical features.

\item When $\Delta<2$, the mass of the light state depends crucially on the dimensionality of the dual operator $\Delta$,
on the normalization of the five-dimensional VEV $\Phi_1$ and on the IR scale $r_1$, but not on $r_2$, the UV cutoff.
With the specific choices we made in the plots, it is clear that even at moderate values of $r_2$ this dependence amounts to
a subleading effect.

\item The mass $m_d^2$ is anomalously light not only for $\Delta\ll 1$, which is a well-known and studied result~\cite{GW3, Kofman:2004tk},
but also when $\Delta\simeq 2$. This is due to the fact that when the limit $\Delta \rightarrow 2$ 
is taken in Eq.~(\ref{Eq:dilatonmass}), the result $m^2_d \sim \frac{8 \Phi_1^2}{3 r_2}$ is suppressed by $1/r_2$.

\item For $\Delta>2$ the mass is very strongly dependent on $r_2$, being exponentially suppressed in the limit of $r_2 \rightarrow \infty$ (taking the limit by holding  $\Phi_1$ fixed) with $m^2_d \sim \frac{4}{3} e^{-2 (\Delta - 2) r_2} (\Delta -2) \Delta^2 \Phi_1^2$. Thus, provided $r_2$ is very large, $m^2_d$ vanishes, for all practical purposes, for all $\Delta>2$. Yet, this statement is very cutoff dependent, and needs to be taken with caution.

\end{itemize}

\begin{figure}[t]
\begin{center}
\begin{picture}(430,150)
\put(0,0){\includegraphics[height=125pt]{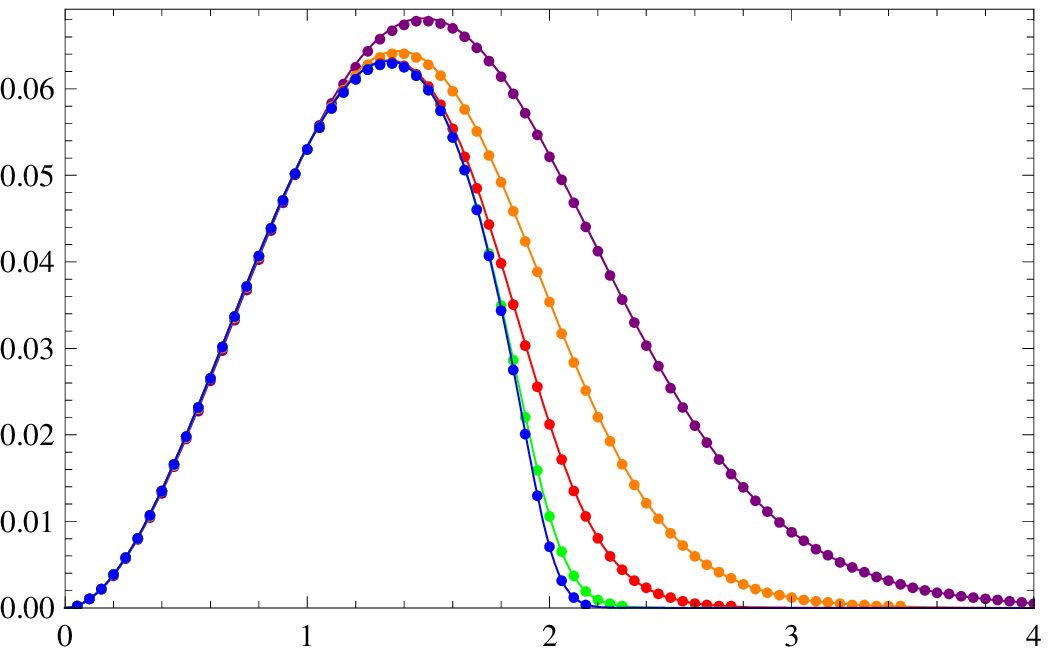}}
\put(230,0){\includegraphics[height=125pt]{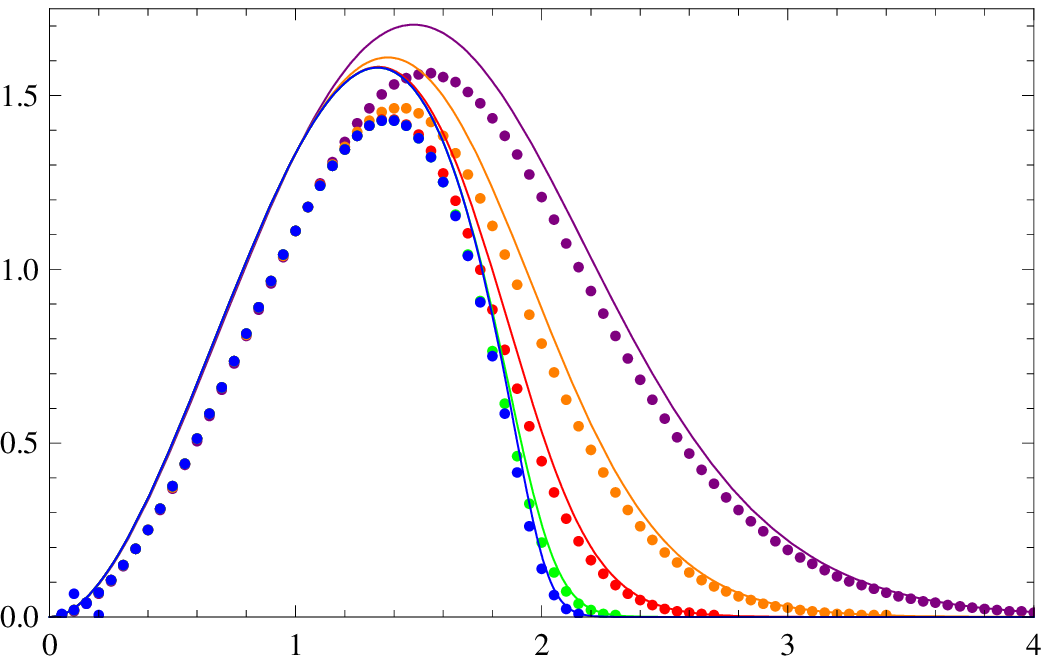}}
\put(3,130){$m_d^2$}
\put(230,130){$m_d^2$}
\put(205,5){$\Delta$}
\put(432,5){$\Delta$}
\end{picture}
\caption{Numerical results. Mass $m_d^2$ of the lightest scalar fluctuation as a function of $\Delta$.
The localized mass terms are divergent ($m_i^2\rightarrow +\infty$).
All plots with $r_1=0$, and $r_2=2,3,5,10,15$ (with faster fall-off for $\Delta > 2$ the higher the UV cutoff). Continuous lines represent  the approximation
in the body of the paper, while the points are numerical results. 
The background is chosen to have $\Phi_1=0.2$ in the left panel, and $\Phi_1=1$ in the right panel.
  }
\label{Fig:GWDelta}
\end{center}
\end{figure}

\begin{figure}[htpb]
\begin{center}
\begin{picture}(470,320)
\put(0,160){\includegraphics[height=140pt]{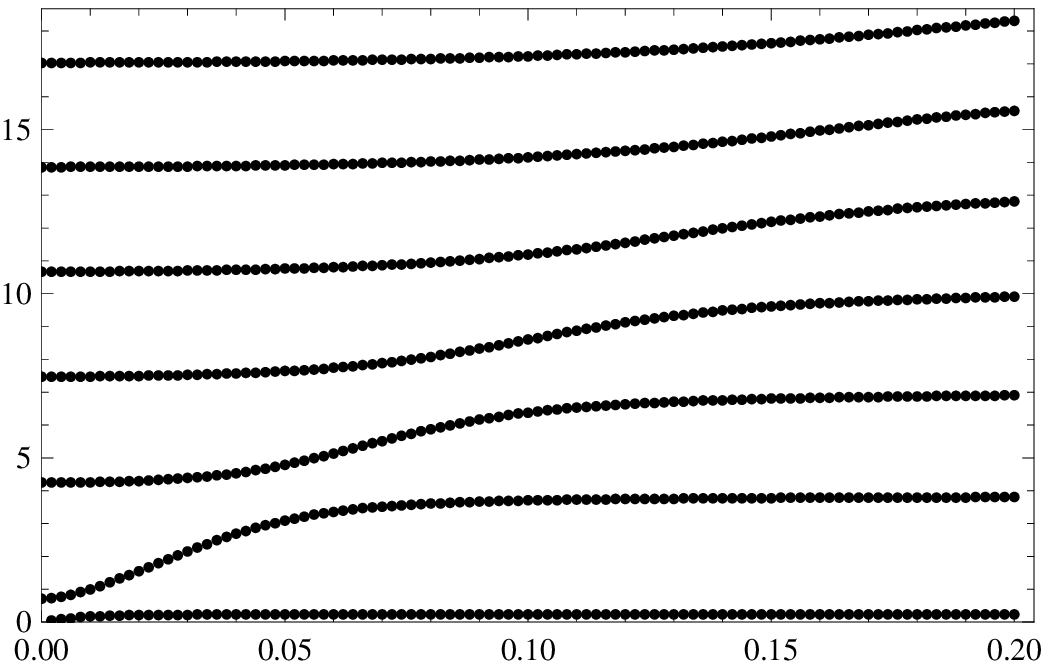}}
\put(240,160){\includegraphics[height=140pt]{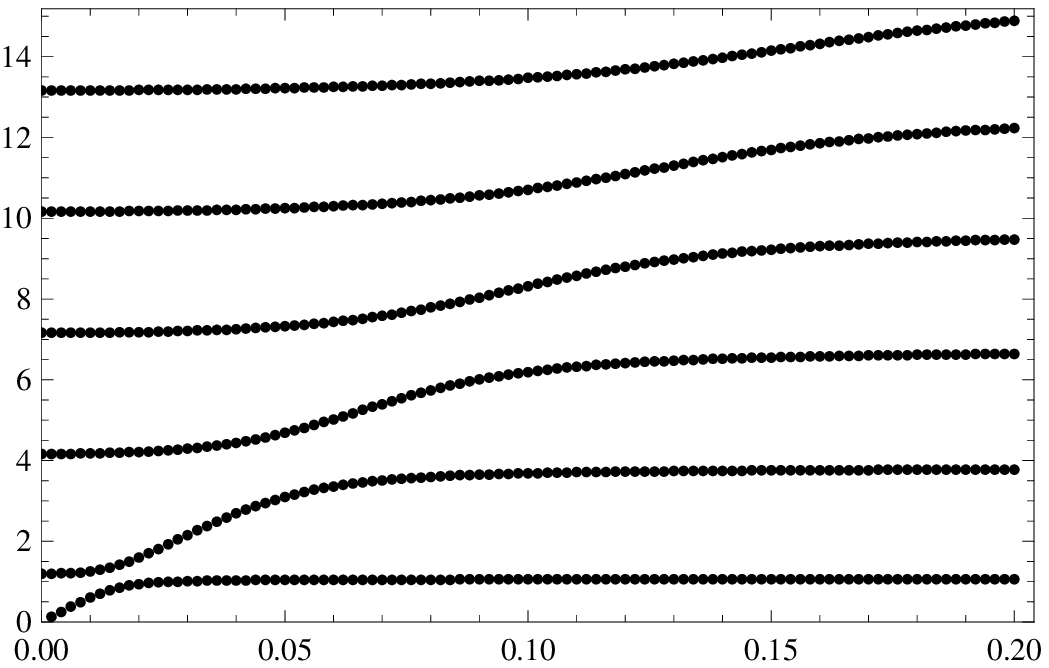}}
\put(0,0){\includegraphics[height=140pt]{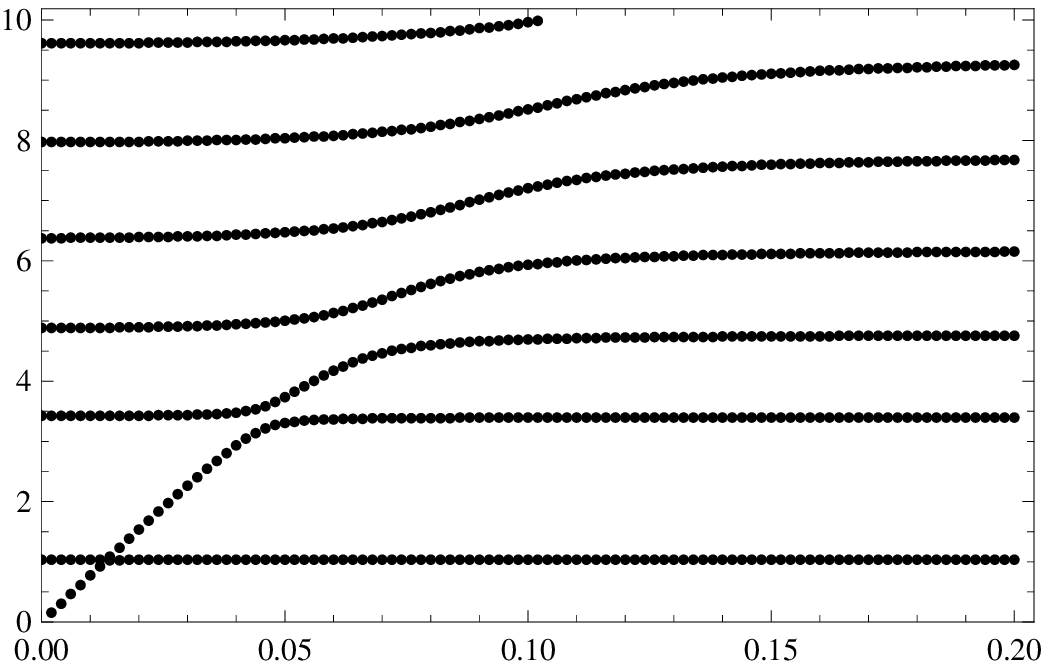}}
\put(240,0){\includegraphics[height=140pt]{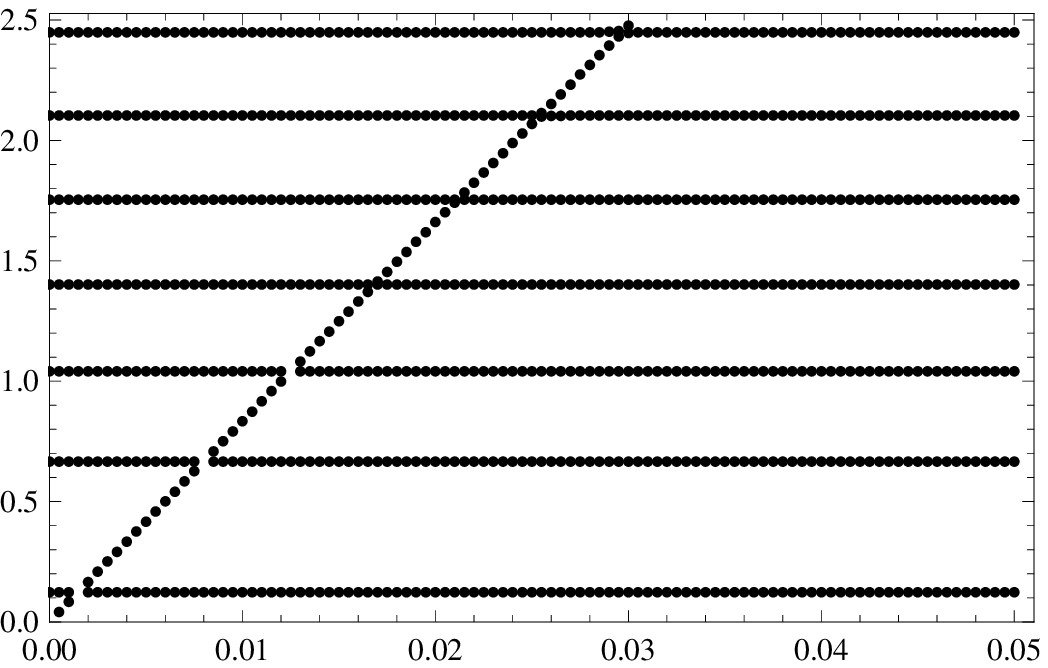}}
\put(5,145){$M$}
\put(5,305){$M$}
\put(245,145){$M$}
\put(245,305){$M$}
\put(220,7){$m_2$}
\put(460,7){$m_2$}
\put(222,167){$m_2$}
\put(462,167){$m_2$}
\end{picture} 
\caption{Numerical results. Mass $M$ of the lightest few scalar states, for $\Delta=1$,  $r_1=0$, $r_2=5$, plotted as a function of the boundary mass $m_2$, for $m_1\rightarrow+\infty$. 
The four plots differ for the choice of $\Phi_1=0.2,1,3,5$ (left to right and top to bottom).
 }
\label{Fig:GWm2}
\end{center}
\end{figure}

\begin{figure}[htpb]
\begin{center}
\begin{picture}(470,320)
\put(0,160){\includegraphics[height=140pt]{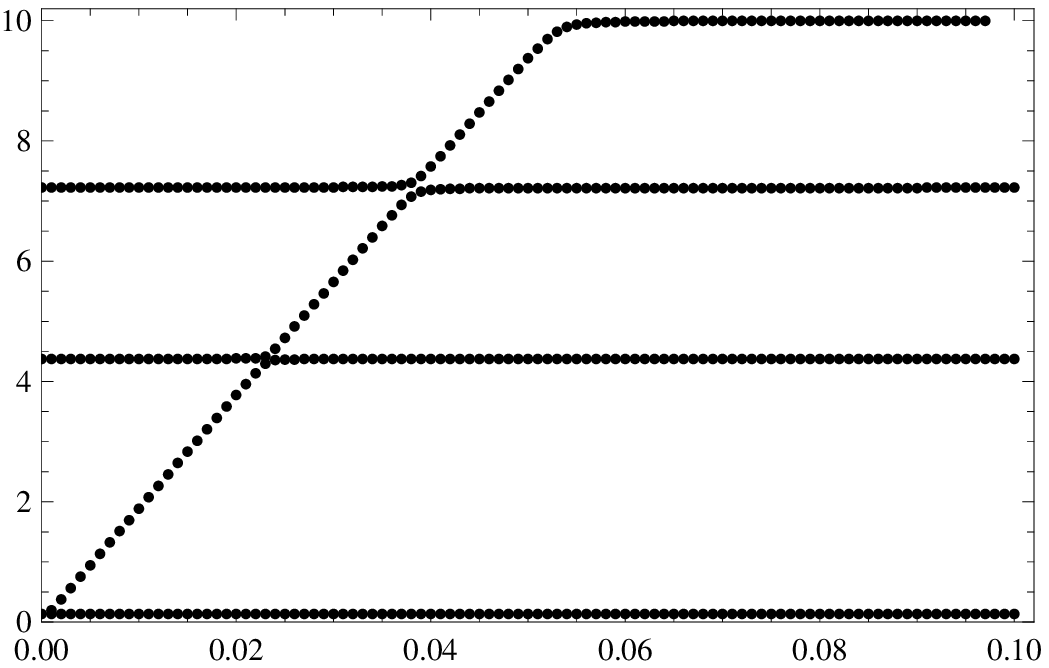}}
\put(240,160){\includegraphics[height=140pt]{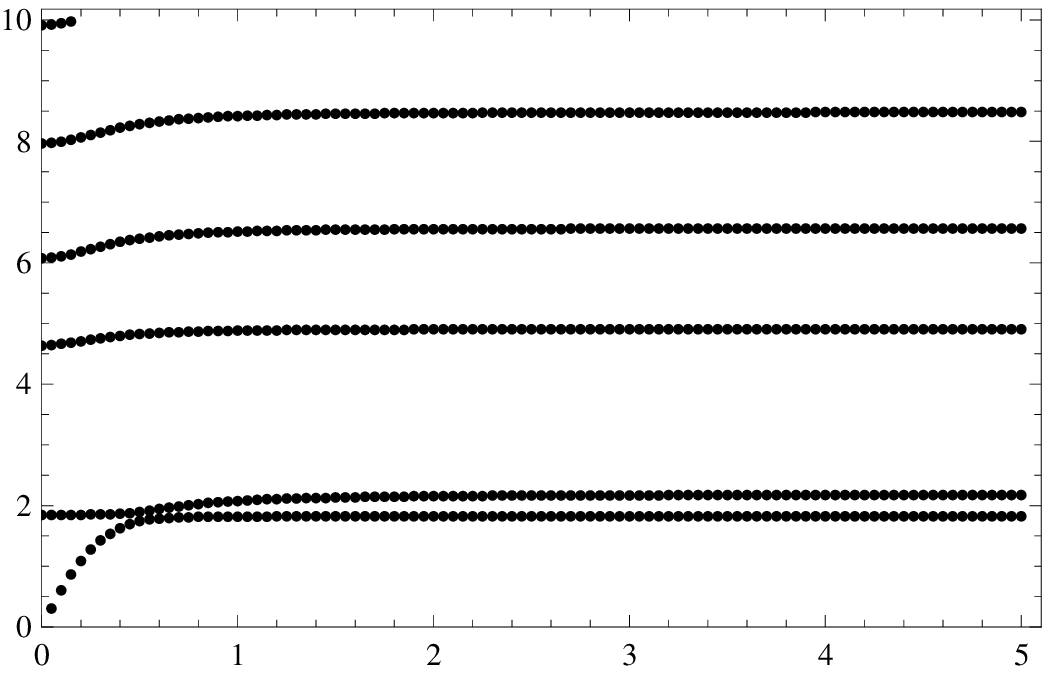}}
\put(0,0){\includegraphics[height=140pt]{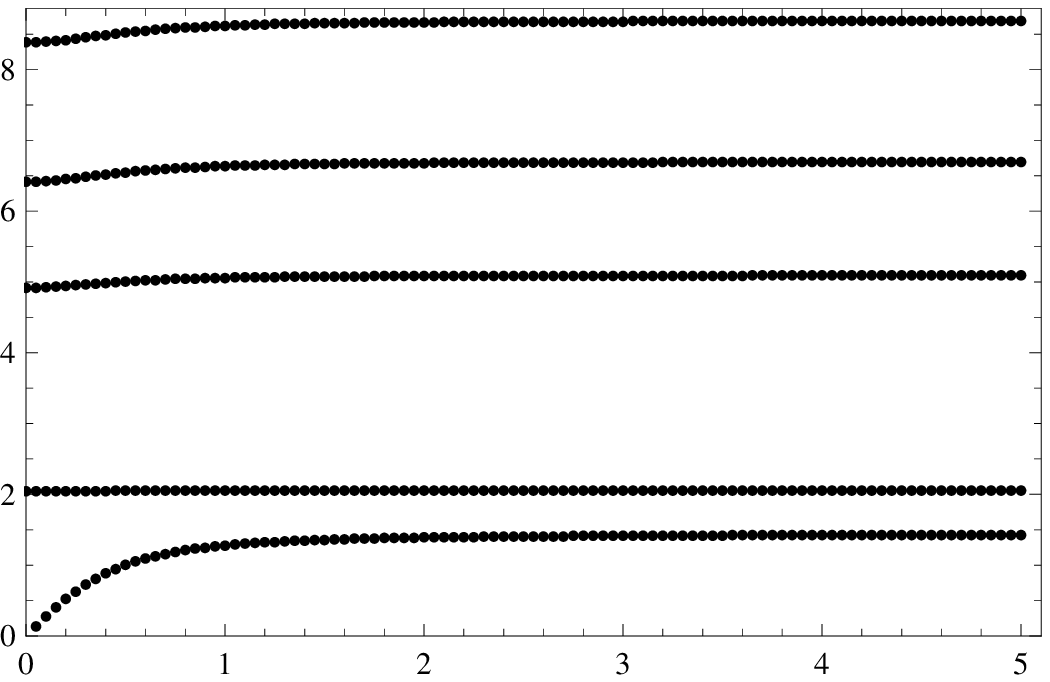}}
\put(240,0){\includegraphics[height=140pt]{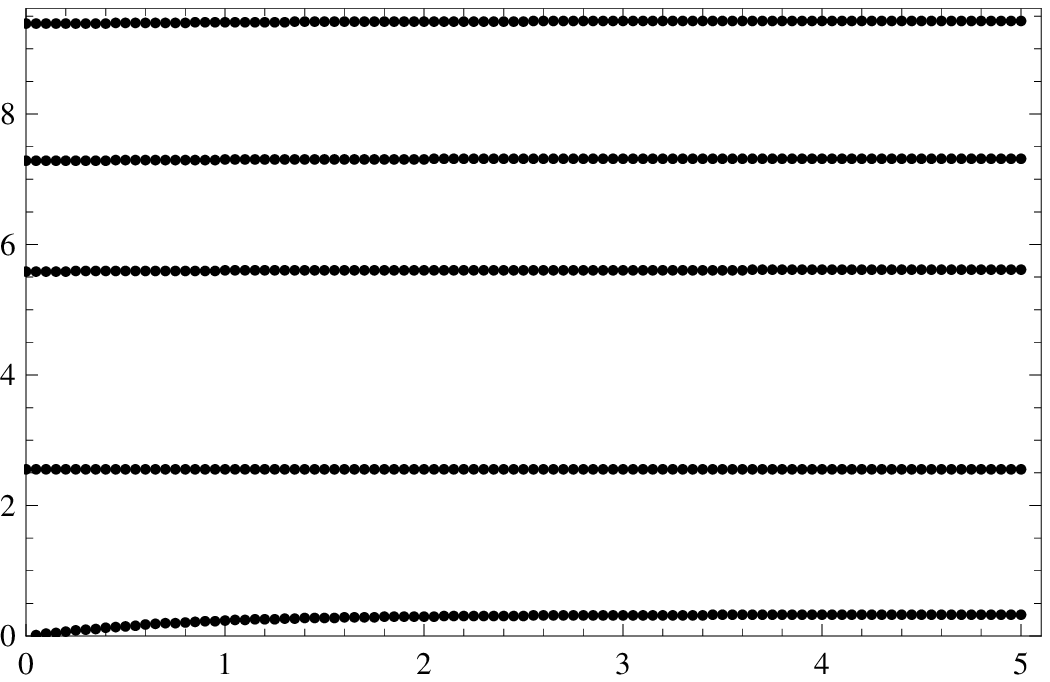}}
\put(3,145){$M$}
\put(3,305){$M$}
\put(243,145){$M$}
\put(243,305){$M$}
\put(220,7){$m_2$}
\put(460,7){$m_2$}
\put(222,167){$m_2$}
\put(462,167){$m_2$}
\end{picture} 
\caption{Numerical results. Mass $M$ of the lightest few scalar states, for $\Phi_1=3$, $r_1=0$ and $r_2=5$, plotted as a function of the boundary mass $m_2$, for $m_1\rightarrow+\infty$. 
The four plots differ for the choice of $\Delta=0.2,1.8,2,2.5$ (left to right and top to bottom).
 }
\label{Fig:GWm2more}
\end{center}
\end{figure}

The more general expression in Eq.~(\ref{Eq:GWgeneral}) taught us a few important subtleties
related to these kinds of systems. In particular, we already explained  that the five-dimensional
sigma-model formalism may yield spectra containing many light states, which have nothing to do with scale invariance.
 They might be related to the light techni-pions and/or techni-axions 
of a generic technicolor model, rather than having to do with the techni-dilaton.
Of course, because all the global symmetries (including scale invariance of the dual theory) are only approximate, 
the spectrum results from non-trivial mixing among all possible scalar bound states. This also implies that 
one has to be very careful in identifying the nature and couplings of the physical (mass eigenstate) states.
It is hence a useful exercise to study this problem within this very simplified model,
by studying explicitly how the boundary terms $m_i^2$ distort the scalar spectrum.

This is done in the four panels of Figures~\ref{Fig:GWm2} and~\ref{Fig:GWm2more}.
Figure~\ref{Fig:GWm2} focuses on values of $\Delta=1$ and $\Phi_1=0.2,1,3,5$, for $r_1=0$ and $r_2=5$.
We keep $m_1^2\rightarrow +\infty$, but vary $m_2$.
A few very interesting results emerge, which are very general.

\begin{itemize}

\item When $m_2\rightarrow 0$, one of the masses vanishes. We already explained the reason for this in Section~\ref{Sec:4a}.

\item An interesting level-crossing pattern develops at intermediate values of $m_2$. In particular, this shows explicitly how
the mixing between the states is very non-trivial. The composition (in terms of the original fluctuations of scalars and metric)
of the states corresponding to the pseudo-dilaton is in general very complicated.

\item The plots are restricted to the physically acceptable region of parameter space in which $m_2^2>0$. However,
notice how the mass of the lightest state vanishes  as a function of $m_2$ for  $m_2\rightarrow 0$.
If one where to look at large negative values of $m_2^2$, the spectrum would contain a tachyon.
This means that in setting up one of these models, some attention has to be given not only to how the (super)potential is chosen,
to what background solutions one studies, but also to which boundary terms are present.

\item It is only for small values of $m_2^2$ that the spectrum differs significantly from that of the $m_i^2 \rightarrow \infty$ limit. Already for $m_2^2 \sim \mathcal O(1)$, the spectrum starts to look the same as for $m_i^2 \rightarrow \infty$, which therefore is a limit that captures the generic behavior. Conversely, taking $m_i^2$ to be small should be thought of as a kind of fine-tuning.

\end{itemize}

Figure~\ref{Fig:GWm2more} shows another remarkable fact.
By varying $\Delta$, one can see that for small choices of $\Delta$ and of $m_2$, there are actually two abnormally light states in the spectrum.
By looking at the superpotential, it is immediately evident that what is happing in this case is that the two are related to the dilaton and a pseudo-Goldstone boson, the latter emerging from the near flatness of the sigma-model scalar potential.
As soon as $m_2$ and $\Delta$ become generic $O(1)$ numbers, both these states will in general be heavy, with masses dictated by the general IR scale
that controls all of the spectrum.
However, when $\Delta>2$ there is always one light state (whose mass is suppressed by the UV cutoff) irrespective of $m_2$.

Let us make a final remark for the reader, who might find it very bizarre that the typical scale of the KK masses in the 
 panels in Figures~\ref{Fig:GWm2} and~\ref{Fig:GWm2more}  are so different.
We said earlier on that the scale controlling the gaps is simply related to $r_1$, and hence one might have
expected the heavy states to have very similar masses, up to an overall shift controlled by $\Phi_1$. While this is not what the figures seem to show, there is indeed no contradiction, for a subtle reason, which ultimately has to do with how we decided to 
perform the numerical comparison between the three different backgrounds, rather than with the physics. The subtlety emerges from the following observation.
In all the backgrounds we are looking at, at asymptotically high values of $r$ the background
is characterized by a vanishing small value for $\bar{\Phi}(r)$, and consequently the warp factor 
$\bar{A}(r)\simeq r + a_0$. We made the choice of setting $a_0=0$ in all cases.
This ensures that the backgrounds are asymptotically all the same, with the same normalizations 
for the 4-dimensional Minkoski space-time variables. In this way, the value of $r_2$ 
always corresponds to the same UV cutoff scale.
However, in the IR the metric is going to change, because $\bar{\Phi}(r)\neq 0$.
The larger the value of $\Phi_1$, the larger the departure from AdS. And, more importantly,
the departure will appear at higher values of $r$. 
As a consequence, it is not true that by keeping the same value of $r_1$ in backgrounds with different $\Phi_1$
one is introducing an IR cutoff at the same scale.
In other words, the numerical value of $r_1$ is not an actual physical scale. It can be converted into such a scale
only as a function of all the other parameters in the background.
We will see a much cleaner example of this later on in Example C, when talking about backgrounds that describe the flow between two fixed points,
for which the geometry interpolates between two AdS spaces with different curvature.

\subsection{Example from consistent truncation}

Here we will study Example B numerically, and focus in particular on the case when the parameter $\Delta$ in the superpotential is fixed to be $\Delta = 3$. As explained before, the model can 
then be thought of as having a stringy origin. The background develops a singularity in the IR at $r=0$. 
However, as shown in Figure~\ref{Fig:SCbackgrounds}, the background fields $\bar{\Phi}$ and $A$ are almost indistinguishable from 
those used in Example A, with the choices $\Delta=3$ and $\Phi_1=e^{-\Delta r_1}$, with the exception of a very narrow region close to the singularity.

We study how the spectrum depends on where we put the IR cutoff. 
In particular we focus on what happens in the limit of letting $r_1 \rightarrow 0$, keeping the UV cutoff fixed. 
The results are plotted in Figure~\ref{Fig:SCIR}, where we also show the behavior of the Example A for comparison. There is a tower of KK modes, and a light scalar. 
Remarkably, despite the presence of an IR singularity, all states in the spectrum converge to their respective (finite) values as the IR cutoff is taken close to the position of the singularity. Also, while a shift in the masses of the heavy states is clearly visible, this effect is very suppressed for the lightest mass. This is a very interesting result, since naively one would expect that the lightest state would be the most sensitive to the IR singularity!

\begin{figure}[t]
\begin{center}
\begin{picture}(430,150)
\put(0,0){\includegraphics[height=132pt]{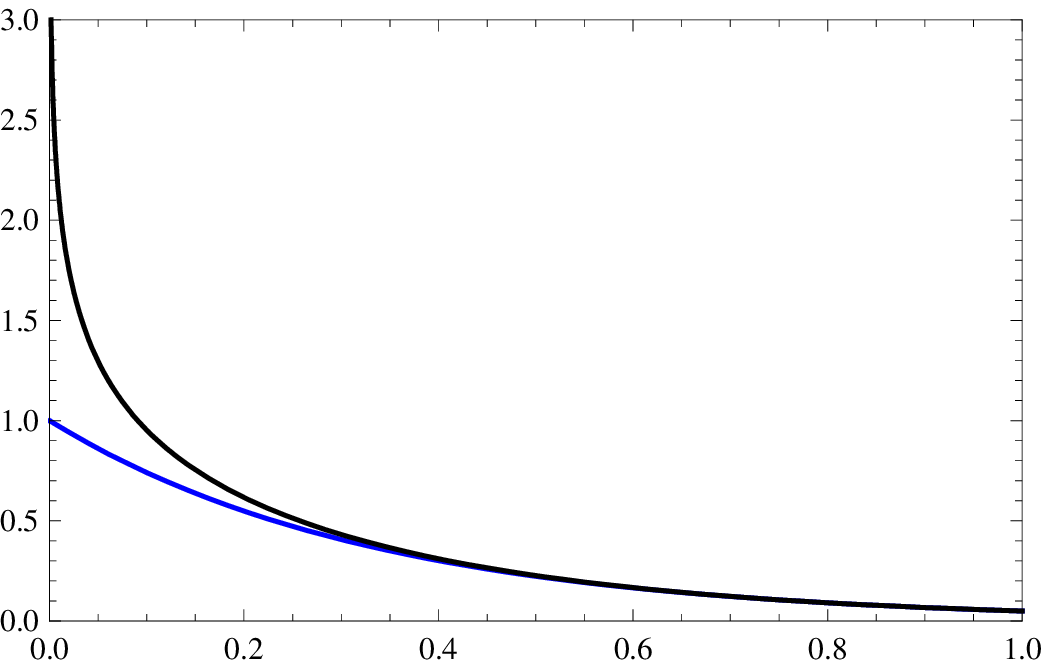}}
\put(230,0){\includegraphics[height=132pt]{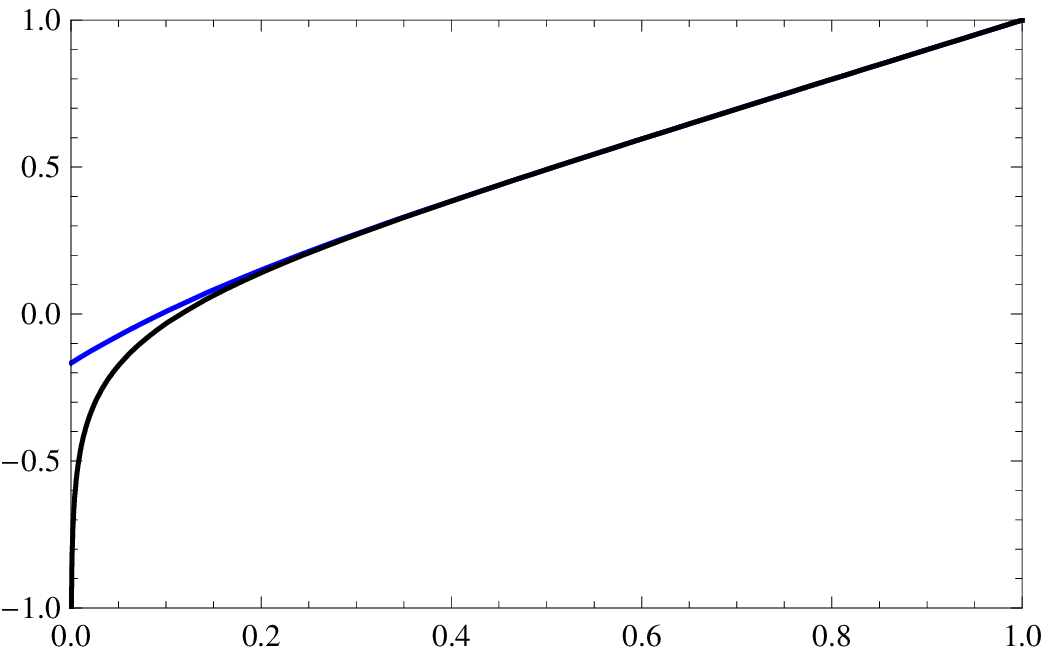}}
\put(6,135){${\bar{\Phi}}$}
\put(240,135){$A$}
\put(209,6){$r$}
\put(444,6){$r$}
\end{picture} 
\caption{The background functions $\Phi$ (left panel) and $A$ (right panel), used in Example B (black lines),
compared to the case discussed in Example A, for the choices $\Delta=3$ and $\Phi_1=e^{-\Delta r_1}$ (blue lines). }
\label{Fig:SCbackgrounds}
\end{center}
\end{figure}

\begin{figure}[htpb]
\begin{center}
\begin{picture}(430,150)
\put(0,0){\includegraphics[height=138pt]{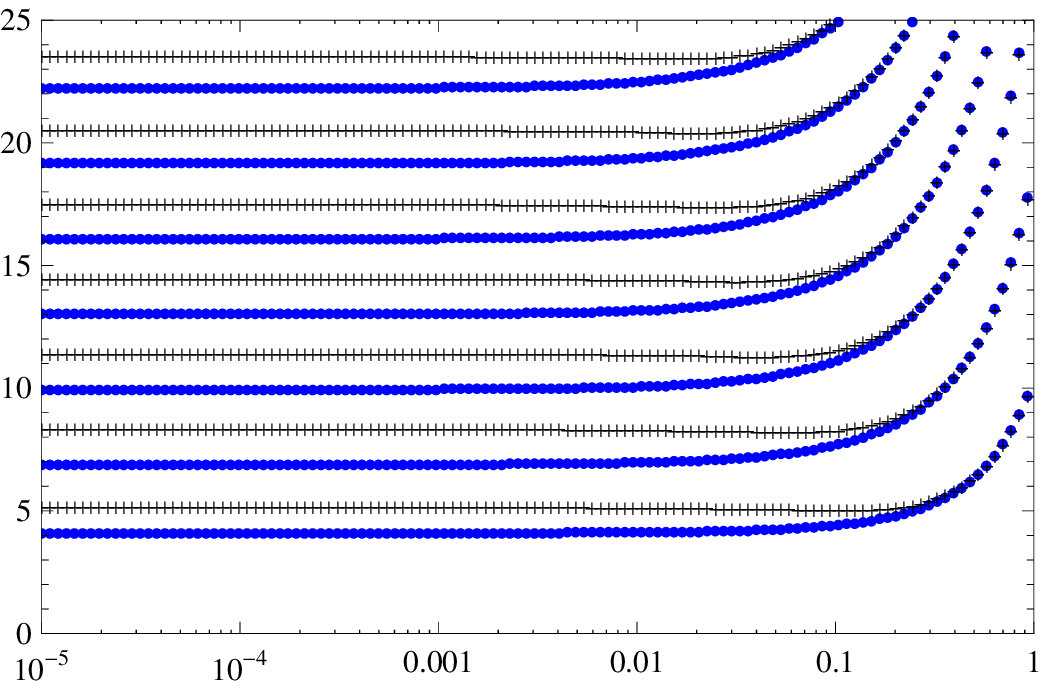}}
\put(230,0){\includegraphics[height=138pt]{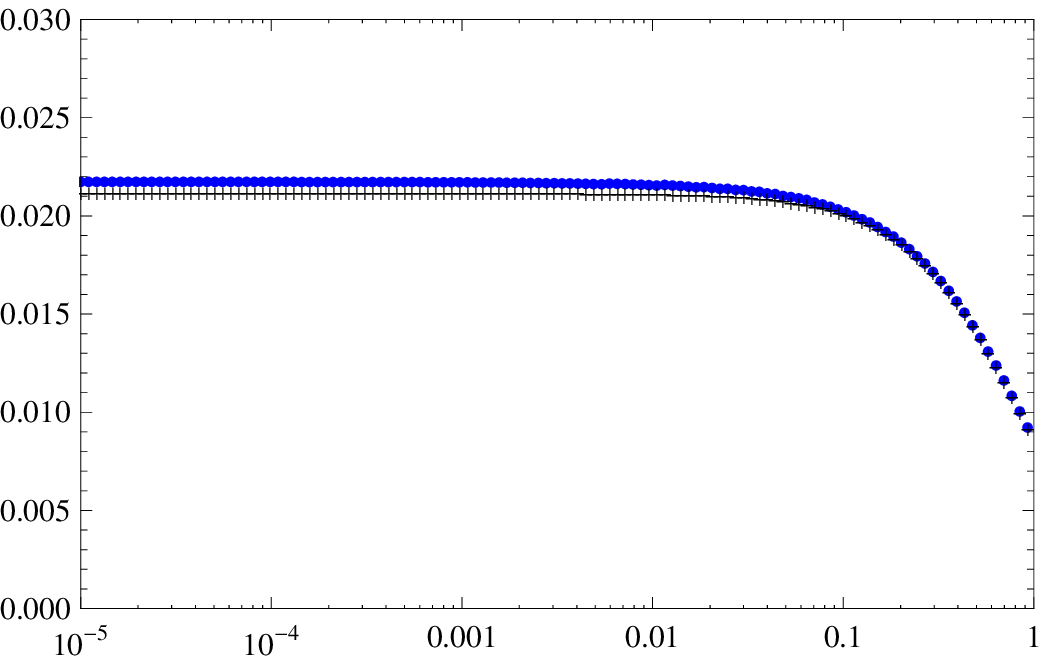}}
\put(4,139){${M}$}
\put(240,139){$M$}
\put(206,9){$r_1$}
\put(444,9){$r_1$}
\end{picture} 
\caption{Numerical results. Mass $M$ of the lightest scalar fluctuations, keeping the UV cutoff fixed at $r_2 = 5$ while varying the IR cutoff $r_1$. The right panel shows a detail of the left one, in which the very lightest state is not visible (notice the scale). The plots show the comparison between the 
results of Example B (black $+$) and of Example A with $\Delta=3$ and $\Phi_1=e^{-\Delta r_1}$ (blue).}
\label{Fig:SCIR}
\end{center}
\end{figure}

\begin{figure}[htpb]
\begin{center}
\begin{picture}(430,150)
\put(0,0){\includegraphics[height=128pt]{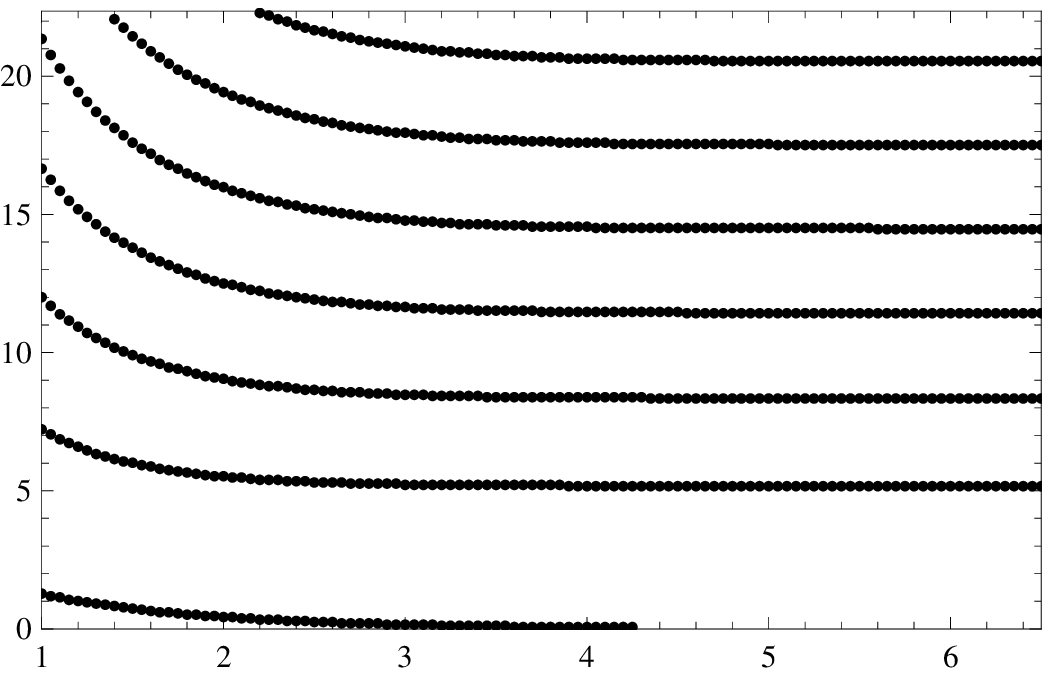}}
\put(230,0){\includegraphics[height=125pt]{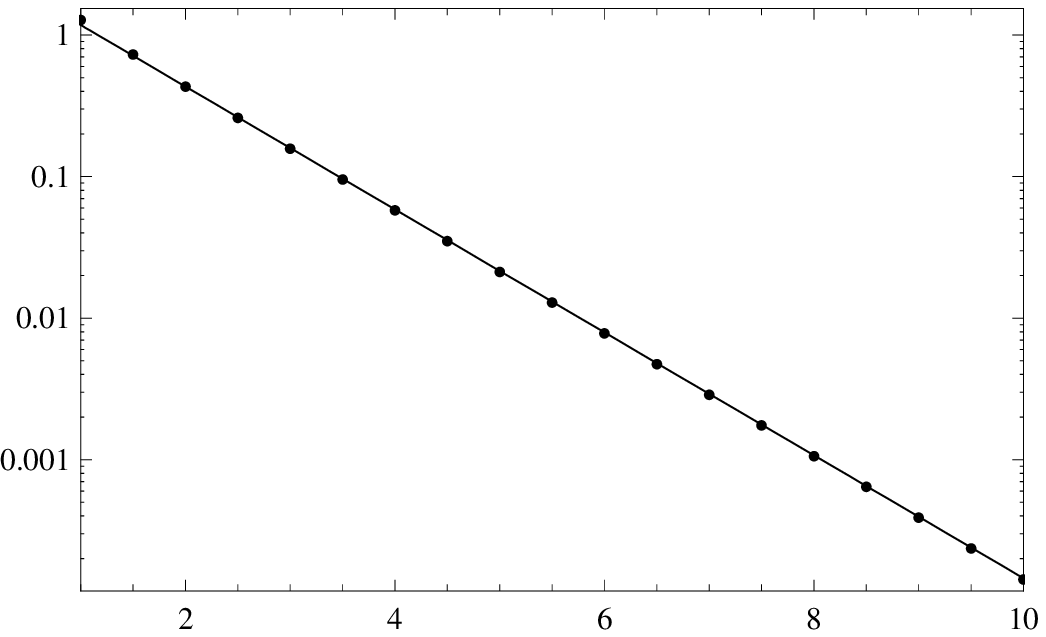}}
\put(5,133){${M}$}
\put(238,133){$M$}
\put(205,8){$r_2$}
\put(438,8){$r_2$}
\end{picture} 
\caption{Numerical results for Example B. Mass $M$ of the lightest scalar fluctuations, 
keeping the IR cutoff fixed at $r_1 = 10^{-6}$ while varying the UV cutoff $r_2$. The right panel shows a detail of the left one, with the line being the appoximation $M \propto e^{-r_2}$. }
\label{Fig:SCUV}
\end{center}
\end{figure}

In Figure~\ref{Fig:SCIR}, the dependence of the spectrum on the UV cutoff is shown. As can be seen, the mass of the light scalar tends to zero in the limit of infinite $r_2$, whereas the rest of the spectrum stabilizes.

\subsection{Phenomenological example: cubic superpotential}

This is to be understood as a toy model describing the RG flow between two fixed points.
In Figure~\ref{Fig:Walkingbg} a set of possible backgrounds for this model 
are shown, with the background functions $\bar{\Phi}$ and ${A}$ explicitly plotted
as a function of $r$. The various backgrounds share the same choices of $\Delta=3$, $\Phi_I=1$,
$r_1=0$ and $r_2=5$. The different curves are obtained by varying $r_{\ast}$,
the value of the radial direction at which the nature of $\bar{\Phi}(r)$ changes. 
For higher values of $\Delta$, the kink becomes more localized.

\begin{figure}[t]
\begin{center}
\begin{picture}(430,150)
\put(0,0){\includegraphics[height=125pt]{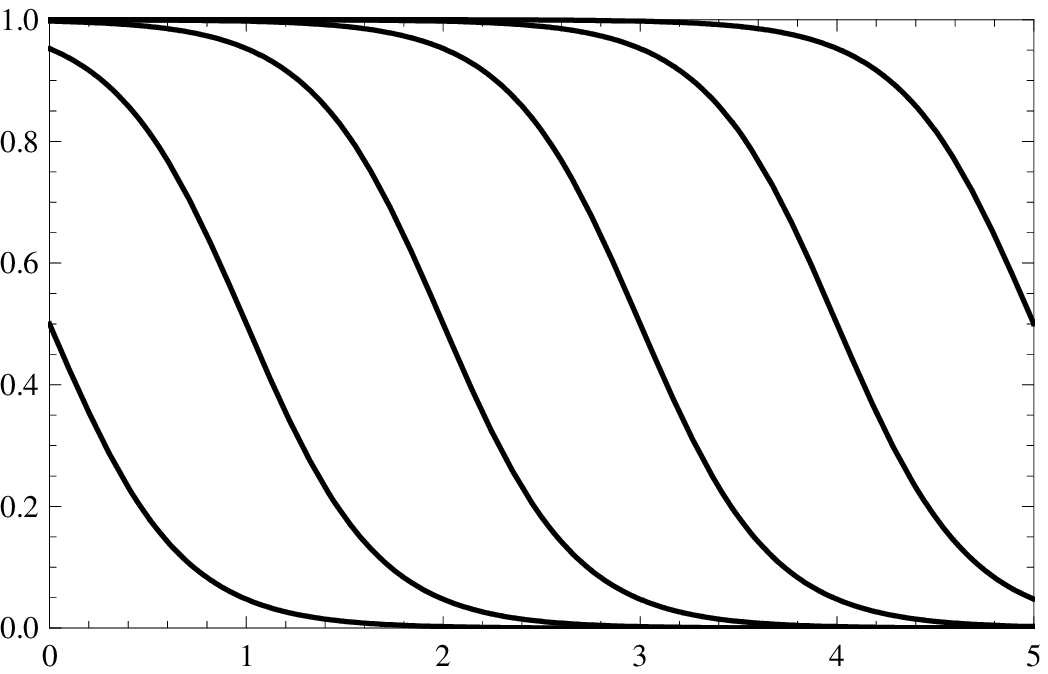}}
\put(230,0){\includegraphics[height=125pt]{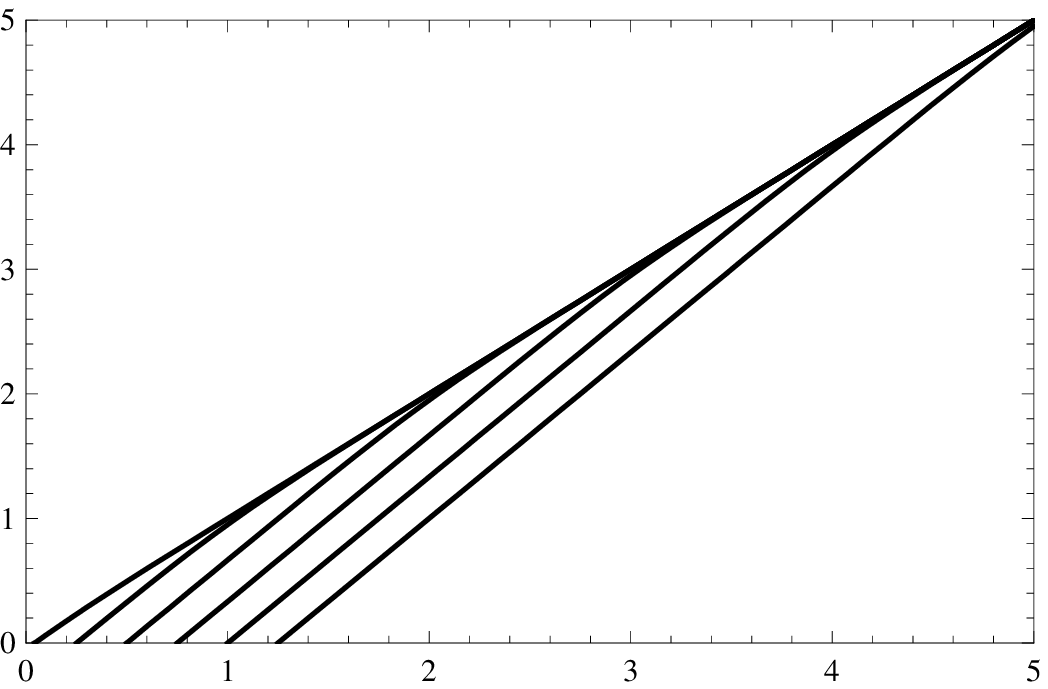}}
\put(5,130){${\bar{\Phi}}$}
\put(232,130){${A}$}
\put(200,6){$r$}
\put(427,6){$r$}
\end{picture} 
\caption{Sampling of the functions determining  the background in Example C, and used in the numerical analysis, plotted against the  
radial coordinate $r$. Plots obtained with $\Delta=3$, $\Phi_I=1$, $r_1=0$ and $r_2=5$.
The backgrounds differ in the choice of $r_{\ast}$. Notice that the integration constant in the warp factor ${A}$ is chosen so that
the warp factor of all the different backgrounds agrees in the far UV ($r\rightarrow \infty$). }
\label{Fig:Walkingbg}
\end{center}
\end{figure}

\begin{figure}[htpb]
\begin{center}
\begin{picture}(430,440)
\put(0,300){\includegraphics[height=125pt]{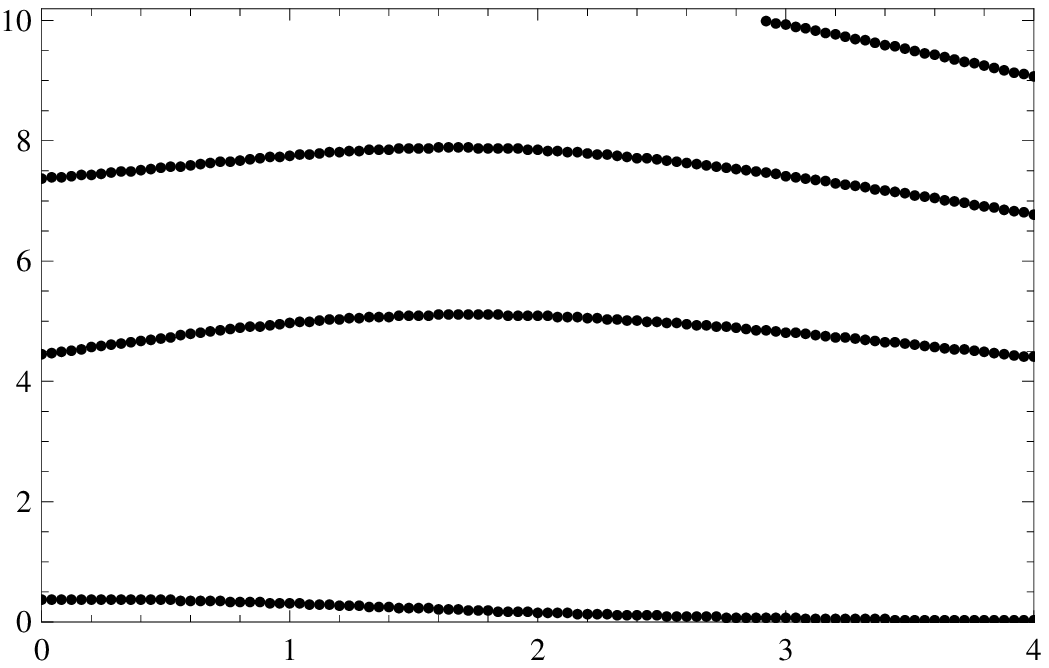}}
\put(230,300){\includegraphics[height=125pt]{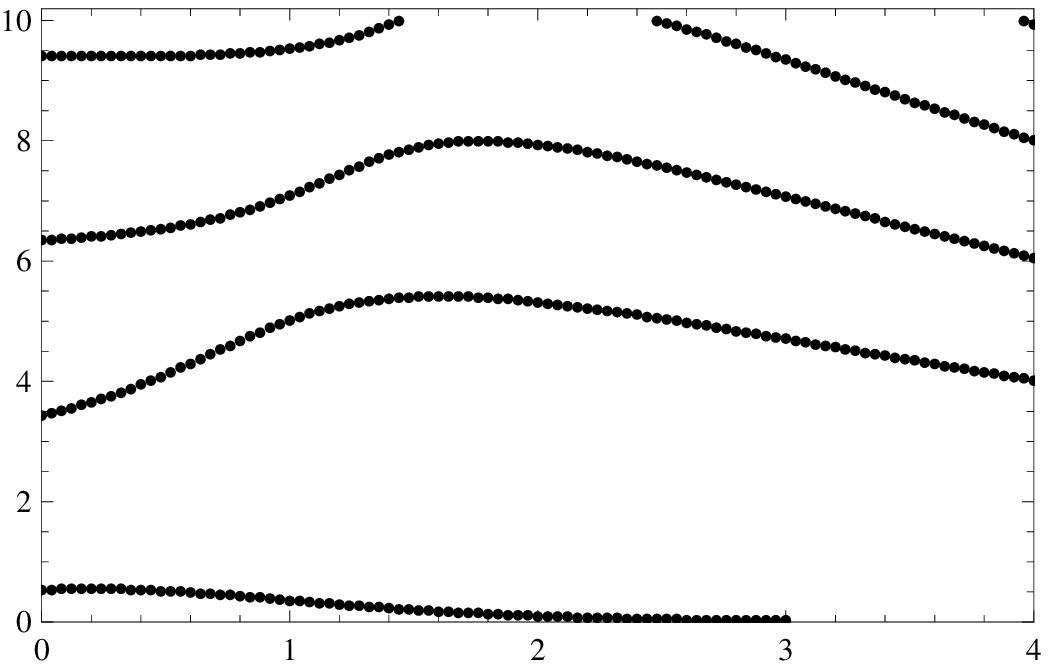}}
\put(0,150){\includegraphics[height=125pt]{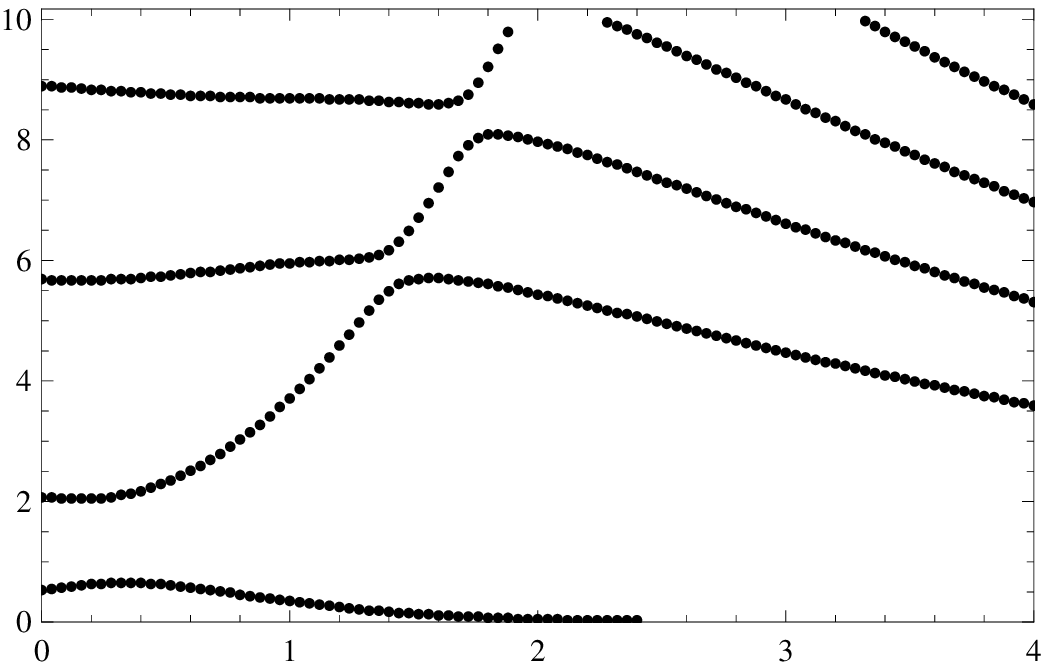}}
\put(230,150){\includegraphics[height=125pt]{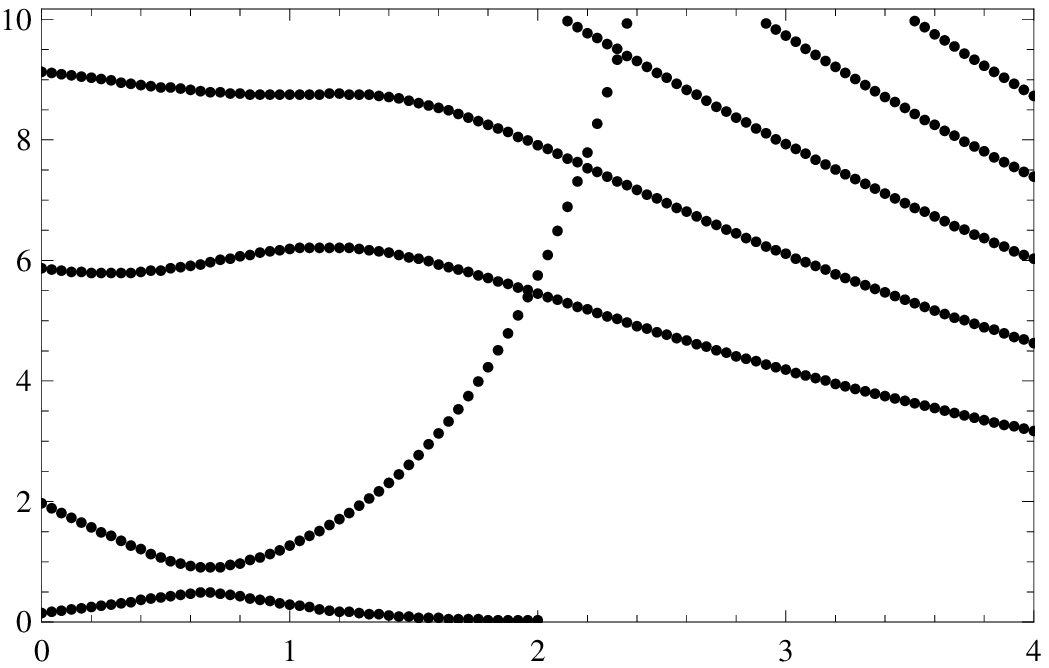}}
\put(0,0){\includegraphics[height=125pt]{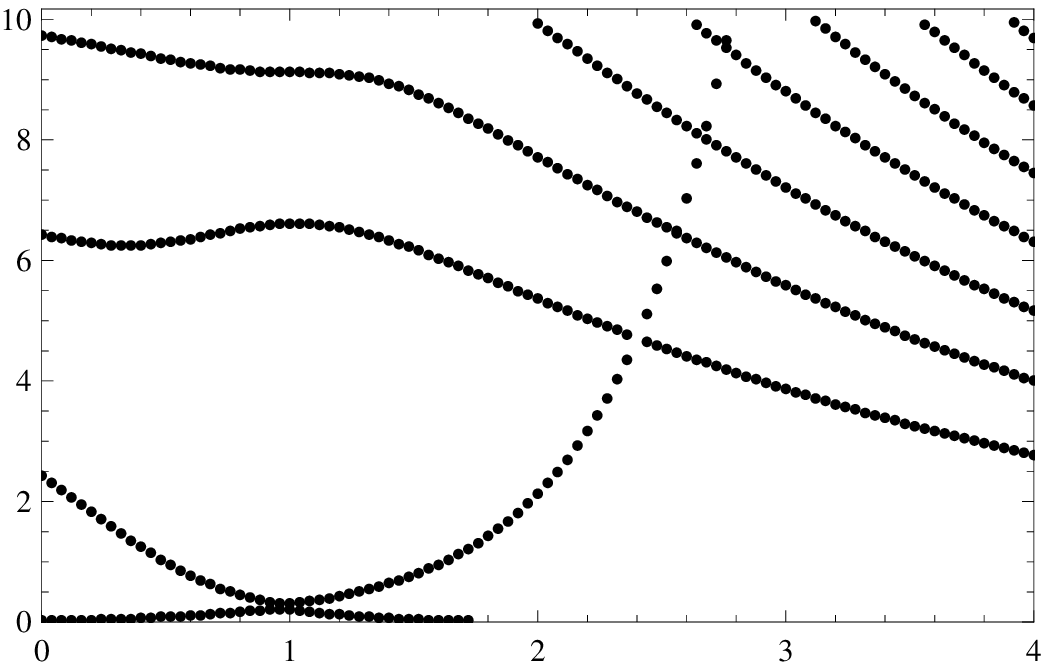}}
\put(230,0){\includegraphics[height=125pt]{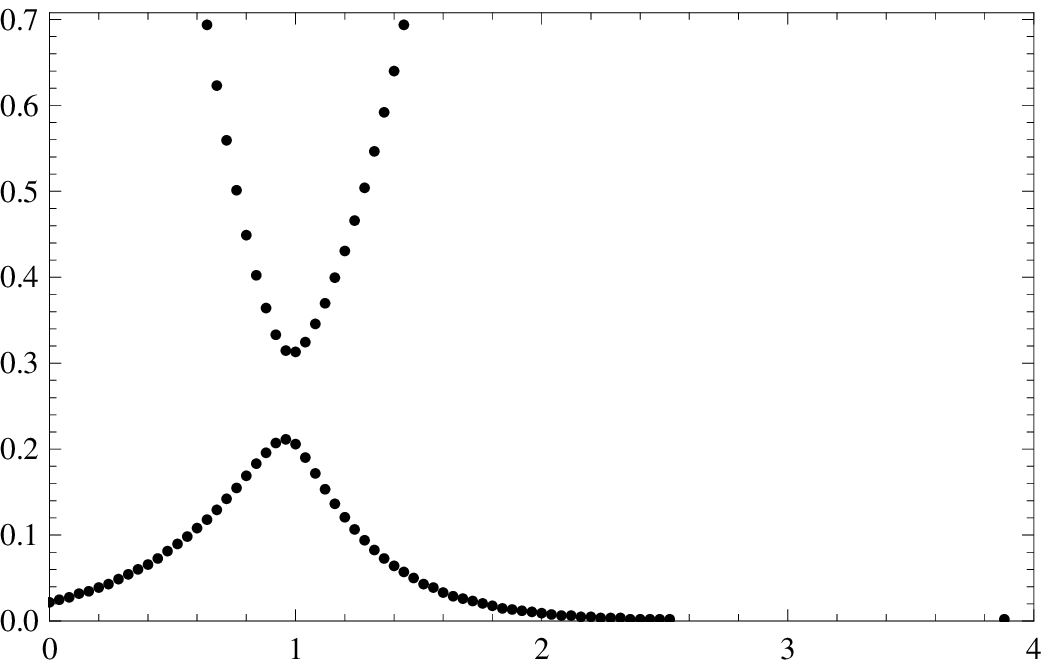}}
\put(5,130){$M$}
\put(232,130){$M$}
\put(202,5){$r_{\ast}$}
\put(432,5){$r_{\ast}$}
\put(5,280){$M$}
\put(232,280){$M$}
\put(202,155){$r_{\ast}$}
\put(432,155){$r_{\ast}$}
\put(5,430){$M$}
\put(232,430){$M$}
\put(202,305){$r_{\ast}$}
\put(432,305){$r_{\ast}$}
\end{picture} 
\caption{Numerical results for Example C. Mass $M$ of the lightest scalar fluctuations, for the choices
$\Delta=1,1.5,2,2.5,3,3$ (left to right, top to bottom), $\Phi_I=1$, $r_1=0$ and $r_2=5$, computed with backgrounds differing in the choice of $r_{\ast}$.
The last panel shows a detail of the left one, with $\Delta=3$. }
\label{Fig:WalkingSpectrumPhi1IR0}
\end{center}
\end{figure}

\begin{figure}[htpb]
\begin{center}
\begin{picture}(430,440)
\put(0,300){\includegraphics[height=125pt]{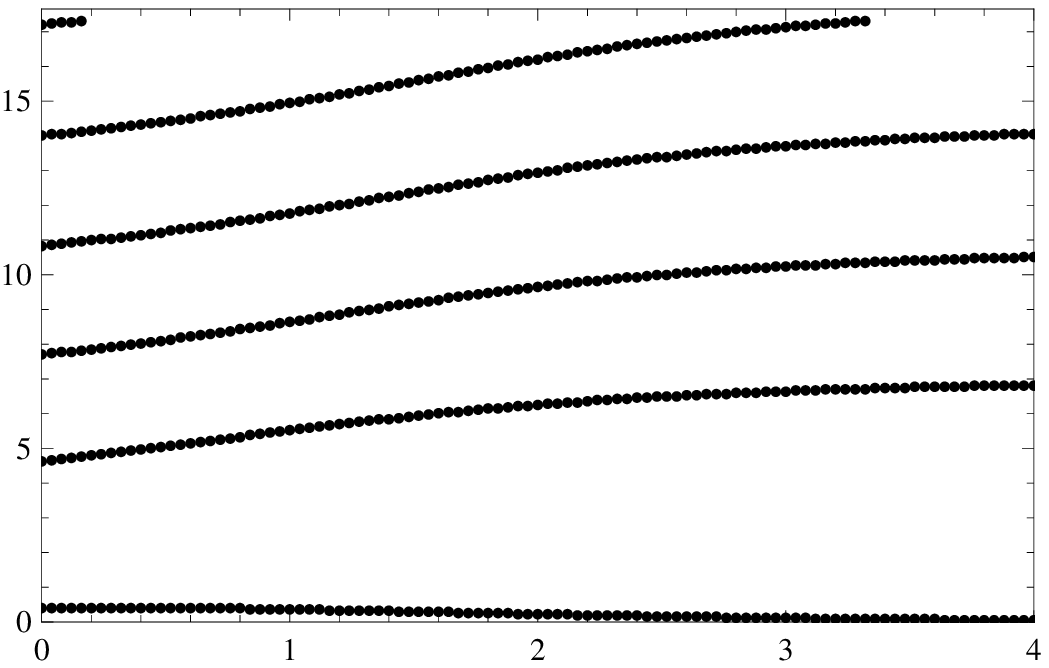}}
\put(230,300){\includegraphics[height=125pt]{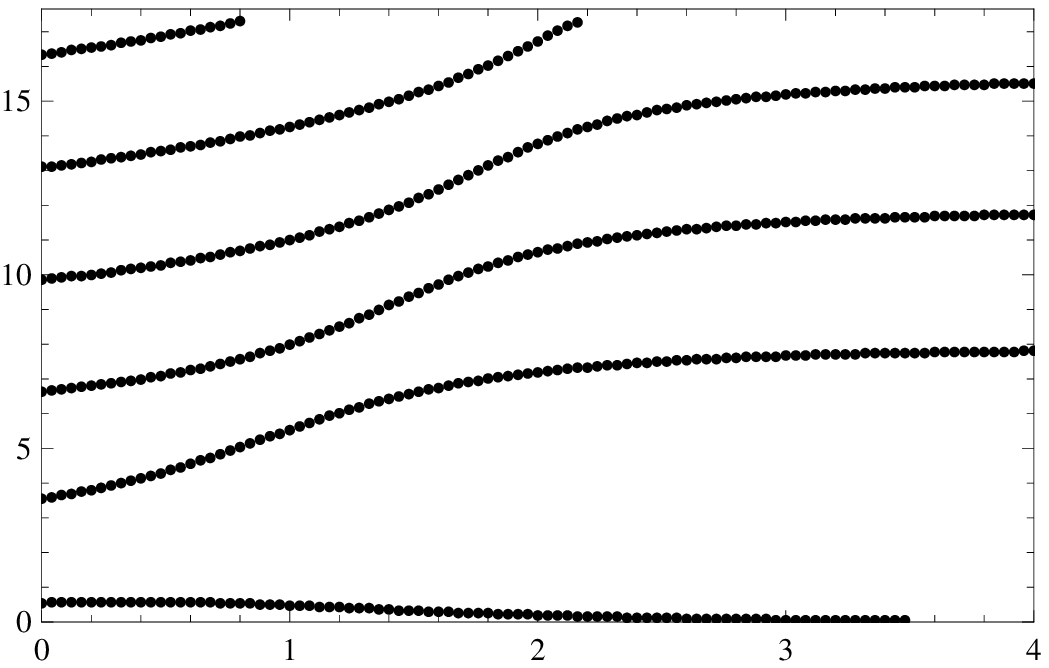}}
\put(0,150){\includegraphics[height=125pt]{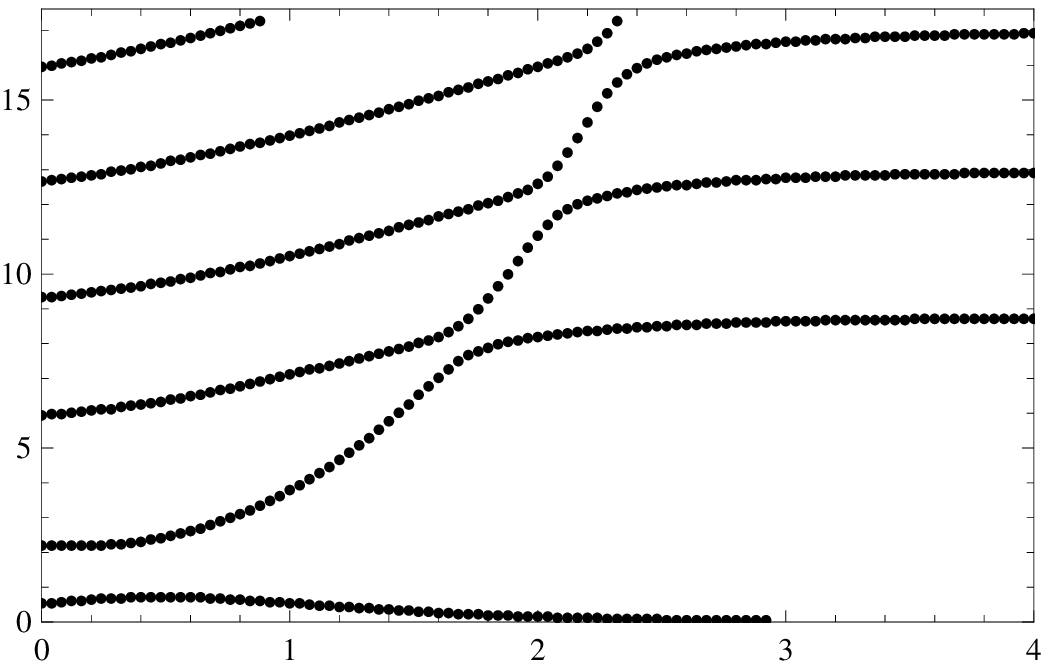}}
\put(230,150){\includegraphics[height=125pt]{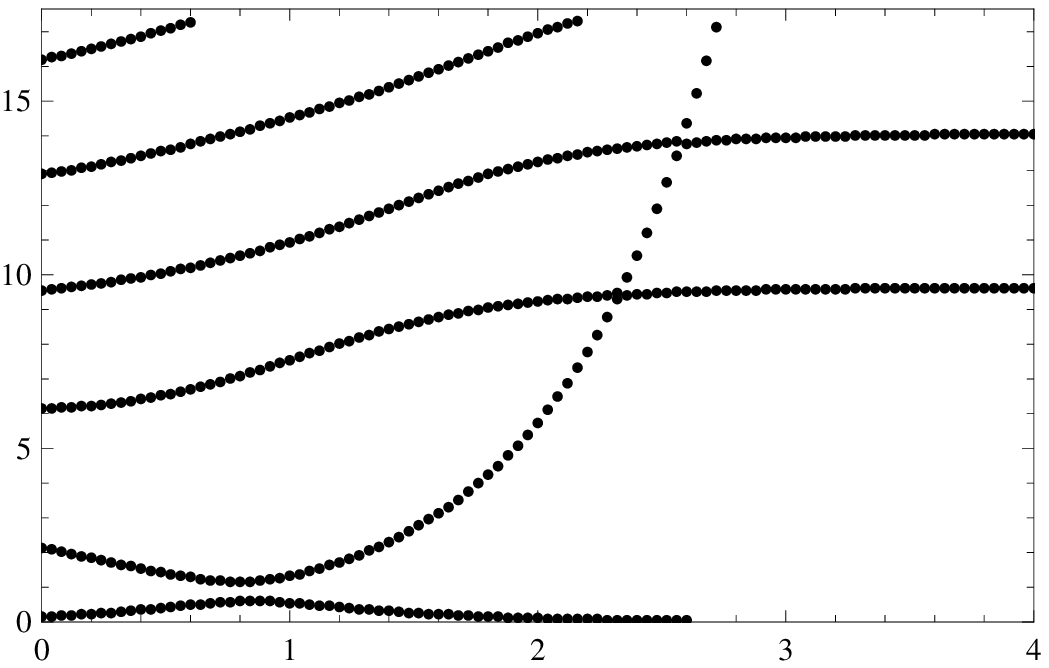}}
\put(0,0){\includegraphics[height=125pt]{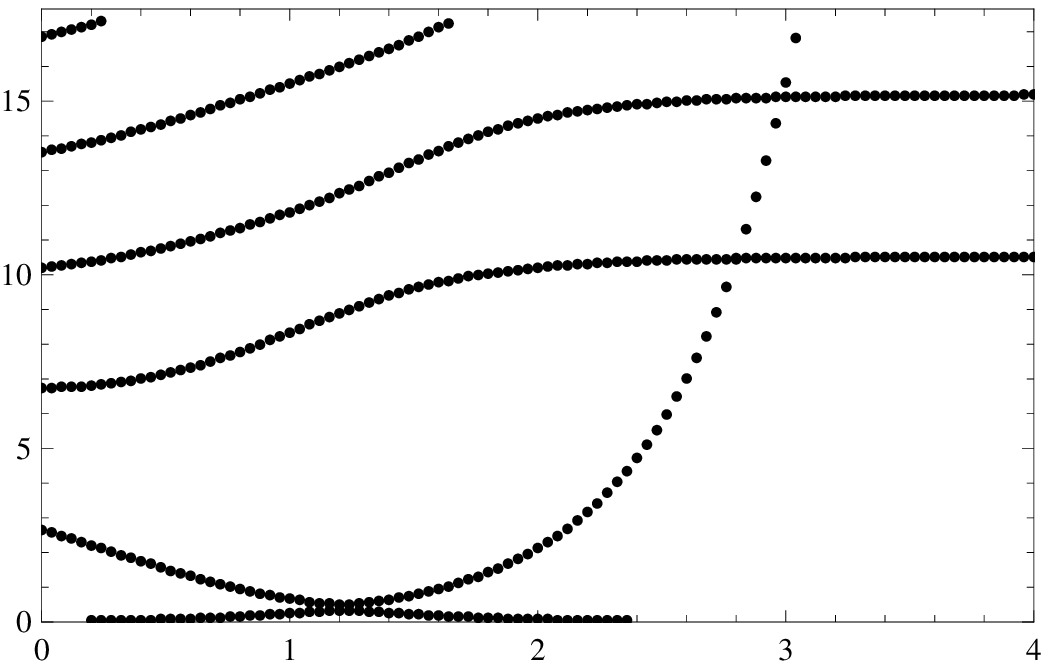}}
\put(230,0){\includegraphics[height=125pt]{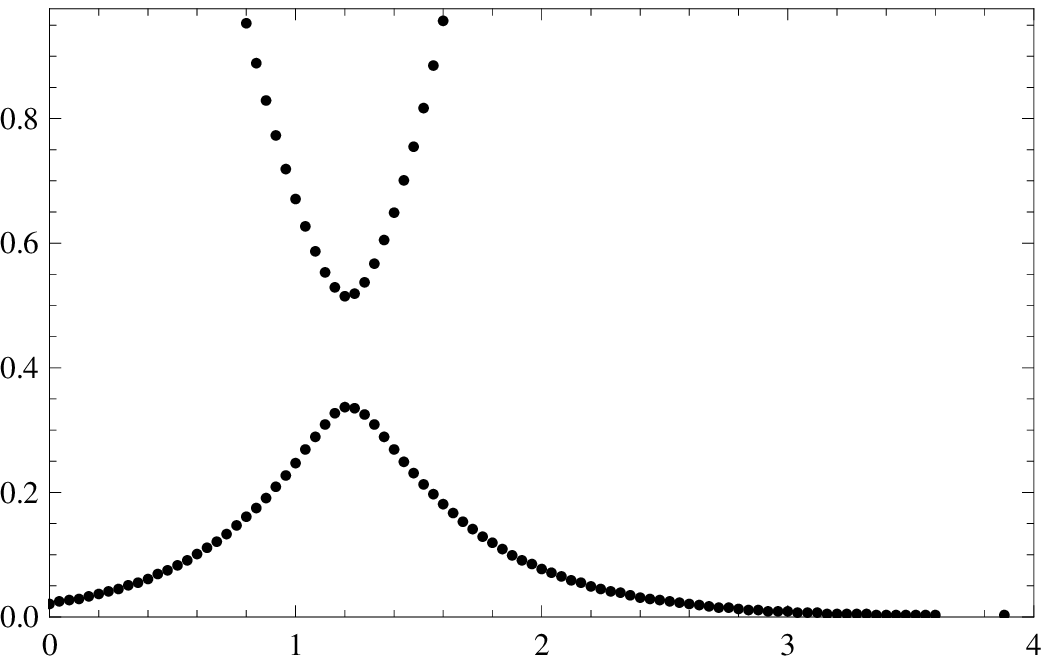}}
\put(5,130){$M$}
\put(232,130){$M$}
\put(202,5){$r_{\ast}$}
\put(432,5){$r_{\ast}$}
\put(5,280){$M$}
\put(232,280){$M$}
\put(202,155){$r_{\ast}$}
\put(432,155){$r_{\ast}$}
\put(5,430){$M$}
\put(232,430){$M$}
\put(202,305){$r_{\ast}$}
\put(432,305){$r_{\ast}$}
\end{picture} 
\caption{Numerical results for Example C. Mass $M$ of the lightest scalar fluctuations, for the choices
$\Delta=1,1.5,2,2.5,3,3$ (left to right, top to bottom), $\Phi_I=1$ and $r_2=5$, computed with backgrounds differing in the choice of $r_{\ast}$,
and by choosing $r_1$ so that $\bar A(r_1)=0$.
The last panel shows a detail of the left one, with $\Delta=3$. }
\label{Fig:WalkingSpectrumPhi1AdependentIR}
\end{center}
\end{figure}

\begin{figure}[htpb]
\begin{center}
\begin{picture}(430,440)
\put(0,300){\includegraphics[height=125pt]{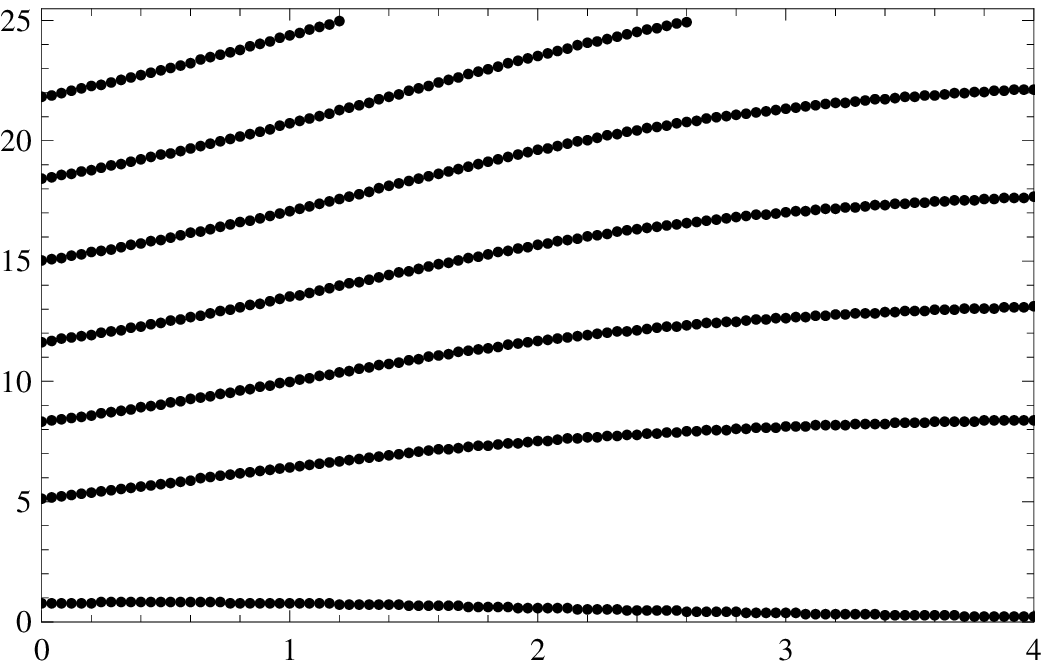}}
\put(230,300){\includegraphics[height=125pt]{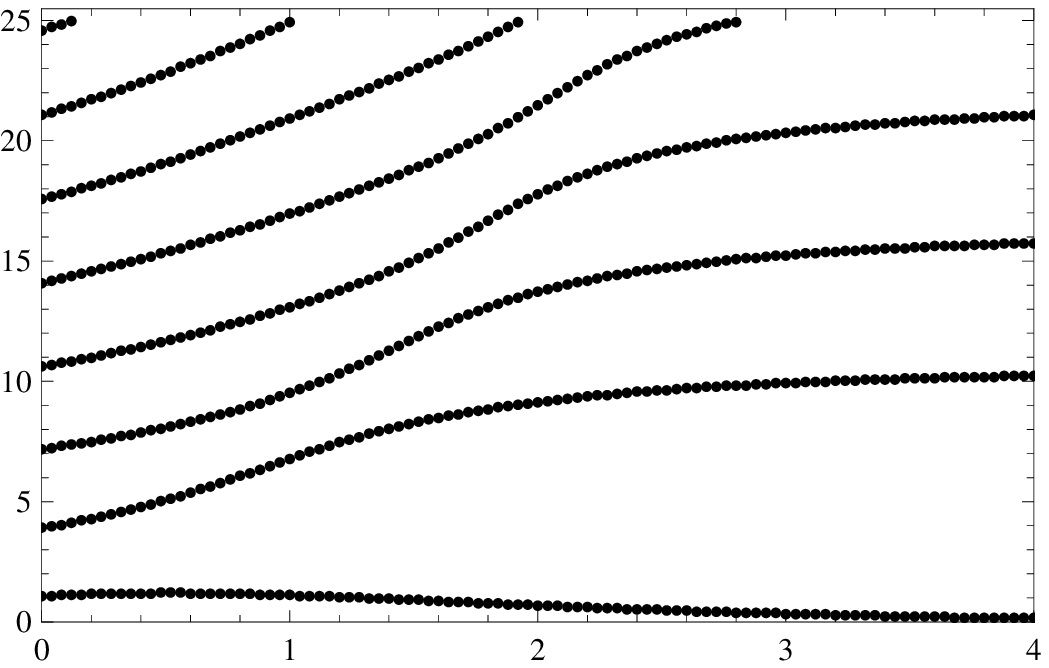}}
\put(0,150){\includegraphics[height=125pt]{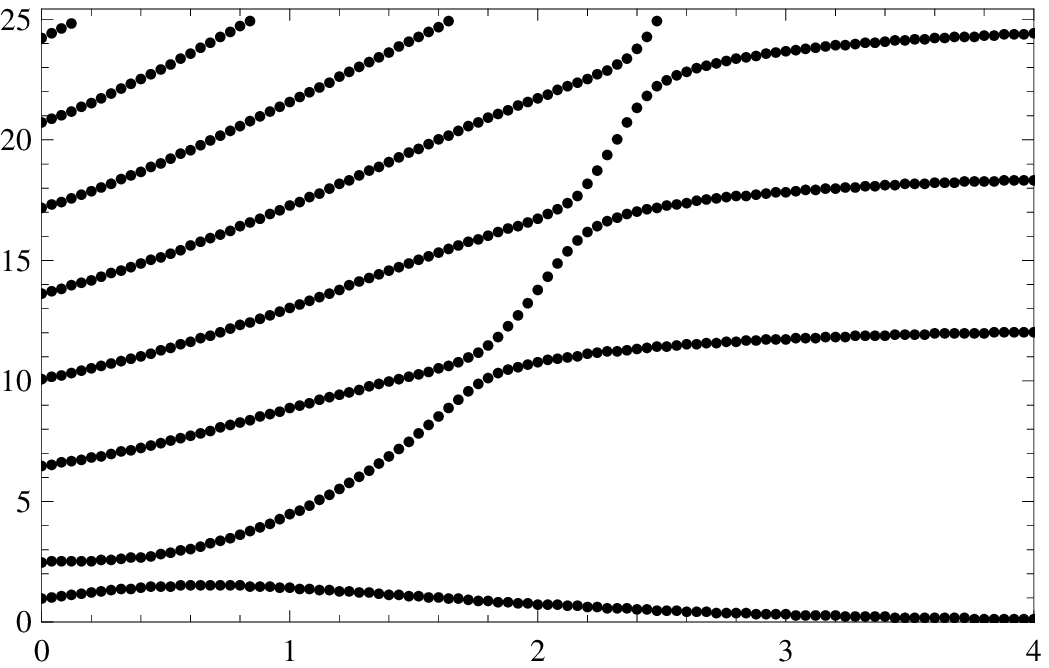}}
\put(230,150){\includegraphics[height=125pt]{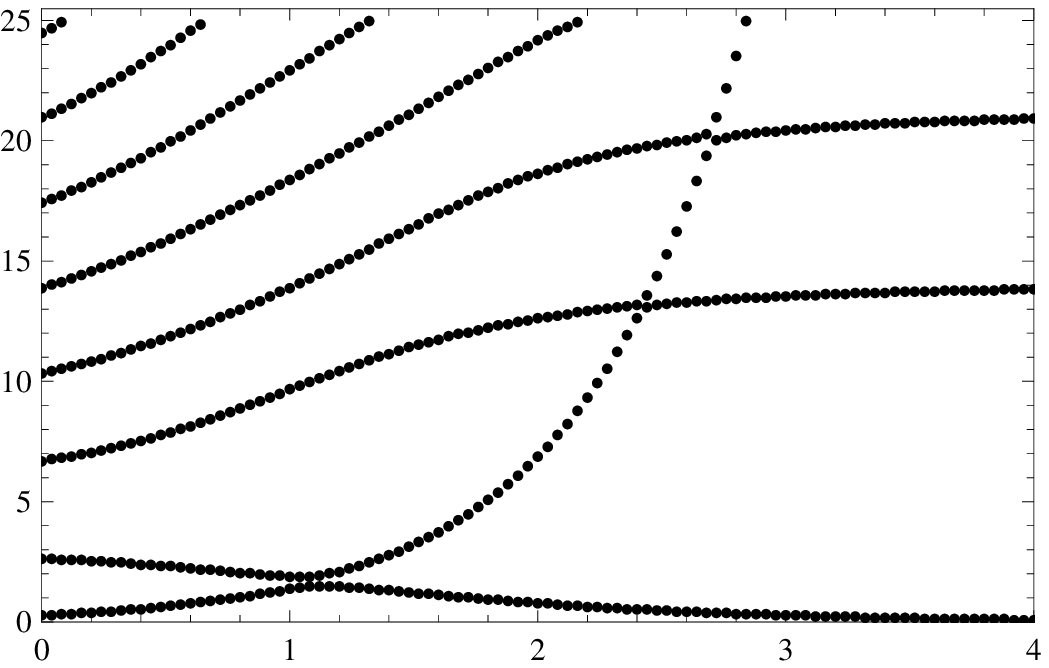}}
\put(0,0){\includegraphics[height=125pt]{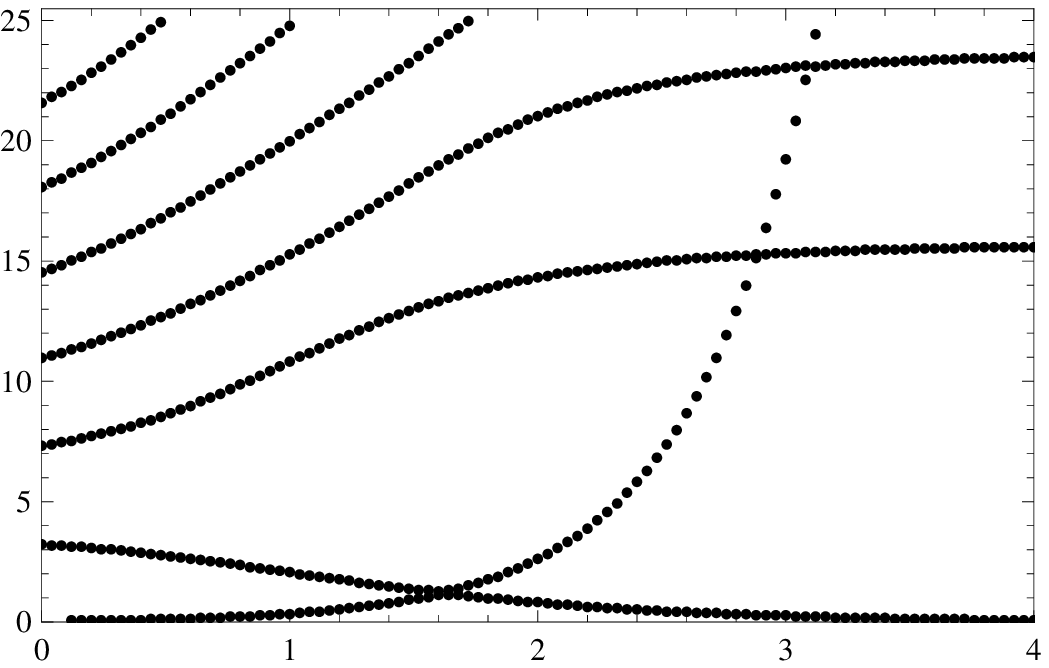}}
\put(230,0){\includegraphics[height=125pt]{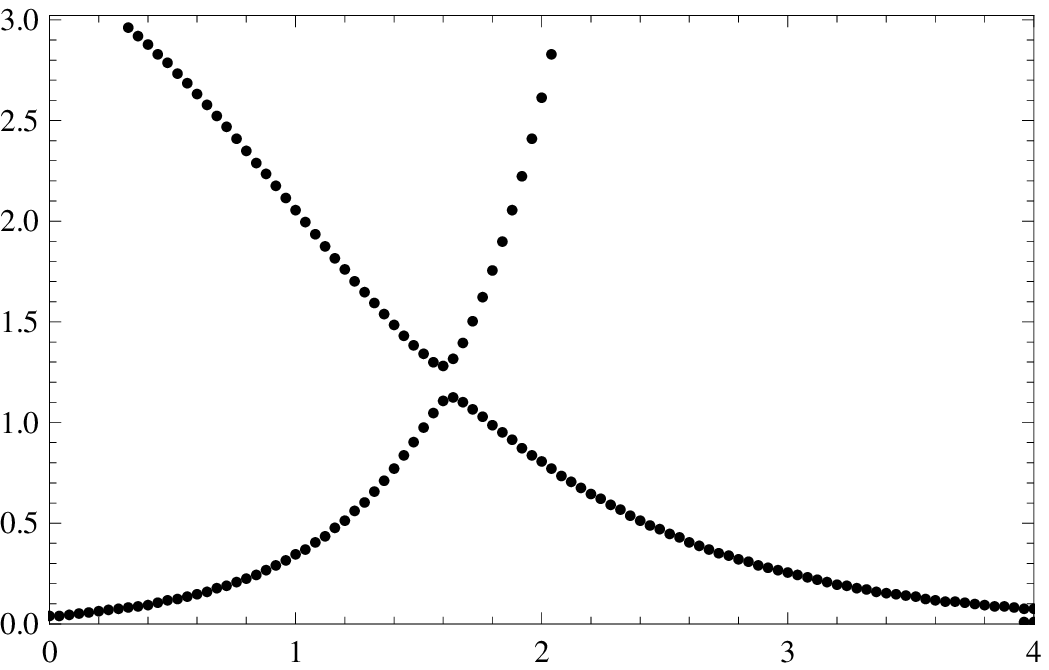}}
\put(5,130){$M$}
\put(232,130){$M$}
\put(202,5){$r_{\ast}$}
\put(432,5){$r_{\ast}$}
\put(5,280){$M$}
\put(232,280){$M$}
\put(202,155){$r_{\ast}$}
\put(432,155){$r_{\ast}$}
\put(5,430){$M$}
\put(232,430){$M$}
\put(202,305){$r_{\ast}$}
\put(432,305){$r_{\ast}$}
\end{picture} 
\caption{Numerical results for Example C. Mass $M$ of the lightest scalar fluctuations, for the choices
$\Delta=1,1.5,2,2.5,3,3$ (left to right, top to bottom), $\Phi_I=2$ and $r_2=5$, computed with backgrounds differing in the choice of $r_{\ast}$,
and by choosing $r_1$ so that $\bar A(r_1)=0$.
The last panel shows a detail of the left one, with $\Delta=3$. }
\label{Fig:WalkingSpectrumPhi2AdependentIR}
\end{center}
\end{figure}

\begin{figure}[htpb]
\begin{center}
\begin{picture}(430,150)
\put(0,0){\includegraphics[height=125pt]{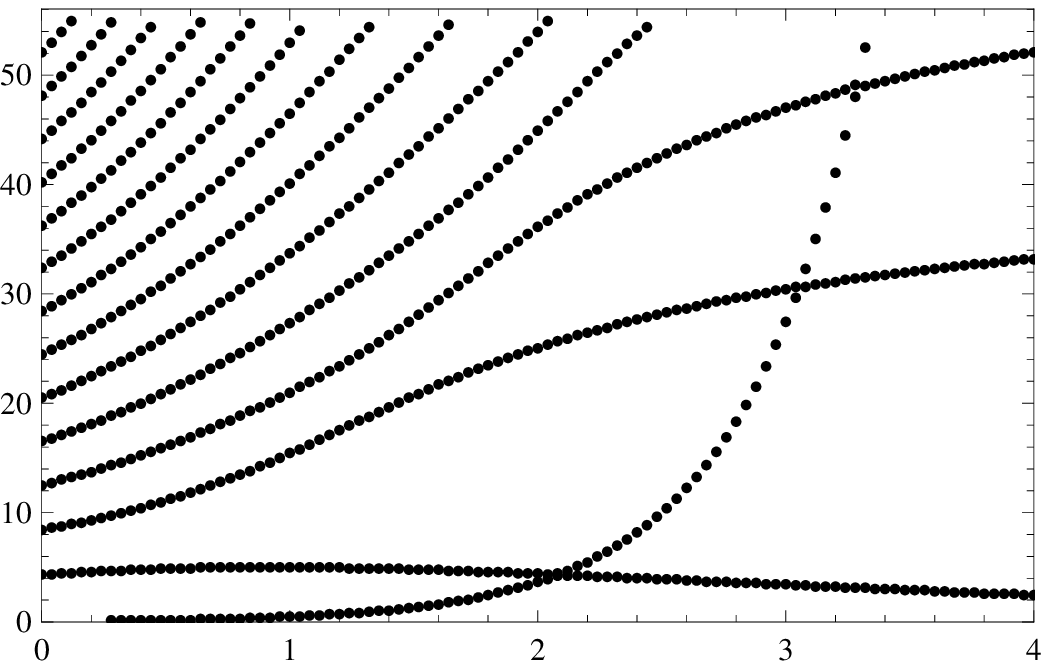}}
\put(230,0){\includegraphics[height=125pt]{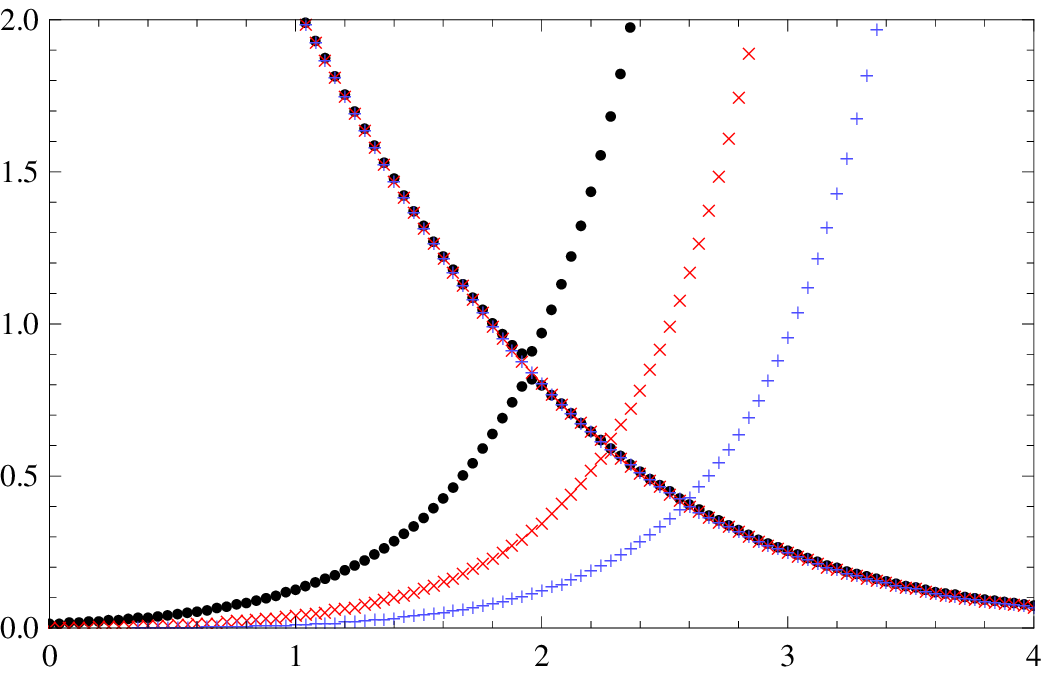}}
\put(5,130){$M$}
\put(232,130){$M$}
\put(202,5){$r_{\ast}$}
\put(432,5){$r_{\ast}$}
\end{picture}
\caption{Numerical results for Example C. The left panel shows the mass $M$ of the lightest scalar fluctuations, for the choice $\Delta=3$, $\Phi_I=4$, and $r_2=5$, computed with backgrounds differing in the choice of $r_{\ast}$, and by choosing $r_1$ so that $\bar A(r_1)=0$. The right panel shows the mass $M$ of the lightest scalar fluctuations, for the choice $\Delta=3$ and $\Phi_I=2$, computed with backgrounds differing in the choice of $r_{\ast}$, by choosing $r_1$ so that $\bar A(r_1)=0$, and varying the UV cutoff $r_2=6$ (black dots), $r_2=7$ (red $\times$), and $r_2=8$ (blue $+$).}
\label{Fig:WalkingSpectrumPhi4andUVdependence}
\end{center}
\end{figure}

Notice that for $r_{\ast}\rightarrow +\infty$, the background becomes purely AdS, with curvature radius determined by the IR fixed point.
Conversely, for $r_{\ast}\rightarrow -\infty$, the AdS background has unit curvature.
Notice also that we have chosen the integration constant in the warp factor in such a way that (asymptotically) 
in the UV all the backgrounds become identical. This means that the same numerical choice for
UV-quantities (such as $r_2$) corresponds to the same physical scale. This is not the case for IR-quantities (such as $r_1$).
We will come back to this comment later on.

In Figure~\ref{Fig:WalkingSpectrumPhi1IR0}, we show the spectrum of masses $M$ 
for the first few composite states as a function of $r_{\ast}$. The spectrum has been computed with $m_i^2 \rightarrow +\infty$. The different figures show the results for different values of $\Delta$. Focusing on the heavy states, we observe the expected behavior: the spectrum of heavy states consists of an infinite tower of evenly spaced KK excitations. There is an artificial suppression of the masses as a function of $r_{\ast}$. This effect was already observed earlier on, when discussing the quadratic potential. Since below $r_{\ast}$ the curvature is not unit, the fact that we chose the warp factors of different backgrounds to agree in the far UV means that the same value of $r_1$ yields a different physical scale. An alternative way of defining the IR cutoff is to make it be at the point where the warp factor $A$ is equal to zero. This then ensures that the IR cutoff is always at the same energy scale, and in this sense it is perhaps more natural from a physical point of view. The resulting plots are shown in Figure~\ref{Fig:WalkingSpectrumPhi1AdependentIR} (for $\Phi_I=1$) and Figure~\ref{Fig:WalkingSpectrumPhi2AdependentIR} (for $\Phi_I=2$). 
There is nothing particularly deep about this, aside from suggesting that we need to exercise some caution when making quantitative statements relating physical scales to each other.

Much more interesting is what happens at the level of light states. The limits of $r_*$ being small or large are associated with the physics of the two different fixed points ($\Phi = 0$ and $\Phi = \Phi_I$). For small values of $r_*$, the backgrounds are very similar to those considered in the case of quadratic superpotential in Sections~\ref{Sec:4a} and \ref{Sec:5a}, and the spectrum behaves qualitatvely the same. For instance, in the case of $\Delta > 2$, there is a light scalar. On the other hand, the backgrounds with larger values of $r_*$ only deviate away from the fixed point $\Phi = \Phi_I$ far in the UV, exhibiting walking behaviour from the IR cutoff up to the scale set by $r_*$. Therefore, in this case the light scalar has the interpretation of being the analog of the dilaton discussed in the context of walking technicolor. This picture is clearest for larger values of $\Delta$ in which case an interesting crossing structure develops. The crossing structure makes it apparent that the nature of the lightest scalar changes radically as $r_*$ is varied, being related to either of the two fixed points, for small or large values of $r_*$, respectively.

In Figure~\ref{Fig:WalkingSpectrumPhi4andUVdependence}, the left panel illustrates that for larger values of $\Phi_I$, the light state whose mass is suppressed by the length of the walking region (i.e. $r_*$) requires a longer such region to become light. The right panel of Figure~\ref{Fig:WalkingSpectrumPhi4andUVdependence} shows the UV cutoff dependence of the light states. As expected, the light scalar associated with the fixed point at $\Phi = 0$ has a mass that is suppressed by the UV cutoff $r_2$ in agreement with the results found for $\Delta > 2$ in Example A. The light scalar associated with the fixed point at $\Phi_I$, however, is unaffected by the value of $r_2$, caring only about $r_*$ which acts as the UV cutoff for this state.

%%%%%%%%%%%%%%%
%\newpage
\section{Field theory interpretation, discussion and general lessons}
\label{Sec:6}

In this section, we discuss the physical meaning and implications of the examples we discussed 
in the paper.
Before we start, we must remind the reader about the two main subjects of the paper.
First of all, we are mostly interested in understanding under which conditions a 
strongly-coupled, (quasi) conformal theory admits anomalously light 
scalars in the spectrum. These can be the result of accidental cancellations,
but more often are the result of the spontaneous breaking of approximate symmetries.
Such symmetries can be internal (giving rise to pseudo-Goldstone bosons, such as the techni-pions and techni-axions of a technicolor model)
or it may happen that scale-invariance is an approximate symmetry, in which case the light scalar is a dilaton.

Second, the framework within which we work is that of gauge-gravity dualities.
What we did was to set up a very flexible formalism, that allows to study the four-dimensional spectrum obtained
from a five-dimensional sigma-model of $n$ scalars coupled to gravity, in the presence of 
UV and IR boundaries in the radial direction.
The formalism exploits the diffeomorphism invariance of the five-dimensional theory in order to 
write the (linearized) equations for the fluctuations and the boundary conditions directly in terms of 
physical states. This fact allows to reduce the complexity of the general problem to a set of 
$n$ (coupled) equations involving only $n$ scalar fields.

We applied this formalism to three examples. Up to now, we focused on the technical aspects,
showing explicitly, on the basis of these three very simple examples,
 how the calculations are carried out, and what 
are the main results. All the examples involve only one scalar field, with trivial sigma-model metric, and 
all admit a description in terms of a superpotential. The latter is quadratic (in Example A),
a simple hyperbolic function (in Example B), or cubic (in Example C).

For particular choices of the boundary potential, all of the examples can be made to contain at least one exactly massless mode. Its composition in terms of the fluctuations of the original scalar and gravity degrees of freedom
always includes a component that couples to the trace of the four-dimensional stress-energy tensor.
In other words, there is always a field that can be identified with the dilaton.
The crucial task is then to identify under which conditions (on the bulk dynamics and on the boundary conditions)
does this state stay light and keeps (at least at leading-order) the appropriate couplings in order
to be identified as a physical dilaton.
In order to do so effectively, one has also to anticipate the effect that coupling the strongly-interacting sector 
(dual to the five-dimensional sigma-model) to an external weakly-coupled sector has on the spectrum.
Hence, one has to make sure that no serious fine-tuning problems are present.

We saw that the boundary conditions are determined by the background solution, up to a certain amount of freedom in the choice of 
the quadratic $m_i^2$ terms.
Special choices of the $m_i^2$ may yield very peculiar results for the spectrum.
In particular, we saw that  setting $m_2^2=0$ automatically implies that an exactly massless state is present.
However, such a special choice is certainly the result of fine-tuning: as soon as coupling to an external sector is added, there is no reason to expect that such a choice is preserved. In general, (perturbative) loop corrections coming from this external sector are going to 
yield corrections to $m_2^2$, which are UV-sensitive (divergent). 
For this reason, it is most interesting to see if a light state exists when $m_i^2\rightarrow +\infty$, in which
case one can be confident that no fine-tuning is hidden in our procedure.
In doing so, one is also guaranteed to break any possible approximate global symmetries
of the sigma-model (and of the dual strongly-interacting theory), so that if a light state exists, 
it cannot be due to such an  internal symmetry.
However, a word of caution is needed here: if for some physical reason the global symmetry of the internal sector happens
to be also a global symmetry of the external, weakly-coupled sector of the full theory, then one has to treat the matrices $m_i^2$
appropriately.

Finally, a completely general comment.
We said in the introduction that the physics of massive states cannot be completely universal,
but rather it is necessarily sensitive to model-dependent details about the dynamics.
If one had the exact dual of a well-known and established technicolor model that fits all the data,
this observation would not matter, one could simply compute the spectrum in the gravity side of the correspondence, 
and conclude with the phenomenological implications of the results.
This is unfortunately not the case, in part because no such a thing 
as a standard technicolor model exists,
but also for a more subtle reason.
Gauge/gravity dualities, in the context of dynamical electro-weak symmetry breaking,
do provide precise and effective calculation techniques, 
in the sense that the results do not depend on uncalculable $O(1)$ coefficients,
as is the case for four-dimensional estimates with strongly-coupled systems.
But one faces the limitation that the gravity description
is dual to models that are not precisely what one wants. 
In particular, all we are going to say is valid only in the strict large-$N$ limit,
and in most of the cases some amount of supersymmetry is present.
Hence, in spite of the fact that one can get actual numbers, rather than order-of-magnitude estimates,
for the relations between masses and couplings of the various states,
 one still needs to consider many different models, and try to understand the parametric dependences,
 rather than the actual numbers.

\subsection{Quadratic superpotential}
\label{Sec:6a}

We start from Example A, in which case the superpotential is simply quadratic.
It must be stressed that all that we are going to say is not restricted to this specific model,
but rather it will hold also for
 any other model that is at least approximated by 
a set of controllable perturbations of a conformal theory,
provided,  in the whole region between the two cutoffs
that we introduce,  the
vicinity to a fixed point controls the dynamics completely.
We will hence use some of the present considerations also in discussing Examples B and C, when appropriate.

The analysis of the spectrum of the scalar fluctuations has revealed the existence of a number of limits in which a parametrically light scalar emerges. Specifically, we found a light scalar if at least one of the conditions $\Delta\ll1$, $m_2^2\ll1$, or $\Delta> 2$ holds. 
The first case has been discussed at length in the literature, and is known to be interpreted in terms of a quasi-marginal deformation
of the CFT. The second case is the result of a fine-tuned choice, as we saw earlier on, and is hence of marginal interest.
We hence focus on the last case, $\Delta> 2$.

When $\Delta>2$, the standard dictionary of the AdS/CFT correspondence implies that 
an operator $\mathcal{O}$ of dimension $\Delta$ has developed a VEV. In which case, 
one clearly expects a massless dilaton to be present.
Because of the finite value of the UV cutoff $r_2$ that we use,
otherwise subleading deformations (such as the insertion of multi-trace operators in the dual theory~\cite{Vecchi:2010dd})
cannot be neglected, and result in a small mass for such a dilaton.
The irrelevant nature of such deformations explains the suppression of
the mass $m^2_d$ in Eq.~(\ref{Eq:dilatonmass}) as a function of $r_2$,
and the fact that a massless state can be recovered in the $r_2\rightarrow +\infty$ limit.

In order to make this more quantitative, let us assume that the dual theory is such
that ${\rm dim} \,\mathcal{O}^2=2\, {\rm dim} \,\mathcal{O}$ (e.~g. in the large-$N$ limit).
In the study performed in the body of the paper, we take the limit $r_2\rightarrow +\infty$
by holding the VEV $\Phi_1$ fixed.
A simple toy-model description of the dual effective potential, due to the multi-trace deformations
that break conformal invariance for finite $r_2$,
can be approximately given by $\Lambda^{4-2\Delta}\left(\mathcal{O}-v\right)^2$, where $v$ is the VEV in the field theory, and the scale $\Lambda$ is related to the UV cutoff.
For $\Delta>2$, this effective potential is suppressed by the UV scale,
because the operator $\mathcal{O}^2$  is irrelevant (together with any even higher order $\mathcal{O}^n$ correction),
hence explaining our result that $m^2_d\propto e^{-2(\Delta-2)r_2}$ from Eq.~(\ref{Eq:dilatonmass}).

\subsection{Example B}

The second example we considered can be thought of as a completion, in the context of 
Type-IIB, of the case with quadratic potential in Example A, for $\Delta=3$ (and $\Phi_1=1$) and for $\Delta=1$
(and $\Phi_1=\sqrt{3}$).
This is nice for several reasons. First of all, it means that the superpotential is known beyond 
the quadratic level, and hence can be used with some degree of confidence also far away from the
UV fixed point. Second,  if one chooses $\di \Omega_5$  in such a way that for $\Phi=0$
it yields the metric on the five-dimensional sphere,
one has  a precise mapping in terms of deformations
of ${\cal N}=4$ super-YM.
In the $\Delta=3$ case, 
we are enforcing a non-trivial gaugino condensate~\cite{GPPZ},
while for $\Delta=1$ we are giving a mass to the fermions.
Yet, the presence of a singularity in the IR of the background means that this is not the complete dual to
the four-dimensional field theory, but that some ingredient is missing, so that 
the physics of the four-dimensional theory is well-captured only away from the singularity.
In practice, this means that the truncated system yielding our very simple superpotential
is actually too simple: it does not capture some non-trivial properties of the strong dynamics 
of ${\cal N}=4$, taking place very far from the UV fixed-point.
This is all well known, and discussed in the literature.

It is therefore interesting to understand what happens when computing the spectrum using our algorithmic procedure,
which implies adding cutoffs both in the UV and in the IR.
There is not much to say about the $\Delta=1$ case. This just provides a very nice cross-check,
showing how the regulator procedure we use yields the correct results, in a simple case in which 
other arguments can be used to discuss the spectrum without introducing any regulators.

We hence focus here on the $\Delta=3$ case, which we extensively studied numerically. Since $\Delta > 2$, for finite UV cutoff 
the interpretation is that we add an irrelevant deformation to a theory and enforce a VEV.
Provided the IR cutoff $r_1$ is far away from the singularity, the VEV is spontaneously breaking 
scale invariance, as in Example A. However, as $r_1$ is chosen to be close to the singularity, 
there is no obvious sense in which the theory is still close to a conformal fixed-point, and hence no a priori
reason to expect the light state to persist.
It is hence very interesting, and somewhat surprising, that it does.
Notice by contrast that the spectrum of heavy states  is shifted, to testify of the fact 
that the singularity is  not a negligible correction to Example A. Even more, remember that
the study of the Wilson loops we briefly sketched shows a very dramatic effect, to the point that the
string probe cannot even approach the region in the 
immediate proximity of the singularity.

The fact that the IR is badly singular clearly signals that the strong dynamics of ${\cal N}=4$
involves other non-trivial effects, which are not captured by this simplest truncated model, as we said earlier.
Yet, the boundary action localized at $r=r_1$ acts as an IR regulator, which effectively removes from the calculation
of the scalar spectrum the pathological effects of the IR singularity.
The resulting spectrum has all the sensible features expected in a healthy field theory.
It would be very nice to know if (and to what degree) the spectrum we computed 
is in quantitative agreement with what is obtained in a modification of the model
such that the IR singularity is resolved. In particular, we found the presence of a very light state (which did not exist for $\Delta=1$), and one should test whether this state is still present in the more complete analysis, rather than being an artifact due to the combined effects of the bad singularity and the IR regulator.

%%%%%%%%%%%%%%%
\subsection{Cubic superpotential}

The field theory motivation for studying this example is that one might be interested in studying 
four-dimensional models in which the RG-flow of a confining (UV-complete) theory 
happens to come very close to a non-trivial fixed point at intermediate energies.
One such example of phenomenological relevance is the class of models 
that embed walking technicolor into extended technicolor, in order to explain
in a unified picture the three (superficially conflicting) requirements of 
generating large masses for standard-model fermions while at the same time suppressing 
new-physics contributions to FCNC processes and preserving the UV-completeness of the theory.

There is no known example of a dynamical model the exact gravity dual of which has all the features required 
in the walking technicolor framework. There exist models that reproduce the flow of a confining theory, 
and there are models that describe the flow between two fixed points, but there are no models in which the
IR fixed point is only approximate, and ultimately the theory confines.
Also, models yielding the flow between two fixed points have a tendency to be very complicated,
such as the Pilch-Warner dual of the flow from ${\cal N}=4$ to the Leigh-Strassler fixed point,
for which the actual background is known only numerically (and within this framework, flows that approach the IR fixed-point but 
do not reach it are described by backgrounds which are badly singular).

Yet, it is of general interest to know whether, in such a theory, it is at all possible that a light dilaton is present.
Notice that the answer to this question is not obvious: even when considering the exact flows between two exact fixed points,
(in which case both the far-UV and deep-IR effective descriptions are provided by CFTs)  
the theory as a whole is not scale-invariant: there exists a physical scale (connected to the $\rho^{\ast}$ in our study)
 that separates  the regimes in which either of the two CFTs provides a sensible approximation for the physics.
So, the actual mass of the lightest scalar (the would-be dilaton) will in general depend on this scale,
on the two CFTs living at the fixed points (the spectrum of  dimensions), and on the specific properties of the flow.

The study we performed shows that indeed there is a light scalar, under quite general conditions,
and that its mass depends on the dimension $\Delta$ and on $\rho^{\ast}$.
Let us start from the dependence on the dimensions.
At the IR fixed point, all the (active) scalars must correspond to irrelevant deformations (otherwise
the flow could not reach such a fixed point). 
The only important distinction comes from the value of the dimension $\Delta$,
which characterizes the flow near the UV fixed point.

What is most remarkable, is that when $\Delta>2$
 the dependence on $r^{\ast}$ of the mass of the lightest state is not a monotonic function.
There exists an actual maximum of the mass, obtained for non-extreme values of $r^{\ast}$.
The reason for this is that when $r^{\ast}$ is so large, or so small, that the theory is effectively always very close to one of the two fixed points, 
the mass of the lightest state is suppressed parametrically, because the theory is effectively very close to conformal.
For intermediate values of $r^{\ast}$, the theory is not well described by a small departure from a CFT, and as a result
a non negligible mass is generated for the lightest state. 
However, the $r^{\ast}$ dependence of such mass must interpolate between the two extremal values of $r^{\ast}$ (the UV and IR cutoffs),
near which the mass practically vanishes. Hence, there is a maximum for such mass.
It is curious to notice that numerically such a maximum is (at least in our examples) still significantly suppressed in respect to the 
typical scale of the heavy states, although the actual physical relevance of such a fact is questionable.

%%%%%%%%%%%%%%%
%\newpage
\section{Conclusions and outlook}
\label{Sec:7}

We conclude the paper by summarizing our main results, by critically reviewing the limitations of this approach
and by suggesting a set of physically interesting applications.

%%%%%%%%%
\subsection{The algorithm}

Let us first summarize the algorithm to be used to compute the spectrum. We start with a five-dimensional sigma-model consisting of $n$ scalars coupled to gravity. For simplicity, suppose that a superpotential $W$ is known, and furthermore that there is no
obstruction to taking $m_i^2\rightarrow \infty$. The spectrum of scalar bound states can then be computed by applying the following steps.

\begin{itemize}

\item Write the action in the form of Eq.~(\ref{Eq:action}), with the bulk dynamics of the $n$ scalars in the form of Eq.~(\ref{Eq:L5}),
and the background metric in the form of Eq.~(\ref{Eq:bgmetric}).

\item Introduce a UV and an IR cutoff, by writing boundary actions as in Eq.~(\ref{Eq:L1}) and Eq.~(\ref{Eq:L2}).

\item Determine the background, by solving Eq.~(\ref{Eq:BPS1}) and  Eq.~(\ref{Eq:BPS2}), subject to the boundary conditions
in Eq.~(\ref{Eq:BOUN1}) and  Eq.~(\ref{Eq:BOUN2}).

\item Obtain the spectrum  by solving Eq.~(\ref{Eq:diffeqWN}), and then identify 
which values of $q^2=\Box$ allow to satisfy the boundary conditions in Eq.~(\ref{Eq:BCWinfty}).

\item If possible, and physically meaningful, take the limits $r_2\rightarrow +\infty$ and 
$r_1\rightarrow r_0$, where $r_0$ is the end-of-space in the IR. 

\end{itemize}

If the superpotential is not known, or it is not legitimate to take $m_i^2\rightarrow \infty$,
all the necessary changes to be implemented in this procedure are explained in the body of the paper.
It is also implicit  that one should familiarize oneself with the notation, which is explained in detail in Section~\ref{Sec:2}, where all the relevant information is provided explicitly.

%%%%%%%%%
\subsection{Limitations of the algorithm}

The algorithm we identified fails, or needs to be partially extended, in the following cases.

\begin{itemize}

\item The bulk action contains terms with more than two derivatives of the scalars and/or the metric.
In this case, the whole procedure has to be rethought from scratch.

\item The dependence of the two-point functions on the UV cutoff requires introducing $q^2$-dependent boundary terms in the UV, in order to regularize the theory and remove the UV cutoff itself. If these terms are polynomial, they just result in a comparatively harmless modification of Eq.~(\ref{Eq:BCWinfty}), to be dealt with via holographic renormalization. If they are  non-polynomial, then the whole physical meaning of the spectrum becomes questionable, and
the best thing one can do is to consider the dual theory 
as some phenomenological model
 with a physical cutoff $r_2$, which cannot be removed.

\item There are exactly flat directions in the supergravity potential, connected with moduli of the field theory.
In this case, there are exactly massless states which have nothing to do with the dilaton,
and at the technical level the $\Box^{-1}$ operator appearing in many equations is badly defined.
One should find a (model-dependent) way to overcome this difficulty, either by adding a perturbation that lifts the flatness of the potential,
or by further truncating the sigma-model in such a way as to decouple the potentially problematic (inactive) fluctuations.

\item In the IR, the space ends in a naked singularity at $r_0$, and the geometry near the singularity 
 is so bad that (super)gravity cannot be trusted.
In this case, one has to keep the IR cutoff $r_1>r_0$, and firm physical conclusions cannot be drawn in full generality,
until a resolution of the singularity within a more general sigma-model is found. However, note that in the one example that we studied where a naked singularity is present, i.e. Example B, our algorithm in fact yields finite results.

\item The IR is not singular, however the end of space is known to be described by extended objects that go beyond
the (super)gravity approximations. Again, in this case one can only keep $r_1>r_0$, using a cutoff that 
chops off the space at a scale where the (super)gravity description still holds.
It is not known (and it would be interesting to know)  whether a 
relation between the spectrum computed with the present algorithmic procedure and the actual physical spectrum
computed by fluctuating the whole background exists.

\end{itemize}

%%%%%%%%%
\subsection{Physics lessons}

The main physical motivation of this work, as explained in the introduction,
is to understand what kind of (confining) strongly-coupled theories yield
potentially light scalars in the spectrum, one (linear-combination) of which
has the couplings of a four-dimensional dilaton.
None of the examples we considered is the exact dual of such a theory, and yet in all the examples a
set of very general lessons appears to emerge coherently.
We summarize here these results.

When the dynamics  of the theory is well described in terms of the properties of 
the RG fixed points and their proximity, a light scalar appears to be present, irrespectively of model-dependent 
details, provided the dimension $\Delta$ (defined in the body of the paper)
satisfies either $\Delta \geq 2$ or $\Delta \ll 1$.
In the latter case, this is the well-known fact that a very small $\Delta$ means that a (quasi-)marginal operator
is deforming the CFT, hence introducing a parametrically small explicit breaking of conformal symmetry.
This is true even in the Higgs sector of the Standard Model, provided all the couplings are very weak,
and is hence not particularly interesting.
A lesser known fact is that this is true also when $\Delta\geq 2$. 
Provided factorization holds, one can interpret 
this particular result in terms of the spontaneous breaking of scale invariance via the VEV of an operator $\cal O$ 
of dimension $\Delta\geq 2$. The fact that the (double-trace) deformation ${\cal O}^2$ is 
irrelevant ensures that the light scalar present has a mass suppressed by the UV cutoff.

However, one has to exert some caution in using these results.
First of all, they hold only in the limit where factorization is exact, and no non-trivial
operator mixing is present. 
Most important, they depend on what is meant by the dynamics being well described in terms of 
its properties in proximity of the fixed-points.
For example, what about QCD (or, maybe better, Yang-Mills)?
It is well known that QCD admits a trivial UV fixed point, and that the departure from it is 
induced by a quasi-marginal deformation. And yet, there is no light scalar state in the glueball spectrum.
The reason is that the RG flow goes very far away from the fixed point, and at the confinement scale
there is no sense in which the theory is still close to it.
Indirectly, this also means that it is very unlikely that the physics near the confinement scale 
is controlled by a non-perturbative fixed point: in this case, as we saw, in spite of the fact that
the whole theory is not scale-invariant, a light state would still be expected, which is not the case for QCD/YM.

A set of minor caveats emerged during the technical calculations.
Summarizing their implications: in the cases where we identified the presence of a light state
in the spectrum, one should not immediately conclude that such a light state is there in general,
but should then ask whether the process of coupling the strongly coupled sector to external observable sectors
still preserves this result.
The whole problem of holographic renormalization is implied.

Finally, we conclude with some important comments about walking technicolor and similar theories.
In this case, one is interested in a model where the RG flow starts near a fixed point, 
evolves towards a different fixed-point, but then never reaches the latter, and ultimately confines, with the formation of
 non-trivial condensates that break the global symmetries of the theory.
What did we learn about this scenario?
While we do not have the gravity dual of a walking theory, and hence a firm conclusion should be postponed 
to a future study, yet by combining the results of Examples A, B and C 
we obtained a set of important indications of the presence of a very light state 
in some cases of walking dynamics. 

Let us summarize here all the elements leading to this conclusion.
In studying Example C, we considered the case where the theory is well-approximated by the dynamics
near the UV fixed point for $r\gg r_{\ast}$, and by the dynamics near the IR fixed-point for $r\ll r_{\ast}$.
We chopped off the space in the IR, assuming this to be the end-of-space of the model.
When $r_{\ast}$ is small, effectively the theory is never really well described by the IR fixed point,
and 
for all practical purposes,  this is not a walking theory.
When $r_{\ast}$ is large enough that a sizable range $0<r<r_{\ast}$ of the fifth-dimension can be thought of
as describing walking, something interesting happens.
There is always a light state, that has a natural interpretation in terms of a dilaton.
The mass of such state is suppressed by $r_{\ast}$: the larger $r_{\ast}$ is, the
lighter the dilaton. Ultimately, this result agrees with the expectations from Example A,
in the sense that if one expands the background near the IR fixed point, the resulting $\Delta$ is always going to be large.
From the field theory point of view, this must be the case: if the RG flow was governed by a relevant coupling, it could not approach the fixed point,
which implies that by expanding near the IR fixed point one must find a large value for $\Delta$.
As a result, $r_{\ast}$ represents the scale of explicit breaking of scale invariance, and enters the effective description 
near the IR fixed point by suppressing the coefficient of the irrelevant operator ${\cal O}^2$ inducing the explicit breaking itself.
This is very interesting for practical purposes: it means that one can write the mass of the light scalar in terms of $r_{\ast}$ 
and $\Delta>2$, hence providing a functional relation between the walking scale associated with $r_{\ast}$ and the mass $m_d^2\propto e^{-2(\Delta-2)r_{\ast}}$.

All of this means that the very fact that the theory walks (in the sense of coming very close to a IR fixed point) 
implies that a potentially light dilaton is present, and that the UV scale at which walking ends effectively suppresses
its mass.
One has then to ask what is the effect on the mass due to the fact that in the IR the theory confines.
Example B yields some very interesting information, by looking at the $\Delta=3$ and $\Delta=1$
cases, and comparing the results with Example A. 
Even when the IR ends in a singularity, the mass of the lightest scalar
is in substantial agreement with what was found in Example A (notice that this is not the case for the heavy KK state, 
for which the shift in mass is significant). The most important parameter appears to be $\Delta$.
If $\Delta>2$, effectively the confining behavior can be attributed to the formation of a condensate that
breaks spontaneously scale invariance, with explicit breaking accounting only for a small effect.
On the contrary, for $\Delta<2$ no light state exists (unless $\Delta \ll 1$ which we already commented about), and scale invariance is broken explicitly. In a realistic walking technicolor theory, the condensate that takes the theory away from the IR fixed point 
is presumably related to the electro-weak scale. On the basis of phenomenological considerations
it is usually believed that the dimension of the chiral-symmetry breaking condensate must be 
$2\leq \Delta \leq 3$, which would imply the existence of a light dilaton in the spectrum.

This opens a possibility that in certain walking technicolor theories, in which the
walking scale is parametrically higher than the electroweak scale, a parametrically light scalar
is present in the spectrum. This light scalar should then be interpreted as a light dilaton, 
and hence have couplings that are very similar to those of the SM light Higgs.
Hence, the discovery of such a light scalar at the LHC might be interpreted as a first indication 
in favor a walking technicolor origin for electroweak symmetry breaking.
Only the (non)detection of somewhat heavier resonances, in the TeV region, would allow to 
solve this possible ambiguity in experimental signals shared by the weakly-coupled Higgs models and
the strongly coupled walking models.
It would be very nice to test this conclusion for an explicit, complete model of the gravity dual of a walking theory,
hence putting this line of arguments of firm grounds.

%%%%%%%%%%%%%
\subsection{Outlook}

The whole machinery we put in place and summarized here has its natural application 
in the study of the spectrum of a confining theory the full dynamics of which is well-captured within
a five-dimensional sigma-model which is obtained by consistently truncating a fundamental, UV-complete theory. 
No phenomenologically useful such constructions exists yet, and hence an obvious direction for further research is to
try to identify such a model.
Doing so, and hence exploiting in full the potential of gauge-gravity dualities, has the great technical advantage that very
non-trivial properties of the dual theory can be computed, and the algorithmic procedure we provided 
renders the extraction of the spectrum a relatively harmless technical matter.

In particular, our comments on walking dynamics and walking technicolor
are not supported by very robust arguments, but rather based on circumstantial evidence emerging from a 
set of toy models, which we chose mostly on the basis of their simplicity. Yet, our work shows that it is
important to find complete duals of walking theories within superstring theory,
and that the study of the spectrum of such models could yield very non-trivial results
of utmost  phenomenological and theoretical importance.

From a more formal point of view, and in the short-term, a large number of interesting questions 
have been left open by this study, and require applying the procedure we outlined to less ambitious
five-dimensional backgrounds.
We mostly concentrated here on semi-realistic toy-models.
It would be very interesting to apply all of the above to the many well-known 
backgrounds that exist in the literature, and whose dual four-dimensional theories are well-understood.
One could then compare the results obtained using the IR and UV  regulators  to the results obtained  
with other techniques, both on the field-theory side and on the gravity side of the correspondence.
This would yield important tests of the correctness of this procedure, and might
help to shed light on the physical implications for models where the field theory is understood only in part.
Also, it would be interesting to implement all of the above within the systematic program
of holographic renormalization.

%%%%%%%%%%%%%%%%%%%%%%%%%%%%%%%%%%%%%%%%%%%%%%%%%%%%%%%%%%%%%%%%%%%%%%
%% Acknowledgments %%%%%%%%%%%%%%%%%%%%%%%%%%%%%%%%%%%%%%%%%%%%%%%%%%%
%%%%%%%%%%%%%%%%%%%%%%%%%%%%%%%%%%%%%%%%%%%%%%%%%%%%%%%%%%%%%%%%%%%%%%
\vspace{1.0cm}
\begin{acknowledgments}
We are grateful to L.~Vecchi for his contribution to the early stages of this study,
in particular for explaining to us
 some of the subtleties related to multi-trace deformations which contributed to shaping Section~\ref{Sec:6a},
and thank him for valuable discussions and comments on the manuscript.
We also thank S.~P.~Kumar, C.~N\'u\~nez, and W.~Mück for discussions. 
The work of MP is supported in part by WIMCS and by the STFC grant ST/G000506/1.
\end{acknowledgments}

%%%%%%%%%%%%%%%%%%%%%%%%%%%%%%%
%\newpage
\appendix
\section{Equations of Motion in the ADM formalism}
\label{App:A}

Here, and in the following appendices, we derive all the equations of motion, and the boundary conditions for the background as well as the fluctuations. Much of the notation is the same as that of~\cite{BHM}.

Starting with the action of the non-linear sigma model with boundary terms (where in the body of the paper $d=4$)
\SP{
	S =& \int d^d x dr \Bigg[ \sqrt{-g} \left(\frac{R}{4} - \frac{1}{2}G_{ab}\partial \Phi^a \partial \Phi^b - V(\Phi)\right) + \\& \sqrt{-\tilde g} \delta (r - r_1) \left( \lambda_{(1)}(\Phi) - \frac{K}{2} \right) - \sqrt{-\tilde g} \delta (r - r_2) \left( \lambda_{(2)}(\Phi) - \frac{K}{2} \right) \Bigg],
}
we derive the equation of motion for the scalars
\SP{
	\nabla^2 \Phi^a + \mathcal{G}^a_{bc} g^{MN} (\partial_M \Phi^b) (\partial_N \Phi^c) - V^a = \sum_i \sqrt{\tilde g g^{-1}} \lambda_{(i)}^a \delta_i
}
and Einstein's equations
\SP{
	- R_{MN} + 2 G_{ab} (\partial_M \Phi^a) (\partial_N \Phi^b) + \frac{4}{d-1} g_{MN} V = 2 \sum_i \sqrt{\tilde g g^{-1}} \left( \hat g_{MN} - \frac{g^{KL} \hat g_{KL}}{d-1} g_{MN} \right) \lambda_{(i)} \delta_i,
}
where $\hat g_{\mu\nu} = g_{\mu\nu}$, and $\hat g_{r\mu} = \hat g_{rr} = 0$ (where the indices $\mu$ and $\nu$ run over the d-dimensional space-time), and tilde is used to refer to $d$-dimensional quantities.

We will now rewrite these equations of motion using the ADM formalism. We start by writing the metric on the form
\SP{
	g_{MN} =
	\left(
	\begin{array}{ll}
	 	\tilde g_{\mu\nu} & n_\nu \\
	 	n_\mu & n_\mu n^\mu + n^2
	\end{array}
\right),
}
where comparing to the notation used in the body of the paper, we have that $n = 1+\nu$ and $n_\mu = \nu_\mu$.
The inverse metric is given by
\SP{
	g^{MN} =
	\frac{1}{n^2}
	\left(
	\begin{array}{ll}
	 	n^2 \tilde g^{\mu\nu} + n^\mu n^\nu & -n^\nu \\
	 	-n^\mu & 1
	\end{array}
\right).
}
The tangent vectors $X^M_\mu$ are given by $X^r_\mu = 0$ and $X^\nu_\mu = \delta^\nu_\mu$. We have a normal vector $N_M = (0,n)$, $N^M = n^{-1} (-n^\mu,1)$. The second fundamental form is
\SP{
	\mathcal{K}_{\mu\nu} = n \Gamma^r_{\mu\nu} = -\frac{1}{2n} (\partial_r g_{\mu\nu} - \tilde \nabla_\mu n_\nu - \tilde \nabla_\nu n_\mu),
}
where we use the notation of~\cite{BHM} (which only differs slightly from the one used in the body of this paper i.e. $K_{MN} \equiv \nabla_M N_N$), so that expressions can be easily compared. One can derive the following relations
\SP{
\label{gammarelations}
	\Gamma^\sigma_{\mu\nu} =& \tilde \Gamma^\sigma_{\mu\nu} - \frac{n^\sigma}{n} \mathcal{K}_{\mu\nu}, \\
	\Gamma^r_{\mu r} =& \frac{1}{n} \partial_\mu n + \frac{n^\nu}{n} \mathcal{K}_{\mu\nu}, \\
	\Gamma^\sigma_{\mu r} =& \tilde \nabla_\mu n^\sigma - \frac{n^\sigma}{n} \partial_\mu n - n \mathcal{K}_{\mu\nu} \left( g^{\nu\sigma} + \frac{n^\nu n^\sigma}{n^2} \right), \\
	\Gamma^r_{rr} =& \frac{1}{n} (\partial_r n + n^\nu \partial_\nu n + n^\mu n^\nu \mathcal{K}_{\mu\nu}), \\
	\Gamma^\sigma_{rr} =& \partial_r n^\sigma + n^\mu \tilde \nabla_\mu n^\sigma - n \tilde \nabla^\sigma n - 2n \mathcal{K}^\sigma_\mu n^\mu - n^\sigma \Gamma^r_{rr}.
}

Finally, we are ready to write down the expressions for the equations of motion using the quantities defined above. The equation of motion for the scalars becomes
\SP{
\label{eq:scalars}
	\Big\{ \partial_r^2 - 2n^\mu \partial_\mu \partial_r + n^2 \tilde \nabla^2 + n^\mu n^\nu \tilde \nabla_\mu \partial_\nu - (n \mathcal{K}^\mu_\mu + \partial_r \ln n - n^\mu \partial_\mu \ln n) \partial_r + &\\
	\left[ n \tilde \nabla^\mu n - \partial_r n^\mu + n^\nu \tilde \nabla_\nu n^\mu + n^\mu (n \mathcal{K}^\nu_\nu + \partial_r \ln n - n^\nu \partial_\nu \ln n) \right] \partial_\mu \Big\} \Phi^a + &\\
	\mathcal{G}^a_{\ bc} \Big[ (\partial_r \Phi^b) (\partial_r \Phi^c) - 2n^\mu (\partial_\mu \Phi^b) (\partial_r \Phi^c) + &\\ (n^2 \tilde g^{\mu\nu} + n^\mu n^\nu) (\partial_\mu \Phi^b) (\partial_\nu \Phi^c) \Big] - n^2 G^{ab} \frac{\partial V }{\partial \Phi^b} = n^2 \sum_i \sqrt{\tilde g g^{-1}} \lambda_{(i)}^a \delta_i &.
}
Einstein's equations separate into normal, mixed, and tangential components, obtained by projecting with $P^{MN} = N^M N^N - \tilde g^{\mu\nu} X^M_\mu X^N_\nu$, $P^{MN}_\mu = N^M X^N_\mu$, and $P^{MN}_{\mu\nu} = X^M_\mu X^N_\nu$, respectively. The normal component reads
\SP{
\label{eq:normal}
	(n \mathcal{K}^\mu_\nu) (n \mathcal{K}^\nu_\mu) - (n \mathcal{K}^\mu_\mu)^2 + n^2 \tilde R - 4n^2 V + 2G_{ab} \Big[ (\partial_r \Phi^a) (\partial_r \Phi^b) - &\\ 2n^\mu (\partial_\mu \Phi^a) (\partial_r \Phi^b) + (n^\mu n^\nu - n^2 \tilde g^{\mu\nu}) (\partial_\mu \Phi^a) (\partial_\nu \Phi^b) \Big] =  4 n^\mu n_\mu \sum_i \sqrt{\tilde g g^{-1}} \lambda_{(i)} \delta_i &.
}
In deriving this expression, the following relations are useful. We have that $P^{\mu\nu} = \frac{n^\mu n^\nu}{n^2} - \tilde g^{\mu\nu}$, $P^{r\mu} = - \frac{n^\mu}{n^2}$, and $P^{rr} = \frac{1}{n^2}$, so that $P^{MN} g_{MN} = 1-d$ and $P^{MN} \hat g_{MN} = \frac{n^\mu n_\mu}{n^2} - d$. Furthermore, $g^{MN} \hat g_{MN} = \frac{n^\mu n_\mu}{n^2} + d$. The mixed component is given by
\SP{
\label{eq:mixed}
	\partial_\mu (n \mathcal{K}^\nu_\nu) - \tilde \nabla_\nu (n \mathcal{K}^\nu_\mu) - n \mathcal{K}^\nu_\nu \partial_\mu \ln n + n \mathcal{K}^\nu_\mu \partial_\nu \ln n - &\\ 2G_{ab} (\partial_r \Phi^a - n^\nu \partial_\nu \Phi^a) \partial_\mu \Phi^b = - 2 n_\mu \sum_j \sqrt{\tilde g g^{-1}} \lambda_{(j)} \delta_j &.
}
and the tangential component is
\SP{
\label{eq:tangential}
	- \partial_r (n \mathcal{K}^\mu_\nu) + n^\sigma \tilde \nabla_\sigma (n \mathcal{K}^\mu_\nu) + n \mathcal{K}^\mu_\nu (n \mathcal{K}^\sigma_\sigma + \partial_r \ln n - n^\sigma \partial_\sigma \ln n) +  n \tilde \nabla^\mu \partial_\nu n + &\\ 
	n \mathcal{K}^\mu_\sigma \tilde \nabla_\nu n^\sigma - n \mathcal{K}^\sigma_\nu \tilde \nabla_\sigma n^\mu - n^2 \tilde R^\mu_\nu + 2n^2 G_{ab} (\tilde \nabla^\mu \Phi^a) (\partial_\nu \Phi^b) + \frac{4 n^2 V}{d-1} \delta^\mu_\nu = &\\ - \frac{2 \delta^\mu_\nu}{d-1} (n^2 + n^\sigma n_\sigma) \sum_l \sqrt{\tilde g g^{-1}} \lambda_{(l)} \delta_l &.
}

At zeroth order in the fluctuations, the equations of motion for the scalars are
\SP{
\label{eq:zerothorderscalar}
	\bar \Phi''^a + d A' \bar \Phi'^a + \mathcal{G}^a_{\ bc} \bar \Phi'^b \bar \Phi'^c - V^a = \sum_i \lambda_{(i)}^a \delta_i,
}
and Einstein's equations yield
\SP{
\label{eq:Einsteinfirstorder}
	d (1-d) A'^2 + 2 G_{ab} \bar \Phi'^a \bar \Phi'^b - 4V =& 0, \\
	A'' + d A'^2 + \frac{4}{d-1} V =& -\frac{2}{d-1} \sum_i \lambda_{(i)} \delta_i.
}
Writing the background as
\SP{
\label{eq:ThetasBackground}
	\bar \Phi'(r) &= (\Theta(r-r_1) - \Theta(r-r_2)) \hat{\bar \Phi}'(r), \\
	A'(r) &= (\Theta(r-r_1) - \Theta(r-r_2)) \hat A'(r),
}
we obtain the boundary conditions
\SP{
\label{eq:BCBackground}
	\bar \Phi'^a \Big|_{r_i} &= \lambda_{(i)}^a \Big|_{r_i}, \\
	A' \Big|_{r_i} &= - \frac{2}{d-1} \lambda_{(i)} \Big|_{r_i},
}
where we have dropped the hats.

\section{Linearized Equations of Motion}

Let us now expand the equations of motion in fluctuations of the metric and the scalar fields to linear order. As explained in Section~\ref{Sec:2d}, we can work in a gauge where $n^\mu = \nu^\mu = 0$. Furthermore, since we want to consider only spin-0 fluctuations, we can put $\epsilon^\mu = {h^{TT}}^\mu_\nu = 0$. This leaves us with the fluctuation variables $(\varphi, \nu, h, H)$. Thus, we expand equations \eqref{eq:scalars}, \eqref{eq:normal}, \eqref{eq:mixed}, and \eqref{eq:tangential} using the following rules:
\SP{
	\Phi^a &= \bar \Phi^a + \varphi^a, \\
	n &= 1 + \nu, \\
	h^\mu_\nu &= \frac{\delta^\mu_\nu}{d-1} h + \frac{\partial^\mu \partial_\nu}{\Box} H.
}
In doing so, we will make use of the relations
\SP{
	\sqrt{\tilde g g^{-1}} &= 1- \nu, \\
	n \mathcal K^\mu_\nu &= - A' \delta^\mu_\nu - \frac{\delta^\mu_\nu}{2(d-1)} h' - \frac{1}{2} \frac{\partial^\mu \partial_\nu}{\Box} H', \\
	R^\mu_\nu &= -\frac{\delta^\mu_\nu}{2(d-1)} e^{-2A} \Box h - \frac{d-2}{2(d-1)} e^{-2A} \partial^\mu \partial_\nu h, \\
	R &= - e^{-2A} \Box h,
}
which are true to first order in the fluctuations.
The linearized equation of motion for the scalar becomes
\SP{
\label{eq:scalarfirstorder}
	&\partial_r^2 \varphi^a + e^{-2A} \Box \varphi^a + d A' \partial_r \varphi^a + 2 \mathcal{G}^a_{\ bc} \bar \Phi'^b \partial_r \varphi^c + \partial_d \mathcal{G}^a_{\ bc} \bar \Phi'^b \bar \Phi'^c \varphi^d - \frac{\partial V^a}{\partial \bar \Phi^c} \varphi^c + \\& \bar \Phi'^a \left( \frac{d}{2(d-1)} \partial_r h + \frac{1}{2} \partial_r H -\partial_r \nu \right) -2 V^a \nu = \sum_i \delta_i (\lambda_{(i)}^a \nu + \partial_c \lambda_{(i)}^a \varphi^c).
}
At first order in the fluctuations, the normal component of Einstein's equations gives
\SP{
\label{eq:normalfirstorder}
	4 \bar \Phi'_a (\mathcal D_r \varphi^a) - 4 V_a \varphi^a -d A' \partial_r h - (d-1) \partial_r H - 8 V \nu - e^{-2A} \Box h = 0,
}
whereas the mixed component gives
\SP{
\label{eq:mixedfirstorder}
	(d-1) A' \nu -\frac{1}{2} \partial_r h - 2 \bar \Phi'_a \varphi^a = 0.
}
From the tangential component of Einstein's equations, we obtain
\SP{
\label{eq:tangentialfirstorder1}
	\frac{\partial^2_r h}{2(d-1)} + \frac{d A'}{d-1} \partial_r h + \frac{A'}{2} \partial_r H - A' \partial_r \nu + \frac{e^{-2A}}{2(d-1)} \Box h + \frac{8V}{d-1} \nu + \frac{4 V_a \varphi^a}{d-1} = &\\ - \frac{2}{d-1} \sum_i \delta_i (\lambda_{(i)} \nu + \partial_a \lambda_{(i)} \varphi^a) & ,
}
and
\SP{
\label{eq:tangentialfirstorder2}
	\frac{1}{2} \partial^2_r H + \frac{d A'}{2} \partial_r H + e^{-2A} \Box \nu + \frac{d-2}{2(d-1)} e^{-2A} \Box h = 0.
}
Equations \eqref{eq:scalarfirstorder}, \eqref{eq:tangentialfirstorder1}, and \eqref{eq:tangentialfirstorder1} lead to the boundary conditions
\SP{
\label{eq:BCscalarfluctuation}
	\varphi'^a \Big|_{r_i} &= \bar \Phi'^a \nu + \partial_c \lambda_{(i)}^a \varphi^c \Big|_{r_i},
}
\SP{
\label{eq:BCh}
	\frac{\partial_r h}{2(d-1)} - A' \nu + \frac{2}{d-1} \bar \Phi'_a \varphi^a \Big|_{r_i} &= 0,
}
and
\SP{
	\partial_r H \Big|_{r_i} &= 0. 
}
Equation \eqref{eq:BCscalarfluctuation} gives the boundary condition for the scalar fluctuations. In the special case of one scalar, this expression agrees with the one given in \cite{Kofman:2004tk}. \eqref{eq:BCh} is actually implied by equation \eqref{eq:mixedfirstorder} and therefore does not give us any new information. Finally, the boundary condition for the variable $H$ shows that we may put
\EQ{
\label{eq:BCc2}
	\mathfrak c_2 \equiv e^{-2A}\partial_\mu \nu^\mu - \frac{1}{2} \partial_r H \Big|_{r_i} = 0
}
at the boundary. The reason is that since the form of the boundary conditions must obey 4d gauge invariance, the expression $H'=0$ must generalize to $\mathfrak c_2 = 0$ had we included the $\nu^\mu$ fluctuations as well (this can be checked explicitly).

\section{Translation to Gauge Invariant Variables}

The significance of that one of the boundary conditions \eqref{eq:BCc2} reads $\mathfrak c_2 = 0$ is that using this relation, and the fact that $c_2$ can be gauged away in the bulk, one now finds a one-to-one map between the gauge-invariant variables $(\mathfrak a^a, \mathfrak b, \mathfrak c)$ (defined in \eqref{eq:gaugeinvariantvariables}) and the fluctuations $(\varphi^a, \nu, h)$. We have that
\SP{
	h &= - 2(d-1) A' e^{2A} \Box^{-1} \mathfrak c, \\
	\varphi^a &= \mathfrak a^a - \bar \Phi'^a e^{2A} \Box^{-1} \mathfrak c, \\
	\nu &= \mathfrak b - e^{2A} \Box^{-1} (2 A' \mathfrak c + \partial_r \mathfrak c).
}
Let us proceed to derive expressions for the equations of motion in the bulk and the boundary conditions in terms of $(\mathfrak a^a, \mathfrak b, \mathfrak c)$. The boundary condition for the scalar fluctuations \eqref{eq:BCscalarfluctuation} becomes
\SP{
\label{eq:BCscalarfluctuationsGI}
	\mathcal D_r \mathfrak a^a \Big|_{r_i} = \bar \Phi'^a \mathfrak b + \left( V^a - d A' \bar \Phi'^a - \lambda^a_{\ |c} \bar \Phi'^c \right) e^{2A} \Box^{-1} \mathfrak c + \lambda^a_{\ |c} \mathfrak a^c \Big|_{r_i}.
}
For the equation of motion for the scalar fluctuations in the bulk \eqref{eq:scalarfirstorder}, we obtain
\SP{
\label{eq:scalarfirstorderGI}
	\Big[ \mathcal D_r^2 + d A' \mathcal D_r + e^{-2A} \Box \Big] \mathfrak{a}^a - \Big[ V^a_{\ |c} - \mathcal{R}^a_{\ bcd} \bar \Phi'^b \bar \Phi'^d \Big] \mathfrak{a}^c - \bar \Phi'^a (\mathfrak{c} + \partial_r \mathfrak{b}) - 2 V^a \mathfrak{b} = 0,
}
whereas the linearized Einstein's equations \eqref{eq:normalfirstorder}, \eqref{eq:mixedfirstorder}, and \eqref{eq:tangentialfirstorder2} lead to
\SP{
\label{eq:normalfirstorderGI}
	\mathfrak{c} = - \frac{2}{(d-1) A'} \left(\bar \Phi'_a \mathcal D_r - \frac{4 V \bar \Phi'_a}{(d-1) A'} - V_a \right) \mathfrak a^a,
}
\SP{
\label{eq:mixedfirstorderGI}
	\mathfrak{b} = \frac{2 \bar \Phi'_a \mathfrak{a}^a}{(d-1) A'},
}
and
\SP{
	\partial_r \mathfrak c + d A' \mathfrak c - e^{-2A} \Box \mathfrak b = 0,
}
respectively (where we have used the latter two equations in deriving the first). We recognize these equations from \cite{BHM} and \cite{Elander:2009bm}. Solving for $\mathfrak b$ and $\mathfrak c$ in \eqref{eq:normalfirstorderGI} and \eqref{eq:mixedfirstorderGI}, we can now rewrite \eqref{eq:BCscalarfluctuationsGI} and \eqref{eq:scalarfirstorderGI} in terms of only the scalar fluctuations $\mathfrak a^a$. We obtain that $\mathfrak a^a$ satisfies the following equation of motion in the bulk
\SP{
	\Big[ \mathcal D_r^2 + d A' \mathcal D_r + e^{-2A} \Box \Big] \mathfrak{a}^a - \Big[ V^a_{\ |c} - \mathcal{R}^a_{\ bcd} \bar \Phi'^b \bar \Phi'^d + \frac{4 (\bar \Phi'^a V_c + V^a \bar \Phi'_c )}{(d-1) A'} + \frac{16 V \bar \Phi'^a \bar \Phi'_c}{(d-1)^2 A'^2} \Big] \mathfrak{a}^c = 0,
}
with boundary conditions
\SP{
	&\left[ \delta^a_{\ b} + e^{2A} \Box^{-1} \left( V^a - d A' \bar \Phi'^a - \lambda^a_{\ |c} \bar \Phi'^c \right) \frac{2 \bar \Phi'_b}{(d-1) A'} \right] \mathcal D_r \mathfrak a^b \Big|_{r_i} = \\&  \left[ \lambda^a_{\ |b} + \frac{2 \bar \Phi'^a \bar \Phi'_b}{(d-1) A'} + e^{2A} \Box^{-1} \frac{2}{(d-1) A'} \left( V^a - d A' \bar \Phi'^a - \lambda^a_{\ |c} \bar \Phi'^c \right) \left( \frac{4 V \bar \Phi'_b}{(d-1) A'} + V_b \right) \right] \mathfrak a^b \Big|_{r_i}.
}
In the special case where there is a superpotential $W$, these expressions can be written as
\SP{
	\Bigg[ \left( \delta^a_b \mathcal D_r + W^a_{|b} - \frac{W^a W_b}{W} - \frac{2d}{d-1} W \delta^a_b \right) \left( \delta^b_c \mathcal D_r - W^b_{|c} + \frac{W^b W_c}{W} \right) + \delta^a_c e^{-2A} \Box \Bigg] \mathfrak{a}^c = 0,
}
and
\SP{
	&\left[ \delta^a_{\ b} + e^{2A} \Box^{-1} \left( \lambda^a_{\ |c} - W^a_{\ |c} \right) \frac{W^c W_b}{W} \right] \mathcal D_r \mathfrak a^b \Big|_{r_i} = \\& \left[ \lambda^a_{\ |b} - \frac{W^a W_b}{W} + e^{2A} \Box^{-1} \left( \lambda^a_{\ |c} - W^a_{\ |c} \right) \frac{W^c W_d}{W} \left( W^d_{\ |b} - \frac{W^d W_b}{W} \right) \right] \mathfrak a^b \Big|_{r_i}.
}

%\newpage
%%%%%%%%%%%%%%%%%%%%%%%%%%%%%%%%%%%%%%%%%%%%%%%%%%%%%%%%%%%%%%%%%%%%%%
%%%  Bibliography  %%%%%%%%%%%%%%%%%%%%%%%%%%%%%%%%%%%%%%%%%%%%%%%%%%%
%%%%%%%%%%%%%%%%%%%%%%%%%%%%%%%%%%%%%%%%%%%%%%%%%%%%%%%%%%%%%%%%%%%%%%


\begin{thebibliography}{99}

\bibitem{WTC}
 B.~Holdom,
  %``Techniodor,''
  Phys.\ Lett.\ B {\bf 150}, 301 (1985);
  %%CITATION = PHLTA,B150,301;%%
    K.~Yamawaki {\it et al.}
    %, M.~Bando and K.~i.~Matumoto,
  %``Scale Invariant Technicolor Model And A Technidilaton,''
  Phys.\ Rev.\ Lett.\  {\bf 56}, 1335 (1986);
  %%CITATION = PRLTA,56,1335;%%
T.~W.~Appelquist {\it et al.}
%, D.~Karabali and L.~C.~R.~Wijewardhana,
  %``Chiral Hierarchies And The Flavor Changing Neutral Current Problem In
  %Technicolor,''
  Phys.\ Rev.\ Lett.\  {\bf 57}, 957 (1986).
  %%CITATION = PRLTA,57,957;%%


\bibitem{dilatonpheno}
W.~D.~Goldberger {\it et al.}
%, B.~Grinstein and W.~Skiba,
  %``Light scalar at LHC: the Higgs or the dilaton?,''
  Phys.\ Rev.\ Lett.\  {\bf 100}, 111802 (2008);
  %[arXiv:0708.1463 [hep-ph]].
  %%CITATION = PRLTA,100,111802;%%
  and 
  L.~Vecchi,
  %``Phenomenology of a light scalar: the dilaton,''
  arXiv:1002.1721 [hep-ph].
  %%CITATION = ARXIV:1002.1721;%%
  
  
\bibitem{dilatonWTC}
  W.~A.~Bardeen {\it et al.}
  %, C.~N.~Leung and S.~T.~Love,
  %``The Dilaton And Chiral Symmetry Breaking,''
  Phys.\ Rev.\ Lett.\  {\bf 56}, 1230 (1986);
  %%CITATION = PRLTA,56,1230;%%
    M.~Bando {\it et al.}
    %, K.~i.~Matumoto and K.~Yamawaki,
  %``TECHNIDILATON,''
  Phys.\ Lett.\  B {\bf 178}, 308 (1986);
  %%CITATION = PHLTA,B178,308;%%
  %``Scale Invariant Technicolor Model And A Technidilaton,''
  Phys.\ Rev.\ Lett.\  {\bf 56}, 1335 (1986);
  %%CITATION = PRLTA,56,1335;%%
    B.~Holdom and J.~Terning,
  %``A Light Dilaton In Gauge Theories?,''
  Phys.\ Lett.\  B {\bf 187}, 357 (1987);
  %%CITATION = PHLTA,B187,357;%%
 %   B.~Holdom and J.~Terning,
  %``NO LIGHT DILATON IN GAUGE THEORIES,''
  Phys.\ Lett.\  B {\bf 200}, 338 (1988);
  %%CITATION = PHLTA,B200,338;%%
  
\bibitem{dilatonWTCrecent}  

D.~D.~Dietrich, F.~Sannino and K.~Tuominen,
  %``Light composite Higgs from higher representations versus electroweak
  %precision measurements: Predictions for LHC,''
  Phys.\ Rev.\  D {\bf 72}, 055001 (2005)
  [arXiv:hep-ph/0505059].
  %%CITATION = PHRVA,D72,055001;%%
 T.~Appelquist and Y.~Bai,
  %``A Light Dilaton in Walking Gauge Theories,''
  arXiv:1006.4375 [hep-ph].
  %%CITATION = ARXIV:1006.4375;%%
  L.~Vecchi,
  %``Technicolor at Criticality,''
  [arXiv:1007.4573 [hep-ph]].
  %%CITATION = ARXIV:1005.4921;%%
  K.~Haba, S.~Matsuzaki and K.~Yamawaki,
  %``Holographic Techni-dilaton,''
  Phys.\ Rev.\  D {\bf 82}, 055007 (2010)
  [arXiv:1006.2526 [hep-ph]].
  %%CITATION = PHRVA,D82,055007;%%
	M.~Hashimoto and K.~Yamawaki,
  %``Techni-dilaton at Conformal Edge,''
  arXiv:1009.5482 [hep-ph].
  %%CITATION = ARXIV:1009.5482;%%

\bibitem{ENP}
D.~Elander, C.~Nunez and M.~Piai,
  %``A light scalar from walking solutions in gauge-string duality,''
  Phys.\ Lett.\  B {\bf 686}, 64 (2010)
  [arXiv:0908.2808 [hep-th]].
  %%CITATION = PHLTA,B686,64;%%

\bibitem{AdSCFTreviews}
For general reviews see for instance O.~Aharony, S.~S.~Gubser, J.~M.~Maldacena, H.~Ooguri and Y.~Oz,
  %``Large N field theories, string theory and gravity,''
  Phys.\ Rept.\  {\bf 323}, 183 (2000)
  [arXiv:hep-th/9905111];
  %%CITATION = PRPLC,323,183;%%
and
M.~Bertolini,
  %``Four Lectures On The Gauge/Gravity Correspondence,''
  Int.\ J.\ Mod.\ Phys.\  A {\bf 18}, 5647 (2003)
  [arXiv:hep-th/0303160].
  %%CITATION = IMPAE,A18,5647;%%

\bibitem{AdSCFT}
 J.~M.~Maldacena,
  %``The large N limit of superconformal field theories and 
%supergravity,''
  Adv.\ Theor.\ Math.\ Phys.\  {\bf 2}, 231 (1998)
  [Int.\ J.\ Theor.\ Phys.\  {\bf 38}, 1113 (1999)]
  [arXiv:hep-th/9711200];
  %%CITATION = IJTPB,38,1113;%%
  S.~S.~Gubser, I.~R.~Klebanov and A.~M.~Polyakov,
  %``Gauge theory correlators from non-critical string theory,''
  Phys.\ Lett.\  B {\bf 428}, 105 (1998)
  [arXiv:hep-th/9802109];
  %%CITATION = PHLTA,B428,105;%%
%\cite{Witten:1998qj}
%\bibitem{Witten:1998qj}
  E.~Witten,
  %``Anti-de Sitter space and holography,''
  Adv.\ Theor.\ Math.\ Phys.\  {\bf 2}, 253 (1998)
  [arXiv:hep-th/9802150].
  %%CITATION = 00203,2,253;%%

\bibitem{conifolds}
 P.~Candelas and X.~C.~de la Ossa,
  %``Comments on Conifolds,''
  Nucl.\ Phys.\  B {\bf 342}, 246 (1990);
  %%CITATION = NUPHA,B342,246;%%
 I.~R.~Klebanov and E.~Witten,
  %``Superconformal field theory on threebranes at a Calabi-Yau  singularity,''
  Nucl.\ Phys.\  B {\bf 536}, 199 (1998)
  [arXiv:hep-th/9807080].
  %%CITATION = NUPHA,B536,199;%%
  I.~R.~Klebanov and A.~A.~Tseytlin,
  %``Gravity Duals of Supersymmetric SU(N) x SU(N+M) Gauge Theories,''
  Nucl.\ Phys.\  B {\bf 578}, 123 (2000)
  [arXiv:hep-th/0002159].
  %%CITATION = NUPHA,B578,123;%%
 I.~R.~Klebanov and M.~J.~Strassler,
  %``Supergravity and a confining gauge theory: Duality cascades and
  %chiSB-resolution of naked singularities,''
  JHEP {\bf 0008}, 052 (2000)
  [arXiv:hep-th/0007191].
  %%CITATION = JHEPA,0008,052;%%
 J.~M.~Maldacena and C.~Nunez,
  %``Towards the large N limit of pure N = 1 super Yang Mills,''
  Phys.\ Rev.\ Lett.\  {\bf 86}, 588 (2001).
  %[arXiv:hep-th/0008001].
  %%CITATION = PRLTA,86,588;%%
  G.~Papadopoulos and A.~A.~Tseytlin,
  %``Complex geometry of conifolds and 5-brane wrapped on 2-sphere,''
  Class.\ Quant.\ Grav.\  {\bf 18}, 1333 (2001)
  [arXiv:hep-th/0012034];
  %%CITATION = CQGRD,18,1333;%%
  A.~Butti, M.~Grana, R.~Minasian, M.~Petrini and A.~Zaffaroni,
  %``The baryonic branch of Klebanov-Strassler solution: A supersymmetric
  %family of SU(3) structure backgrounds,''
  JHEP {\bf 0503}, 069 (2005)
  [arXiv:hep-th/0412187].
  %%CITATION = JHEPA,0503,069;%%
See also the recent developments in
 J.~Maldacena and D.~Martelli,
  %``The unwarped, resolved, deformed conifold: fivebranes and the baryonic
  %branch of the Klebanov-Strassler theory,''
  JHEP {\bf 1001}, 104 (2010)
  [arXiv:0906.0591 [hep-th]];
  %%CITATION = JHEPA,1001,104;%%
   J.~Gaillard, D.~Martelli, C.~Nunez and I.~Papadimitriou,
  %``The warped, resolved, deformed conifold gets flavoured,''
  arXiv:1004.4638 [hep-th];
  %%CITATION = ARXIV:1004.4638;%%
  D.~Cassani and A.~F.~Faedo,
  %``A supersymmetric consistent truncation for conifold solutions,''
  arXiv:1008.0883 [hep-th];
  %%CITATION = ARXIV:1008.0883;%%
I.~Bena, G.~Giecold, M.~Grana, N.~Halmagyi and F.~Orsi,
  %``Supersymmetric Consistent Truncations of IIB on T(1,1),''
  arXiv:1008.0983 [hep-th].
  %%CITATION = ARXIV:1008.0983;%%


\bibitem{deformingN4}
See for instance
  J.~Polchinski and M.~J.~Strassler,
  %``The string dual of a confining four-dimensional gauge theory,''
  arXiv:hep-th/0003136;
  %%CITATION = HEP-TH/0003136;%%
 K.~Pilch and N.~P.~Warner,
  %``N = 1 supersymmetric renormalization group flows from IIB supergravity,''
  Adv.\ Theor.\ Math.\ Phys.\  {\bf 4}, 627 (2002)
  [arXiv:hep-th/0006066].
  %%CITATION = 00203,4,627;%%
See also
D.~Z.~Freedman, S.~S.~Gubser, K.~Pilch and N.~P.~Warner,
  %``Renormalization group flows from holography supersymmetry and a
  %c-theorem,''
  Adv.\ Theor.\ Math.\ Phys.\  {\bf 3}, 363 (1999)
  [arXiv:hep-th/9904017].
  %%CITATION = 00203,3,363;%%
and references there in.

%\cite{Freedman:2003ax}
\bibitem{Freedman:2003ax}
  D.~Z.~Freedman, C.~Nunez, M.~Schnabl {\it et al.},
  %``Fake supergravity and domain wall stability,''
  Phys.\ Rev.\  {\bf D69}, 104027 (2004).
  [hep-th/0312055].

\bibitem{RS}
 L.~Randall and R.~Sundrum,
  %``A large mass hierarchy from a small extra dimension,''
  Phys.\ Rev.\ Lett.\  {\bf 83}, 3370 (1999)
  [arXiv:hep-ph/9905221].
  %%CITATION = PRLTA,83,3370;%%

\bibitem{GW}
W.~D.~Goldberger and M.~B.~Wise,
  %``Modulus stabilization with bulk fields,''
  Phys.\ Rev.\ Lett.\  {\bf 83}, 4922 (1999)
  [arXiv:hep-ph/9907447].
  %%CITATION = PRLTA,83,4922;%%
  
  \bibitem{GW2}
O.~DeWolfe, D.~Z.~Freedman, S.~S.~Gubser and A.~Karch,
  %``Modeling the fifth dimension with scalars and gravity,''
  Phys.\ Rev.\  D {\bf 62}, 046008 (2000)
  [arXiv:hep-th/9909134].
  %%CITATION = PHRVA,D62,046008;%%

\bibitem{RSreview}
For reviews and discussions on the subject see for instance
 N.~Arkani-Hamed, M.~Porrati and L.~Randall,
  %``Holography and phenomenology,''
  JHEP {\bf 0108}, 017 (2001)
  [arXiv:hep-th/0012148];
  %%CITATION = JHEPA,0108,017;%%
R.~Rattazzi and A.~Zaffaroni,
  %``Comments on the holographic picture of the Randall-Sundrum model,''
  JHEP {\bf 0104}, 021 (2001)
  [arXiv:hep-th/0012248].
  %%CITATION = JHEPA,0104,021;%%

\bibitem{Higgsless}
 C.~Csaki, C.~Grojean, L.~Pilo and J.~Terning,
  %``Towards a realistic model of Higgsless electroweak symmetry breaking,''
  Phys.\ Rev.\ Lett.\  {\bf 92}, 101802 (2004)
  [arXiv:hep-ph/0308038];
  %%CITATION = PRLTA,92,101802;%%
 Y.~Nomura,
  %``Higgsless theory of electroweak symmetry breaking from warped space,''
  JHEP {\bf 0311}, 050 (2003)
  [arXiv:hep-ph/0309189].
  %%CITATION = JHEPA,0311,050;%%

\bibitem{AdSQCD}
  J.~Erlich, E.~Katz, D.~T.~Son and M.~A.~Stephanov,
  %``QCD and a Holographic Model of Hadrons,''
  Phys.\ Rev.\ Lett.\  {\bf 95}, 261602 (2005)
  [arXiv:hep-ph/0501128];
  %%CITATION = PRLTA,95,261602;%%
   L.~Da Rold and A.~Pomarol,
  %``Chiral symmetry breaking from five dimensional spaces,''
  Nucl.\ Phys.\  B {\bf 721}, 79 (2005)
  [arXiv:hep-ph/0501218].
  %%CITATION = NUPHA,B721,79;%%

\bibitem{compositeHiggs}
See for instance
R.~Contino, Y.~Nomura and A.~Pomarol,
  %``Higgs as a holographic pseudo-Goldstone boson,''
  Nucl.\ Phys.\  B {\bf 671}, 148 (2003)
  [arXiv:hep-ph/0306259];
  %%CITATION = NUPHA,B671,148;%%
 K.~Agashe, R.~Contino and A.~Pomarol,
  %``The Minimal Composite Higgs Model,''
  Nucl.\ Phys.\  B {\bf 719}, 165 (2005)
  [arXiv:hep-ph/0412089].
  %%CITATION = NUPHA,B719,165;%%

\bibitem{AdSTC}
   D.~K.~Hong and H.~U.~Yee,
  %``Holographic estimate of oblique corrections for technicolor,''
  Phys.\ Rev.\  D {\bf 74}, 015011 (2006)
  [arXiv:hep-ph/0602177];
  %%CITATION = PHRVA,D74,015011;%%
 M.~Piai,
 %   ``Precision electro-weak parameters from AdS(5), localized kinetic terms and
  %anomalous dimensions,''
  arXiv:hep-ph/0608241,
  %%CITATION = HEP-PH 0608241;%%
 % M.~Piai,
  %``Walking in the third millennium,''
  arXiv:hep-ph/0609104,
  %%CITATION = HEP-PH/0609104;%%
  %``Vector mesons from AdS/TC to the LHC,''
  arXiv:0704.2205 [hep-ph];
  %%CITATION = ARXIV:0704.2205;%%
    K.~Haba, S.~Matsuzaki and K.~Yamawaki,
  %``$S$ Parameter in the Holographic Walking/Conformal Technicolor,''
  arXiv:0804.3668 [hep-ph];
  %%CITATION = ARXIV:0804.3668;%%
M.~Round,
  %``Holographic Renormalisation and the Electroweak Precision Parameters,''
  arXiv:1003.2933 [hep-ph].
  %%CITATION = ARXIV:1003.2933;%%
See also
J.~Hirn and V.~Sanz,
  %``A negative S parameter from holographic technicolor,''
  Phys.\ Rev.\ Lett.\  {\bf 97}, 121803 (2006)
  [arXiv:hep-ph/0606086],
  %%CITATION = PRLTA,97,121803;%%
%J.~Hirn and V.~Sanz,
  %``The fifth dimension as an analogue computer for strong interactions at the
  %LHC,''
  JHEP {\bf 0703}, 100 (2007)
  [arXiv:hep-ph/0612239];
  %%CITATION = JHEPA,0703,100;%%
     C.~D.~Carone, J.~Erlich and J.~A.~Tan,
  %``Holographic bosonic technicolor,''
  arXiv:hep-ph/0612242;
  %%CITATION = HEP-PH/0612242;%%
M~Fabbrichesi, M.~Piai, L.~Vecchi
arXiv:0804.0124 [hep-ph];
  %%CITATION = ARXIV:0804.0124;%%
  J.~Hirn, A.~Martin and V.~Sanz,  
  %``Describing viable technicolor scenarios,''
  arXiv:0807.2465 [hep-ph];
  %%CITATION = ARXIV:0807.2465;%%

\bibitem{super}
C.~P.~Herzog,
  %``Lectures on Holographic Superfluidity and Superconductivity,''
  J.\ Phys.\ A  {\bf 42}, 343001 (2009)
  [arXiv:0904.1975 [hep-th]].
  %%CITATION = JPAGB,A42,343001;%%

\bibitem{NPP}
 C.~Nunez {\it et al.}
  %, I.~Papadimitriou and M.~Piai,
  %``Walking Dynamics from String Duals,''
  arXiv:0812.3655 [hep-th].
  %%CITATION = ARXIV:0812.3655;%%

\bibitem{NPR}
 C.~Nunez, M.~Piai and A.~Rago,
  %``Wilson Loops in string duals of Walking and Flavored Systems,''
  arXiv:0909.0748 [hep-th].
  %%CITATION = ARXIV:0909.0748;%%

\bibitem{P}
M.~Piai,
  %``Lectures on walking technicolor, holography and gauge/gravity dualities,''
  arXiv:1004.0176 [hep-ph].
  %%CITATION = ARXIV:1004.0176;%%



\bibitem{GW3}
C.~Csaki, M.~L.~Graesser and G.~D.~Kribs,
  %``Radion dynamics and electroweak physics,''
  Phys.\ Rev.\  D {\bf 63}, 065002 (2001)
  [arXiv:hep-th/0008151].
  %%CITATION = PHRVA,D63,065002;%%

%\cite{Kofman:2004tk}
\bibitem{Kofman:2004tk}
  L.~Kofman, J.~Martin and M.~Peloso,
  %``Exact identification of the radion and its coupling to the observable
  %sector,''
  Phys.\ Rev.\  D {\bf 70}, 085015 (2004)
  [arXiv:hep-ph/0401189].
  %%CITATION = PHRVA,D70,085015;%%

\bibitem{BHM}
 M.~Bianchi, M.~Prisco and W.~Mueck,
  %``New results on holographic three-point functions,''
  JHEP {\bf 0311}, 052 (2003)
  [arXiv:hep-th/0310129];
  %%CITATION = JHEPA,0311,052;%%
  M.~Berg, M.~Haack and W.~Mueck,
  %``Bulk dynamics in confining gauge theories,''
  Nucl.\ Phys.\  B {\bf 736}, 82 (2006)
  [arXiv:hep-th/0507285];
  %%CITATION = NUPHA,B736,82;%%
    M.~Berg, M.~Haack and W.~Mueck,
  %``Glueballs vs. gluinoballs: Fluctuation spectra in non-AdS/non-CFT,''
  Nucl.\ Phys.\  B {\bf 789}, 1 (2008)
  [arXiv:hep-th/0612224].
  %%CITATION = NUPHA,B789,1;%%

\bibitem{Giovannini:2001fh}
  M.~Giovannini,
  %``Gauge-invariant fluctuations of scalar branes,''
  Phys.\ Rev.\  D {\bf 64}, 064023 (2001)
  [arXiv:hep-th/0106041].
  %%CITATION = PHRVA,D64,064023;%%

%\cite{Elander:2009bm}
\bibitem{Elander:2009bm}
  D.~Elander,
  %``Glueball Spectra of SQCD-like Theories,''
  JHEP {\bf 1003}, 114 (2010)
  [arXiv:0912.1600 [hep-th]].
  %%CITATION = JHEPA,1003,114;%%

\bibitem{HR}
K.~Skenderis,
  %``Lecture notes on holographic renormalization,''
  Class.\ Quant.\ Grav.\  {\bf 19}, 5849 (2002)
  [arXiv:hep-th/0209067];
  %%CITATION = CQGRD,19,5849;%%
  I.~Papadimitriou and K.~Skenderis,
  %``AdS / CFT correspondence and geometry,''
  arXiv:hep-th/0404176.
  %%CITATION = HEP-TH/0404176;%%

\bibitem{MS}
V.~F.~Mukhanov,
  %``Gravitational Instability Of The Universe Filled With A Scalar Field,''
  JETP Lett.\  {\bf 41} (1985) 493
  [Pisma Zh.\ Eksp.\ Teor.\ Fiz.\  {\bf 41} (1985) 402];
  %%CITATION = ZFPRA,41,402;%%
 M.~Sasaki,
  %``Large Scale Quantum Fluctuations in the Inflationary Universe,''
  Prog.\ Theor.\ Phys.\  {\bf 76}, 1036 (1986).
  %%CITATION = PTPKA,76,1036;%%


\bibitem{Gubser}
  S.~S.~Gubser, C.~P.~Herzog, S.~S.~Pufu and T.~Tesileanu,
  %``Superconductors from Superstrings,''
  Phys.\ Rev.\ Lett.\  {\bf 103}, 141601 (2009)
  [arXiv:0907.3510 [hep-th]].
  %%CITATION = PRLTA,103,141601;%%
See also
  %%CITATION = JHEPA,1006,081;%%
  D.~Cassani, G.~Dall'Agata and A.~F.~Faedo,
  %``Type IIB supergravity on squashed Sasaki-Einstein manifolds,''
  JHEP {\bf 1005} (2010) 094
  [arXiv:1003.4283 [hep-th]] and
  %%CITATION = JHEPA,1005,094;%%
  J.~P.~Gauntlett and O.~Varela,
  %``Universal Kaluza-Klein reductions of type IIB to N=4 supergravity in five
  %dimensions,''
  JHEP {\bf 1006}, 081 (2010)
  [arXiv:1003.5642 [hep-th]].

\bibitem{GPPZ}
 J.~Distler and F.~Zamora,
  %``Non-supersymmetric conformal field theories from stable anti-de Sitter
  %spaces,''
  Adv.\ Theor.\ Math.\ Phys.\  {\bf 2}, 1405 (1999)
  [arXiv:hep-th/9810206].
  %%CITATION = 00203,2,1405;%%
 L.~Girardello, M.~Petrini, M.~Porrati and A.~Zaffaroni,
  %``Novel local CFT and exact results on perturbations of N = 4 super
  %Yang-Mills from AdS dynamics,''
  JHEP {\bf 9812}, 022 (1998)
  [arXiv:hep-th/9810126].
  %%CITATION = JHEPA,9812,022;%%
  L.~Girardello, M.~Petrini, M.~Porrati and A.~Zaffaroni,
  %``Confinement and condensates without fine tuning in supergravity duals  of
  %gauge theories,''
  JHEP {\bf 9905}, 026 (1999)
  [arXiv:hep-th/9903026].
  %%CITATION = JHEPA,9905,026;%%
 L.~Girardello, M.~Petrini, M.~Porrati and A.~Zaffaroni,
  %``The supergravity dual of N = 1 super Yang-Mills theory,''
  Nucl.\ Phys.\  B {\bf 569}, 451 (2000)
  [arXiv:hep-th/9909047].
  %%CITATION = NUPHA,B569,451;%%
 K.~Pilch and N.~P.~Warner,
  %``N = 1 supersymmetric renormalization group flows from IIB supergravity,''
  Adv.\ Theor.\ Math.\ Phys.\  {\bf 4}, 627 (2002)
  [arXiv:hep-th/0006066].
  %%CITATION = 00203,4,627;%%

\bibitem{G}
S.~S.~Gubser,
  %``Curvature singularities: The good, the bad, and the naked,''
  Adv.\ Theor.\ Math.\ Phys.\  {\bf 4}, 679 (2000)
  [arXiv:hep-th/0002160];
  %%CITATION = 00203,4,679;%%
J.~M.~Maldacena and C.~Nunez,
  %``Supergravity description of field theories on curved manifolds and a no  go
  %theorem,''
  Int.\ J.\ Mod.\ Phys.\  A {\bf 16}, 822 (2001)
  [arXiv:hep-th/0007018].
  %%CITATION = IMPAE,A16,822;%%

\bibitem{GPPZspectrum}
 R.~Apreda, D.~E.~Crooks, N.~J.~Evans and M.~Petrini,
  %``Confinement, glueballs and strings from deformed AdS,''
  JHEP {\bf 0405}, 065 (2004)
  [arXiv:hep-th/0308006];
  %%CITATION = JHEPA,0405,065;%%
 W.~Mueck and M.~Prisco,
  %``Glueball scattering amplitudes from holography,''
  JHEP {\bf 0404}, 037 (2004)
  [arXiv:hep-th/0402068].
  %%CITATION = JHEPA,0404,037;%%

\bibitem{wilson}
J.~M.~Maldacena,
  %``Wilson loops in large N field theories,''
  Phys.\ Rev.\ Lett.\  {\bf 80}, 4859 (1998)
  [arXiv:hep-th/9803002].
  %%CITATION = PRLTA,80,4859;%%
S.~J.~Rey and J.~T.~Yee,
  %``Macroscopic strings as heavy quarks in large N gauge theory and  anti-de
  %Sitter supergravity,''
  Eur.\ Phys.\ J.\  C {\bf 22}, 379 (2001)
  [arXiv:hep-th/9803001].
  %%CITATION = EPHJA,C22,379;%%
J.~Sonnenschein,
  %``What does the string / gauge correspondence teach us about Wilson loops?,''
  arXiv:hep-th/0003032.
  %%CITATION = HEP-TH/0003032;%%

%\cite{Vecchi:2010dd}
\bibitem{Vecchi:2010dd}
  L.~Vecchi,
  %``Multitrace deformations, Gamow states, and Stability of AdS/CFT,''
  [arXiv:1005.4921 [hep-th]].


\end{thebibliography}
\end{document}